\documentclass[pra,aps,reprint,superscriptaddress,
a4paper,twocolumn,nofootinbib,english,longbibliography]{revtex4-1}

\usepackage{amsmath}
\usepackage{amssymb}
\usepackage{xcolor}
\usepackage{graphicx}
\usepackage{bm}

\usepackage{enumitem}
\usepackage{comment}

\usepackage{bbold} %for identity operator (changes E too)

\newcommand{\beq}{\begin{equation}}
\newcommand{\eeq}{\end{equation}}

\definecolor{mygreen}{rgb}{0.0,0.55,0.3}

\definecolor{calumblue}{RGB}{10, 62, 169}

\definecolor{calumpurple}{RGB}{100, 20, 230}

\definecolor{calumred}{RGB}{180,24,24}%{200, 30, 30}

\newcommand{\beqOpt}{\begin{equation}}
\newcommand{\eeqOpt}{\end{equation}}

\definecolor{level1}{RGB}{48, 110, 29}
\newcommand{\lvla}[1]{{\color{level1}#1}}
\definecolor{level2}{RGB}{29, 77, 124}
\newcommand{\lvlb}[1]{{\color{level2}#1}}
\definecolor{level3}{RGB}{176, 45, 25}
\newcommand{\lvlc}[1]{{\color{level3}#1}}

\newcommand{\vac}{\text{vac}}
\newcommand{\He}{H_\text{eff}}
\renewcommand{\l}{\langle}
\renewcommand{\r}{\rangle}

\newcommand{\Tr}{\text{Tr}}

\newcommand{\Ucal}{{\mathcal{U}}}

\newcommand{\I}{I}%{I(R)}
\newcommand{\IIb}{II(R)}%{II(R)}
\newcommand{\deph}{R}%{\rho^\textsc{f}}
\newcommand{\UUE}{\mathbb{U}}%{\mathcal{U}_2}%{\mathcal{U\otimes U}_E}
\newcommand{\WF}{\mathcal{W}_{\textsc f}} %{\mathbb{W}}%{\mathcal{W}_F}
\newcommand{\WFt}{\tilde{\mathcal{W}}_{\textsc f}} %{\tilde{\mathcal{W}}_F}
\newcommand{\WC}{\mathcal{W}_{\textsc c}} %{\mathbb{W}_C}%{\mathcal{W}_C}
\newcommand{\dL}{d\mathbb{L}}%{d\mathcal{L}^\textsc{f}}
\newcommand{\SJED}{S_\alpha}%{A_\alpha} %SJED
\newcommand{\cgo}{\mathcal{A}}%{\mathcal{X}} %jump operator for SJED
\newcommand{\dC}{d_{\textsc c}}
\newcommand{\dCt}{d_{\textsc c}}
\newcommand{\dS}{d_S} %{d_\textsc{s}}
\newcommand{\Yc}{\Upsilon_{\textsc c}}
\newcommand{\Yf}{\Upsilon_{\textsc f}}

\newcommand{\IE}{\mathbb{1}_{\textsc e}}
\newcommand{\UE}{U_{\textsc e}}
\newcommand{\UEt}{U_{{\textsc e},t}}
\newcommand{\UEtp}{U_{{\textsc e},t'}}
\newcommand{\superUE}{\mathcal{U}_{\textsc e}}
\newcommand{\superUEt}{\mathcal{U}_{{\textsc e},t}}

\renewcommand{\IE}{\mathbb{1}_{E}}
\renewcommand{\UE}{U_{E}}
\renewcommand{\UEt}{U_{{E},t}}
\renewcommand{\UEtp}{U_{{E},t'}}
\renewcommand{\superUE}{\mathcal{U}_{E}}
\renewcommand{\superUEt}{\mathcal{U}_{{E},t}}

\newcommand{\sytr}{\tilde}%{\tilde}{\breve} %symmetry transform notation

\newcommand{\pic}{\pi_{\textsc c}}
\newcommand{\Pic}{\bm{\pi}_{\textsc c}}

\newcommand{\Nc}{N_{\textsc c}}

\newcommand{\Btraj}{T1}
\newcommand{\Bcgtraj}{T2}
\newcommand{\Bftraj}{T3}
\newcommand{\Bcgcmps}{M1}
\newcommand{\Bdcmps}{M2}
\newcommand{\Bpdcmps}{M3}
\newcommand{\Bcmps}{M4}
\newcommand{\Bqme}{QME}

\newcommand{\rs}{\chi}%{\phi} %reset state
\newcommand{\coh}{\text{coh}}
\newcommand{\disp}{D}%{T}%{D}%{D(\{a_j\}_j)}
\newcommand{\Trans}{\mathcal{U}_T}%{\mathcal{T}_\text{trans}}%{\mathcal{T}}%{\mathcal{D}}
\newcommand{\trans}{U_T}%{T}%{\mathcal{D}}
\newcommand{\Urot}{\mathcal{U}_\text{R}} %{\mathcal{U}_\text{rot}}
\newcommand{\urot}{U_\text{R}} %{U_\text{rot}}

\newcommand{\nocontentsline}[3]{}
\newcommand{\tocless}[2]{\bgroup\let\addcontentsline=\nocontentsline#1{#2}\egroup}

\usepackage[normalem]{ulem}
\usepackage{tikz}
\usetikzlibrary{shapes.misc}

\usepackage{hyperref}
\usepackage{cleveref}
\hypersetup{
    colorlinks=true,
    linkcolor=black,
    citecolor=blue,
}

\makeatletter
\def\l@subsubsection#1#2{}
\makeatother

\begin{document}
 
\title{Weak unitary symmetries of open quantum dynamics: \\ beyond quantum master equations}
\author{Calum A. Brown}
\affiliation{Department of Applied Mathematics and Theoretical Physics, University of Cambridge, Wilberforce Road, Cambridge CB3 0WA, United Kingdom}
\author{Robert L. Jack}
\affiliation{Department of Applied Mathematics and Theoretical Physics, University of Cambridge, Wilberforce Road, Cambridge CB3 0WA, United Kingdom}
\affiliation{Yusuf Hamied Department of Chemistry, University of Cambridge, Lensfield Road, Cambridge CB2 1EW, United Kingdom}
\author{Katarzyna Macieszczak}
\affiliation{Department of Physics, University of Warwick, Coventry CV4 7AL, United Kingdom}

\begin{abstract} 
We consider Markovian open quantum dynamics with weak unitary symmetries.  Starting from the quantum master equation for the system alone, it is known that the joint dynamics of the system and its environment can be obtained by dilation, leading to a closed dynamics for a continuous matrix product state. Performing counting measurements on the environment gives rise to stochastic dynamics of quantum trajectories for the system, which when averaged yield back the quantum master equation.  In this work, we identify necessary and sufficient conditions under which the dynamics of these different descriptions retain the weak symmetry of the quantum master equation and we characterise the resulting symmetries of the different descriptions in terms of their generators. We find that the joint dynamics always features a separable symmetry directly related to that of the quantum master equation, but for quantum trajectories the corresponding symmetry is present only if the counting measurement satisfies certain conditions.  
%We finish by discussing implications of these symmetries for dynamical simulations.
\end{abstract}

\maketitle

\hypersetup{
    colorlinks=true,
    linkcolor=magenta,
    citecolor=blue,
}

\section{Introduction}

The importance of symmetries and their associated conservation laws for closed quantum systems is well established. They provide physical insight, reduce the degrees of freedom in high-dimensional problems (due to conservation laws), and supply spectral information about the equation of motion for the system, which in turn simplifies its solutions. Understanding symmetries is therefore crucial in the development of practical quantum  technologies. However, experimental quantum systems are never fully isolated from their environments, so the study of symmetries in open quantum systems \cite{Breuer_and_Petruccione,Gardiner2004} is of vital importance for quantum technology platforms.  Noether's theorem does not apply to such systems: they exhibit  (so-called) weak symmetries, which are not associated with any conservation law~\cite{Buca2012,Albert2014}. 
Such symmetries play a vital role in simplification of master operators~\cite{Albert2014,Macieszczak2021,McDonald2022}, suppression of decoherence 
\cite{Santos2020}, and engineering of steady states \cite{Baumgartner2008a,Albert2014}, %,Wei23}, 
among others.
They are also of fundamental importance in resource theories of asymmetry and coherence in quantum information~\cite{Streltsov2017,Chitambar2019,Baumgratz2014,Yadin2016,Chitambar2016}.  

A canonical framework for the dynamics of open quantum system is the (Markovian) quantum master equation (QME) \cite{Lindblad1976,Gorini1976}.  This is the most general Markovian evolution for a system density matrix that respects the physical constraints of complete positivity and trace-preservation. There has been significant recent progress in understanding symmetry properties of these dynamics, including unitary symmetries~\cite{Buca2012,Albert2014,Minganti2018,Lieu_breaking20,Altland21} and beyond \cite{roberts21,lieu_quad20,gneiting22,Sa2023,Paszko2024}.  However, the QME only determines the dynamics of the system itself, and all information about its environment is discarded.
In this work, we go beyond this limitation and clarify the presence of unitary symmetries
of {more informative} dynamical descriptions, where environmental information is retained.

A complete description of the joint dynamics {of the system and its environment can be obtained by modelling the environment as a set of bosonic fields.  One obtains a {pure} state that describes both the system and its environment: it is represented as a continuous matrix product state (cMPS) \cite{Verstraete2010,Osborne2010}, {which evolves with a so-called stochastic Hamiltonian}~\cite{Hudson1984,Parthasarathy1992}.
We show here that the weak symmetry of the QME actually stems from a separable symmetry of the joint dynamics, which appears as a symmetry of the stochastic Hamiltonian. This result is an extension of Stinespring's theorem for symmetric quantum channels~\cite{Stinespring,Keyl1999,Marvian2012} from discrete-time to continuous-time dynamics.

Now suppose that the system is monitored by counting quanta in the environment~\cite{Gardiner2004,Wiseman2010}.
{This} effectuates dephasing of the cMPS in the measurement basis, leading to a mixed joint state, which is described by a continuous matrix product operator (cMPO).  The evolution of system states conditioned on these (random) measurement {outcomes are} described by stochastic quantum trajectories~\cite{Breuer_and_Petruccione,Daley2014}: these are encoded in the dephased cMPOs, together with the associated measurement outcomes.  Quantum trajectory dynamics are therefore an intermediate level between the density matrix of the system alone, and the full cMPO description of the system and environment together. 
This work explains that counting measurements may break the symmetry of the closed joint dynamics; this also depends on how detailed are the measurements, which effectuates either full- or partial-dephasing of the cMPO.
We determine the conditions under which the symmetry
% of the QME is present in the continuously-monitored dynamics,
remains under the continuous monitoring,
according to the level of detail of the measurements, and the associated (quantum or classical) information.

These results are of fundamental interest as characterisations of symmetric quantum systems.  In addition, quantum trajectories -- and other stochastic descriptions of quantum open systems -- are also of practical importance.  This is due in part to the existence of efficient numerical simulation methods \cite{Dum1992,Dalibard1992,Molmer93,Carmichael1993,Plenio1998,Daley2014}, 
as well as recent experimental progress in realising continuously-monitored open systems \cite{Gleyzes2007,Guerlin2007,Deleglise2008,Murch2013,Hofmann2016,Kurzmann2019,minev19}, such that individual stochastic trajectories are accessible in experiments.  This allows investigation of phenomena beyond the reach of the QME \cite{Chan2019,Skinner2019,Turkeshi2021,LeGal24,Garrahan10,Lesanovsky13,cabot23,Brown24}, and higher experimental control can be achieved in quantum processing by exploiting feedback \cite{abdelhafez19,propson21,grigoletto22,herasymenko23}; {including quantum error correction \cite{Li2017,Temme2017}, as well  advanced quantum state preparation using conditional states \cite{Choi2023,Cotler2023,Ippoliti2023}. Finally, we note that while properties of quantum trajectories  have been investigated at length for system with strong symmetries~\cite{Munoz19,Buca2019,Tindall_2020,Zhang_2020}, this work addresses the largely unexplored realm of weak symmetries, beyond the QME description.

The following Section outlines the scope and main results of this work, after which we embark on the detailed analysis.

\section{Outline}
\label{sec:outline}

\begin{figure*}
    \centering
    \includegraphics[width=0.88\linewidth]{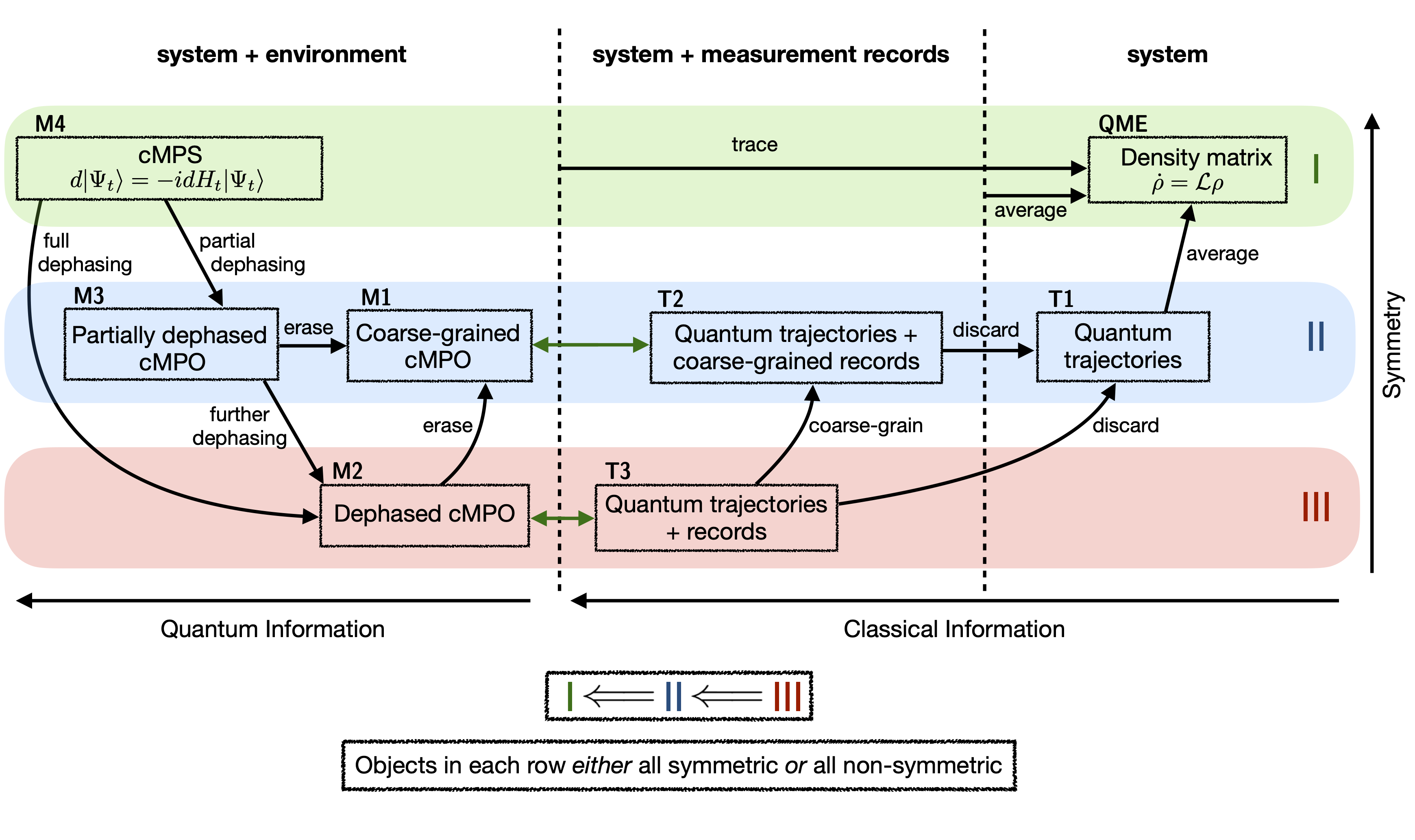}
    \vspace{-8pt}  %% there is some whitespace at bottom of figure so we avoid excessive spacing
    \caption{\textbf{Outline of main results:}  An open quantum system can be described in different ways, for example as a joint cMPS for the system and its environment (top left), or as a density matrix {for the system alone} (top right).  When performing measurements {on the environment},  the resulting dynamics of different levels of description may be symmetric or non-symmetric, depending on the measurement choice.  Each row of the diagram corresponds to distinct conditions: Symmetry Condition I is the least restrictive while Symmetry Condition III is the most.  If the relevant condition is fulfilled then all objects in the row undergo symmetric dynamics. Green arrows indicate a one-to-one correspondence between objects, while black arrows indicate information being reduced, either by losing quantum information in the joint state or neglecting classical information obtained from measurements. Loss of quantum information may break the symmetry, while discarding classical information may restore it.
    Boxes have alphanumeric labels with T referring to quantum trajectories, and M referring to cMPOs; larger numbers indicate more information content.
    }
    \label{fig:outline}
\end{figure*}

As already noted in the Introduction, open quantum {dynamics} can be described at different levels of detail, according to  what measurements are made, and how the environment is described.
{In this work, we show how} symmetries of the dynamics manifest differently according to the level of description.  Fig.~\ref{fig:outline} outlines the various objects of interest, the connections between them, and their symmetries.  This Section highlights the salient features of Fig.~\ref{fig:outline} {and presents the symmetry conditions, which are the main results of this work}. Full definitions of all objects are given in later sections, as well as more detailed results, and further discussion of the connections between the different objects.  [The labelling of the various objects (as M4, T3, etc) is described in the caption.]
Some relevant equations from this outline are repeated in Secs.~\ref{sec:qme_symm}-\ref{sec:cMPS}, to make their discussions self-contained.

\subsection{Descriptions of open quantum {dynamics}}

The three columns of Fig.~\ref{fig:outline} indicate descriptions of the system and its environment at different levels of detail: the left column (system + environment) considers the joint Hilbert space of system and environment; the right column considers states in the system's Hilbert space; the central column (system + measurement records) considers system {states} supplemented by records of measurements performed on the environment.
To illustrate this, we  highlight two objects which appear in the right and left columns.  The system \emph{density matrix} $\rho_t$ at time $t$ evolves deterministically according to the {\emph{quantum master equation} (QME)}
\begin{equation}\label{eq:rho_dot}
    %dot\rho = {\cal L}\rho
    \frac{d}{dt}\rho_t= \mathcal{L}( \rho_t),
\end{equation}
where ${\cal L}$ is the \emph{quantum master operator} that acts as a superoperator on the system space {(see Sec.~\ref{sec:qme_symm})}.  By contrast, a complete description of the system and environment is {a pure joint state, denoted by $|\Psi_t\rangle$.  It evolves unitarily according to %the stochastic equation
\begin{equation}\label{eq:Psi_dot}
     d |\Psi_t\rangle = -i  d H_t|\Psi_t\rangle,
\end{equation}
where $d |\Psi_t\rangle$ is the (small change) in $|\Psi_t\rangle$ between times $t$ and $t+dt$, and $ d H_t$ is the \emph{stochastic Hamiltonian} (see Sec.~\ref{sec:cMPS}).
The state $|\Psi_t\rangle$ may be represented as a continuous matrix product state (cMPS).

The rows 
of Fig.~\ref{fig:outline} classify objects according to their amount of environmental/measured information.
There is no such information in the top row: the QME describes the evolution of the system alone; the cMPS in box \Bcmps\ evolves unitarily, which reflects that no measurements are being made.
Other rows of Fig.~\ref{fig:outline} correspond to (detailed or coarse-grained) {continuous} measurements on the environment, whose outcomes may be summarised in \emph{measurement records}, further discussed in Sec.~\ref{sec:traj}.  The effects of measurements may also enter through their back-action on the system state (in quantum trajectories) or through (full or partial) dephasing {of} the cMPS.

In this work, we denote the \emph{conditional state} {of the system} for a measurement record up to time $t$ by $\psi_t$, and we refer to the path followed by the conditional state over a time interval $[0,t)$ as a \emph{quantum trajectory}. This distinction between measurement records and quantum trajectories is important, because these trajectories can be viewed as sample paths of a {(classical)} piecewise-deterministic Markov process over the {(quantum)} Hilbert space of {the system} \cite{Breuer_and_Petruccione}. {We  refer this process  as the \emph{unravelled quantum dynamics}}.  The corresponding probability density for $\psi_t$ is denoted by $P_t(\psi)$; it evolves {deterministically} by an equation similar to the Fokker-Planck equation,
\begin{equation}\label{eq:P_dot-intro}
%	\frac{\partial}{\partial t}{P}_t(\psi) = {\cal W}^\dag (P_t)(\psi),
	\frac{\partial}{\partial t}{P}_t = {\cal W}^\dag (P_t),
\end{equation}
where ${\cal W}^\dag$ is the generator of the unravelled quantum dynamics (see Sec.~\ref{sec:traj}).
This description (box \Btraj) {does not include any measurement records and} is found in the right column of Fig.~\ref{fig:outline}. The {combined evolutions} of conditional states and measurement records can also be captured by stochastic processes, and belong in the central column (boxes \Bftraj\ and \Bcgtraj), the difference between full and coarse-grained records is explained in Sec.~\ref{sec:traj}.

Measurement {outcomes} can also be analysed by following the joint (quantum) state of the system and the environment.  The set of all possible outcomes is described (together with their probabilities) by a mixed state in the joint Hilbert space of system and environment.  This object is described by a \emph{continuous matrix product operator} (cMPO) which we denote by $R_t$. Such objects belong in the left column of Fig.~\ref{fig:outline}, and evolve deterministically as
\begin{equation}\label{eq:dL_intro}
	dR_t = d\mathbb{L}_t (R_t),
\end{equation}
where  $d\mathbb{L}_t$ is a superoperator on the joint Hilbert space (see Sec.~\ref{sec:cMPS}).
The cMPO $R_t$ may also be obtained from the pure cMPS (box~\Bcmps) by a projective measurement which, as usual in quantum measurement theory, leads to its dephasing in the measurement basis (box~\Bdcmps).  Depending on the amount of measured information, some coherence may be retained (partial dephasing, box~\Bpdcmps).  See  Sec.~\ref{sec:cMPS} for a full discussion of the different dephasing and measurement operations that lead to boxes \Bcgcmps-\Bdcmps.

The arrows in Fig.~\ref{fig:outline} indicate the relationships between different objects.  Specifically, black arrows indicate reduction of {(quantum or classical) information}.  Quantum information is lost by dephasing, while classical information is reduced by discarding/averaging measurement outcomes, or, in both cases, by tracing out the environment.  The cMPS is the most {informative} description while the density matrix of the system is the least. By contrast, the green arrows  in Fig.~\ref{fig:outline} indicate that descriptions in terms of cMPO or quantum trajectories contain equivalent information, see Sec.~\ref{sec:discussion1} for a detailed discussion of this point.

\subsection{Symmetries of different descriptions}

This work presents two sets of main results.
We summarise here the first of these sets.
Having outlined several different descriptions of open quantum dynamics, we explain how their various symmetries are defined.

We start with a {\emph{weak unitary symmetry} of the quantum master dynamics}. This means that for some unitary operator $U$ we have
\begin{equation}\label{eq:symmL}
\mathcal{ULU}^\dag = \cal L,
\end{equation}
where the unitary superoperator ${\cal U}$ acts on density matrices $\rho$ as ${\cal U} (\rho)=U\rho U^\dag$, cf.~Refs.~\cite{Buca2012,Albert2014} and see Sec.~\ref{sec:qme_symm}.
We show in this paper that this symmetry is equivalent to a \emph{separable unitary symmetry} of the joint dynamics of the system and environment, specifically  
\begin{equation}\label{eq:symmdH}
    (U\otimes U_{E})dH_t'(U\otimes U_{E})^\dag = dH_t' \;,
\end{equation}
where the environmental symmetry operator $U_{E}$ and the stochastic Hamiltonian $d H_t'$ are both defined in Sec.~\ref{sec:cMPS}.  (The dynamics is symmetric when viewed in an appropriate reference frame; this enters via the definition of $d H_t'$.)

We also explain that a symmetry of the {unravelled quantum dynamics for conditional states of the system} corresponds to
\begin{equation}\label{eq:symmW}
    \Upsilon\mathcal{W}^\dag\Upsilon^\dag=\mathcal{W}^\dag
\end{equation}
for a suitable unitary {operation} $\Upsilon$, defined in Sec.~\ref{sec:traj}.  This operation is determined by the unitary super-operator ${\cal U}$ that acts on system conditional states.  Similarly, 
a symmetry of cMPO dynamics is
\begin{equation}\label{eq:symm_dL}
    \mathbb{U} \, d\mathbb{L}_t \,\mathbb{U}^\dag = d\mathbb{L}_t,
\end{equation}
{where} the unitary superoperator $\mathbb{U}=\mathcal{U}\otimes\mathcal{U}_E$ acts on the joint Hilbert space of the system and the environment and is defined in terms of $U$ and $\UE$, see Sec.~\ref{sec:cMPS}.

In contrast to the cMPS symmetry in Eq.~\eqref{eq:symmdH}, the symmetries in Eqs.~(\ref{eq:symmW},\ref{eq:symm_dL}) do not follow directly from Eq.~\eqref{eq:symmL}, this is discussed in the following Sec.~\ref{sec:hier}.
Note also that Eq.~\eqref{eq:symmdH} is a symmetry of closed (unitary) dynamics, so there is an associated conservation law.  However, Eqs.~(\ref{eq:symmW},\ref{eq:symm_dL}) do not prescribe any conserved quantity: we interpret them as weak symmetries in analogy with Eq.~\eqref{eq:symmL}.

{All these symmetries imply that any initially symmetric description remains symmetric at all times. For example,} 
 an initial density matrix {$\rho_0$} is symmetric if it is an eigenmatrix of ${\cal U}$ with eigenvalue $1$.  Eq.~\eqref{eq:symmL} implies a commutation relation $[{\cal U},{\cal L}]=0$, so evolution by Eq.~\eqref{eq:rho_dot} means that $\rho_t$ {stays symmetric} as an eigenmatrix with the same eigenvalue.  
Symmetries of other dynamical operators (such as {$dH_t$, ${\cal W}^\dag$, or $d\mathbb{L}_t$}) imply similar results for the evolution of the relevant objects [such as {$|\Psi_t\rangle$},  $P_t(\psi)$, or $R_t$]. 
%{This aspect of symmetric evolution is helpful for efficient numerical simulation, see Sec.~\ref{sec:implications}.}

\subsection{Hierarchy of Symmetry Conditions}
\label{sec:hier}

This Section summarises the second set of results of this work: we present necessary and sufficient conditions under which the various objects in Fig.~\ref{fig:outline} are symmetric.  The central physical insight behind these conditions is 
that performing measurements on the environment can break symmetries of the joint dynamics, but these symmetries may also be restored by neglecting or discarding some {of the measured} classical information.  
 This possibility is reflected in Fig.~\ref{fig:outline} by objects in the same row being either all symmetric (symmetry preserved) or all non-symmetric (symmetry broken).  Each row forms a symmetry class, labelled I, II, III.  
 
Since different choices of measurement correspond to different representations (unravellings) of the quantum master operator, those determine the presence or the absence of symmetries.  This means that each symmetry class is associated with a Symmetry Condition that applies to the representation.
These conditions are increasingly restrictive and form a hierarchy: if a representation obeys Symmetry Condition III then it also obeys Symmetry Conditions II and I; similarly Symmetry Condition II implies Symmetry Condition~I, but the reverse implications do not hold.
{For any given representation, this hierarchy means that the symmetry of  can only decrease if the quantum information is reduced, see Fig.~\ref{fig:outline}.  Similarly, discarding classical information can only increase the symmetry.}

To formulate the conditions, we recall~\cite{Lindblad1976,Gorini1976} that
\begin{equation}
\label{eq:LL-outline}
        \mathcal{L}(\rho) = -i[H,\rho] + \sum_{j=1}^d \left(J_j\rho J_j^\dag-\frac{1}{2}\{J_j^\dag J_j , \rho\}\right),
\end{equation}
{where} $H$ is the \emph{system  Hamiltonian} and $J_1,J_2,\dots, J_d$
% $\{J_j\}_{j}$
are \emph{jump operators}.  Together, these form a \emph{representation} {(with $d$ jump operators)} of the quantum master operator $\mathcal{L}$, {which we will also refer as a representation of the QME}.  

The Symmetry Conditions can then be summarised as:
\begin{enumerate}[label=\textcolor{red}{\arabic*.}, ref=\arabic*]
	\item[(I)] \label{Cond:I} \emph{The weak symmetry of the master operator is present, i.e., Eq.~\eqref{eq:symmL} holds.} 
	The equivalent condition to be satisfied by any representation of $\mathcal{L}$ is given in Sec.~\ref{sec:qme_symm}. This condition holds if and only if Eq.~\eqref{eq:symmdH} holds. 
	
	\item[(II)] \label{Cond:II} \emph{The Hamiltonian is symmetric and the action of the symmetry on the jump operators gives rise to a permutation $\pic$ of the composite action operators $\cgo_\alpha$  defined in Sec.~\ref{sec:traj},} 
		$${\cal U}(H)=H , \qquad {\cal U} \cgo_\alpha {\cal U}^\dag= \cgo_{\pic(\alpha)}  \quad \forall{\alpha}.$$     
    The equivalent condition to be satisfied by a representation of $\mathcal{L}$ is given in Sec.~\ref{sec:traj}.  This representation-dependent condition hold if and only if Eq.~\eqref{eq:symmW} holds. 

    All super-operators ${\cal A}_\alpha$ have the property that if $\psi$ is pure then so is ${\cal A}_\alpha(\psi)$.
    
	\item[(III)] \label{Cond:III} \emph{The Hamiltonian is symmetric and the action of the symmetry on the jump operators gives rise to their permutation $\pi$ up to a real phase,}
	$${\cal U}(H)=H , \qquad {\cal U}(J_j)=J_{\pi(j)}{\rm e}^{i\phi_j} \quad \forall{j}.$$ 
    This representation-dependent condition holds if and only if Eq.~\eqref{eq:symm_dL} holds for the dynamics of the fully-dephased cMPO.  

Representations that satisfy Symmetry Condition III with $\pi(j)=j$ are known as \emph{weakly symmetric representations}~\cite{Macieszczak2021}. 

\end{enumerate}

Physically,
Symmetry Condition~I ensures the {separable} symmetry of the joint dynamics of the system and environment. Symmetry Condition III corresponds to the situation where all environmental measurements respect that symmetry.  Symmetry Condition II corresponds to an intermediate situation where trajectories of the conditional state remain symmetric even though the measurements may break that symmetry, see Sec.~\ref{sec:traj} for a detailed discussion.

\subsection{Outline of the paper}

In the remainder of the paper,  Sec.~\ref{sec:qme_symm} reviews {weak symmetries of quantum master dynamics and corresponding properties of quantum master operator representations}, Sec.~\ref{sec:traj} discusses symmetries {in the dynamics} of conditional states and associated measurement records, while Sec.~\ref{sec:cMPS} discusses symmetries in the dynamics of the cMPS and cMPOs. We {relate these results} in Sec.~\ref{sec:discussion1} and we provide illustrative examples in Sec.~\ref{sec:examples}.
Sec.~\ref{sec:implications} describes consequences of these symmetries, including for eigenfunctions of generators for the unravelled quantum dynamics, the structure of the Liouville and Choi matrices for the quantum master operators.  We conclude in Sec.~\ref{sec:outlook}, with an outlook.

\section{Weak symmetries of quantum master dynamics}\label{sec:qme_symm}

This Section reviews
open quantum systems' dynamics, {as} governed by quantum master equations with weak unitary symmetries.   We present Symmetry Condition~I, which
for a given representation determines the presence of a weak symmetry of the corresponding quantum master operator.

\subsection{Quantum master dynamics}

Throughout this work, we consider systems with a finite-dimensional Hilbert spaces, denoted $\mathcal{H}$.  The system density matrix $\rho$ evolves deterministically
{according to} the quantum master equation (QME)~\cite{Lindblad1976,Gorini1976,davies1974}
\begin{equation}\label{eq:QME}
    \frac{d}{dt}\rho_t
   =\mathcal{L}(\rho_t),
\end{equation}
with ${\cal L}$ as in Eq.~\eqref{eq:LL-outline}.
For a given initial state $\rho_0$, the evolution is given formally by 
\begin{equation}\label{eq:rho_t}
\rho_t = e^{t\cal L}(\rho_0).
\end{equation}

In the following, we often consider \emph{paths} followed by the system state between times $0$ and $t$.\footnote{%
We use the word \emph{path} for any kind of curve parametrised by time.  A \emph{quantum trajectory} will be a stochastic path followed by a conditional state $\psi_t$.}  
For an initial state $\rho_0$ the deterministic path followed by the density matrix up time $t$ is
\begin{equation}\label{eq:path-rho}
\rho_{[0,t)}=(  \rho_\tau )_{\tau\in[0,t)} \; .
\end{equation}
Such smooth paths arise as solutions of the QME, which is a differential equation for the deterministic evolution.  This may be contrasted with stochastic paths, e.g., quantum trajectories, which generically feature discontinuities, due to stochastic transitions.}

\subsection{Weak symmetries of quantum master dynamics}

 Consider a unitary operator $U$ on $\mathcal{H}$
  and define a corresponding linear superoperator that acts on  matrices $\rho$ as
\begin{equation}
\mathcal{U}(\rho)=U \rho U^\dag.
\end{equation}
The {quantum master dynamics} features a weak unitary symmetry~\cite{Baumgartner2008a,Buca2012,Albert2014} if 
\begin{equation}\label{eq:weakU_QME}
    \mathcal{ULU}^\dag = \mathcal{L}
\end{equation}
or, equivalently, $\mathcal{UL} = \mathcal{LU}$. %$[\mathcal{U},\mathcal{L}]=0$.
This gives rise to a symmetry group because Eq.~\eqref{eq:weakU_QME} also holds with ${\cal U}$ replaced by ${\cal U}^n$ for any integer $n$. If $U^{N}=\mathbb{1}$ for a positive integer $N$, the group is finite and its order is the smallest such $N$}.

A recurring theme in this work is that symmetries can be expressed either in terms of generators {of open quantum dynamics}, or by equivalent statements for paths.  In the current context, {the weak symmetry in Eq.~\eqref{eq:weakU_QME} is defined in terms of the quantum master operator, and this is equivalent to 
    $\mathcal{U}e^{t\mathcal{L}} 
=e^{t\mathcal{L}}\mathcal{U}$, recall Eq.~\eqref{eq:rho_t}.
Therefore,} the weak symmetry  implies that the symmetry-transformed path 
\begin{equation}
\sytr\rho_{[0,t)}=[  {\cal U}(\rho_\tau) ]_{\tau\in[0,t)} \; ,
\label{eq:tilde-path-rho}
\end{equation}
solves the QME with initial condition ${\cal U}(\rho_0)$, cf.~Eq.~\eqref{eq:path-rho}.
Moreover, if this property holds for all paths then Eq.~\eqref{eq:weakU_QME} also holds.  Hence the symmetry can be characterised either in terms of the generator $\cal L$ or in terms of the paths $\rho_{[0,t)}$ and $\sytr\rho_{[0,t)}$.
The tilde in Eq.~\eqref{eq:tilde-path-rho} indicates that $\sytr\rho_{[0,t)}$ is the symmetry-transformed counterpart of $\rho_{[0,t)}$; we continue to use this convention in later Sections.

{When the initial state is symmetric, $\sytr\rho_0 = \rho_0$,  the symmetry-transformed path coincides with the original one, $\sytr\rho_{[0,t)}=\rho_{[0,t)}$. Therefore, an initially-symmetric state remains symmetric for all time,} 
\begin{equation}	\label{eq:rho-symm}
	{\cal U}(\rho_0)=\rho_0 \quad \Rightarrow \quad	{\cal U}(\rho_t)=\rho_t.
\end{equation}
{In particular, if the system state is asymptotically unique}~\cite{Davies1970,Evans1977,Schirmer2010,Nigro2019}, it must be symmetric, ${\cal U}(\rho_\infty)=\rho_\infty$. 
As already mentioned in Sec.~\ref{sec:outline}, these last results can also be obtained by analysing the spectral properties of $\cal L$ in the presence of the weak symmetry in Eqs.~\eqref{eq:weakU_QME}. 

\subsection{Symmetry Condition I}

There are different choices of Hamiltonian and jump operators that lead to the same QME in Eq.~\eqref{eq:QME}.  These are called   representations of the QME.  Our hierarchy of symmetry conditions is framed in terms of these representations.
Starting from the representation of ${\cal L}$ in Eq.~\eqref{eq:LL-outline},
note that the corresponding 
symmetry-transformed master operator $\mathcal{U}\mathcal{L}\mathcal{U}^\dag$ has the same structure, except that one replaces
\begin{equation}
	H \mapsto {\cal U}(H), \qquad J_j \mapsto {\cal U}(J_j) \quad\forall j\;.
	\label{eq:UJH0}
\end{equation}
That is, 
 $\mathcal{U}(H), %\{\mathcal{U}(J_j)\}_{j=1}^d$
\,\mathcal{U}(J_1),\dots, \mathcal{U}(J_d)$
is a representation of $\mathcal{U}\mathcal{L}\mathcal{U}^\dag$. Hence, the weak symmetry holds if and only if $\mathcal{U}(H),\, \mathcal{U}(J_1),\dots \mathcal{U}(J_d)$ %$\{\mathcal{U}(J_j)\}_{j=1}^d$
is also a representation of $\mathcal{L}$, cf.~Eq.~\eqref{eq:weakU_QME}.  
We interpret this statement as an implicit condition on $H,J_1,\dots,J_d$, which can be made explicit using standard results from Ref.~\cite{Wolf2012}, as follows. 

For any representation   $H, \,J_1,\dots, J_d$, %\{J_j\}_{j=1}^d$,
the corresponding \emph{traceless representation}  is $H',J_1',\dots,J_d'$ %$\{J'_j\}_{j=1}^d$
with
\begin{subequations}\label{eq:traceless_rep}
	\begin{align}\label{eq:traceless_rep_a}
		H' & = H + \frac{i}{2\dS}\sum_{j=1}^d\left[J_j\Tr(J_j^\dag)-J_j^\dag\Tr(J_j)\right],
		\\
%	\end{equation}
%	\begin{equation}
	\label{eq:traceless_rep_b}
		J_j' & = J_j - \frac{\mathbb{1}}{\dS}\Tr(J_j) \quad \forall{j},
	\end{align}
\end{subequations}
where $\dS = {\rm dim}(\mathcal{H})$. 
Then, Proposition 7.4 of~\cite{Wolf2012} yields:
\par\smallskip \noindent $\bullet\; $
\textbf{Theorem~I} 
\par\noindent
\emph{For any two representations of a given master operator, $H,J_1\dots J_d$ %\{J_k\}_{k=1}^d$
and $\tilde H,
\tilde{J}_1,\dots \tilde{J}_{\tilde{d}}$
with $ \tilde d\geq d$, 	there exists an isometric $ \tilde d\times  d$ matrix $\mathbf{V}$ such that}
\begin{subequations}\label{eq:theorem_I}
	\begin{align}
{\tilde H'}  &= {H'},\\ 
{\tilde J_{j}'} &= \sum_{k=1}^{d} \mathbf{V}_{jk} \,{J'_{k}}  \quad \forall{j}.
\end{align}
\end{subequations}\\
(An isometric matrix $\mathbf{V}$ obeys $\mathbf{V}^\dag \mathbf{V}=\mathbb{1}$; square isometric matrices are unitary.)  We emphasise that $H,J_1\dots J_d$ %\{J_k\}_{k=1}^d$
and $\tilde H,
\tilde{J}_1,\dots \tilde{J}_{\tilde{d}}$ are generic representations of the same master operator, no symmetry is assumed.

With these results in hand, recall that weak unitary
symmetry of ${\cal L}$ requires that 
 $H, J_1,\dots, J_d$ and $\mathcal{U}(H),\mathcal{U}(J_1),\dots ,\mathcal{U}(J_d)$ are representations of the same ${\cal L}$.  
 Noting that $\mathcal{U}(H')=[\mathcal{U}(H)]'$ and $\mathcal{U}(J_j')=[\mathcal{U}(J_j)]'$ for all $j$, and also that 
 both representations have $d$ jump operators,
 Theorem I then implies~\cite{Macieszczak2021}:\\
\par\smallskip \noindent $\bullet\; $
 \textbf{Symmetry Condition~I}
\\ %\par\noindent
\emph{For any representation $H, J_1,\dots, J_d$ %\{J_j\}_{j=1}^d$
of $\mathcal{L}$, the weak symmetry $\mathcal{U}\mathcal{L}\mathcal{U}^\dag=\mathcal{L}$ holds if and only if
\begin{subequations}\label{eq:symm_cond_I}
	\begin{equation}\label{eq:symm_cond_Ia}
\mathcal{U}(H') = H'
\end{equation}
and there exists a unitary $d\times d$ matrix $\mathbf{U}$ such that
\begin{equation}
\label{eq:symm_cond_Ib}
 \mathcal{U}(J_j') = \sum_{k=1}^{d} \mathbf{U}_{jk}J_k' \quad \forall{j}.
\end{equation}
\end{subequations}
}%
The matrix $\mathbf{U}$ in Eq.~\eqref{eq:symm_cond_Ib} is not {generally} unique,
Appendix~\ref{app:boldU} discusses the freedoms in its construction.  

Recalling Sec.~\ref{sec:hier}, Symmetry Condition~I is the least restrictive symmetry condition, corresponding to the first row in Fig.~\ref{fig:outline}.  It is important that  the weak symmetry {in Eq.~\eqref{eq:weakU_QME} only involves ${\cal L}$, so Symmetry Condition \I\ in Eq.~\eqref{eq:symm_cond_I}} is either satisfied for all representations of ${\cal L}$, or it is broken by all.

\subsection{Weakly symmetric representations}

It has been shown~\cite{Macieszczak2021} that for a master operator with a weak symmetry, there exist \emph{weakly symmetric representations},
where the system Hamiltonian and  jump operators are all eigenmatrices of the symmetry superoperators. That is,
\begin{subequations} \label{eq:weakR}
		\begin{align}
			\label{eq:weakRa}
	\Ucal ({H}) &= {H},\\  
		\label{eq:weakRb}
\Ucal({J}_j) &= e^{i\delta_j} {J}_j \quad \forall{j},
	\end{align}
\end{subequations}  
where $e^{i\delta_j}$ is an eigenvalue of ${\cal U}$, {so} that $e^{i\delta_j}=e^{i(\phi_k-\phi_l)}$ where $e^{i\phi_k}$ and $e^{i\phi_l}$ are eigenvalues of $U$,  {with} $\phi_k,\phi_l\in \mathbb{R}$.  

It is easily verified that weakly symmetric representations satisfy Symmetry Condition \I.  We show below that they also satisfy the stronger Symmetry Conditions II and III, but those conditions can be also satisfied by representations which are not weakly symmetric.

\section{Weak symmetries of unravelled quantum dynamics}\label{sec:traj}

This Section describes quantum trajectories and the symmetries of their dynamics.  An overview is given in Fig.~\ref{fig:records}.  We {first} recall the construction of the trajectories and the associated {measurement} records.  We {then} explain in what sense the dynamics of these objects can be symmetric and present {in detail} Symmetry Conditions II and III, which are necessary and sufficient for symmetry of the objects in Boxes \Btraj, \Bcgtraj, and \Bftraj\,of Fig.~\ref{fig:records}.  

\subsection{Quantum trajectories}\label{sec:traj_intro}

The material in this subsection is standard~\cite{Belavkin1990,Dalibard1992,Dum1992,Carmichael1993,Molmer93,Wiseman2010,Daley2014} but different conventions exist in the literature.  We take inspiration from Ch.~6 of Ref.~\cite{Breuer_and_Petruccione}, but we focus throughout on pure density matrices instead of state vectors, see also Refs.~\cite{Carollo2019,Carollo2021}.

\subsubsection{Measurement records and conditional states}
\label{sec:measure-intro}

Quantum trajectories describe the time evolution of the quantum state of the system, conditioned on outcomes of measurements performed on the environment.  For a given representation, each jump operator $J_j$ is assumed to be associated with emission of a quantum (e.g., a photon) of type $j$, which can be detected.
The information about emitted quanta up to time $t$ is contained in a \emph{measurement record} 
\begin{equation}\label{eq:full_record}
    \bm{m}_t = (t_1,j_1;\dots;t_n,j_n),
\end{equation}
which indicates that a quantum of type $j_i$ was emitted at $t_i$, with $0\leq t_1<\dots<t_n< t$. These records are stochastic as they {capture all} outcomes of quantum measurements. We sometimes refer to them as ``full measurement records'' to distinguish situations where some classical information about the outcomes has been discarded.
	
The probability density for {measurement records} depends on the density matrix $\psi_0$ of the initial system state (which is assumed pure), it is given by 
\begin{equation}
	p(\bm{m}_t | \psi_0) =\Tr[\varphi_t(\bm{m}_t)] \; ,
	\label{eq:path_prob_m}
\end{equation}   
where
\begin{subequations}	\label{eq:varphi}
\begin{equation}
	\varphi_t(\bm{m}_t) 
	=        {\cal G}_{t-t_n} {\cal J}_{j_n} \cdots {\cal G}_{t_2-t_1} {\cal J}_{j_1} {\cal G}_{t_1}(\psi_{0}),
	\label{eq:varphi_t}
\end{equation}
with
\begin{equation}\label{eq:superJ}
	\mathcal{J}_j (\varphi) = J_j \varphi J_j^\dag \quad {\forall j}
\end{equation}
and
\begin{equation}\label{eq:G_t}
{\cal G}_{t} (\varphi) = e^{-i\He t} \varphi \,e^{i\He^\dag t} \;,
\end{equation}
where
\begin{equation}\label{eq:H_eff}
\He=H-\frac{i}{2}\sum_{j=1}^d J_j^\dag J_j.
\end{equation}
\end{subequations}
The \emph{conditional state} {of the system} for this record is
\begin{equation}
    \psi_t(\bm{m}_t) 
=      \frac{	\varphi_t(\bm{m}_t) }{\Tr[\varphi_t(\bm{m}_t)] }
        \label{eq:psi_t}
\end{equation}
(outcomes with $\Tr[\varphi_t(\bm{m}_t)]=0$ have probability zero, so they can be neglected).
Eq.~\eqref{eq:varphi_t} depends on the initial state $\psi_0$, but we suppress this dependence in the argument of most objects, including  in Eq.~\eqref{eq:psi_t}. % In the following it is implicit that quantities depend on this initial state.
Note that the conditional state $\psi_t$ is pure, because  $\psi_0$ is pure.

In this work, we define a \emph{quantum trajectory} up to time $t$ as the path followed by the conditional state for $0\leq\tau<t$. The quantum trajectory associated with the measurement record $\bm{m}_t$ is  then\footnote{The state $\psi_\tau$  in Eq.~\eqref{eq:m-to-psi} is considered as a function of $\bm{m}_t$, since any record $\bm{m}_\tau$ for ${\tau\in|0,t)}$ is contained in $\bm{m}_t$.}
\begin{equation}
\psi_{[0,t)}{(\bm{m}_t)} = \left[ \psi_\tau(\bm{m}_\tau) \right]_{\tau\in|0,t)}.
\label{eq:m-to-psi}
\end{equation}
From Eq.~\eqref{eq:varphi}, quantum trajectories consist of  deterministic segments (effectuated by ${\cal G}_t$), interspersed with {stochastic} transitions (due to $\{{\cal J}_j\}_j$).  %\\
Note that a quantum trajectory does not generically allow reconstruction of its associated measurement record because different records, $\bm{m}_t, \bm{m}'_t$, may yield the same quantum trajectory, $\psi_{[0,t)}{(\bm{m}_t)}=\psi_{[0,t)}{(\bm{m}'_t)}$. We explain below that this significantly affects the symmetries of the corresponding dynamics. 

The QME is recovered by \emph{averaging} over {measurement records}. That is, the density matrix
\begin{equation}\label{eq:rho_t_p}
	\rho_t = %\mathbb{E}\left[\psi_t(\bm{m}_t)  \right],
	\int d\bm{m}_t\, p(\bm{m}_t | \psi_0) \,\psi_t(\bm{m}_t),
\end{equation}%\\
solves Eq.~\eqref{eq:QME}.  (This system state $\rho_t$ is generically mixed.)
Here and below, the integral  $\int d\bm{m}_t$ with respect to the measurement record $\bm{m}_t$ in Eq.~\eqref{eq:full_record} should be interpreted as $\sum_{n=0}^\infty \sum_{j_1,\dots,j_n=1}^d\int_{0\leq t_1<\dots<t_n< t} dt_1 \cdots dt_n$.

\begin{figure}
    \centering
    \includegraphics[width=\linewidth]{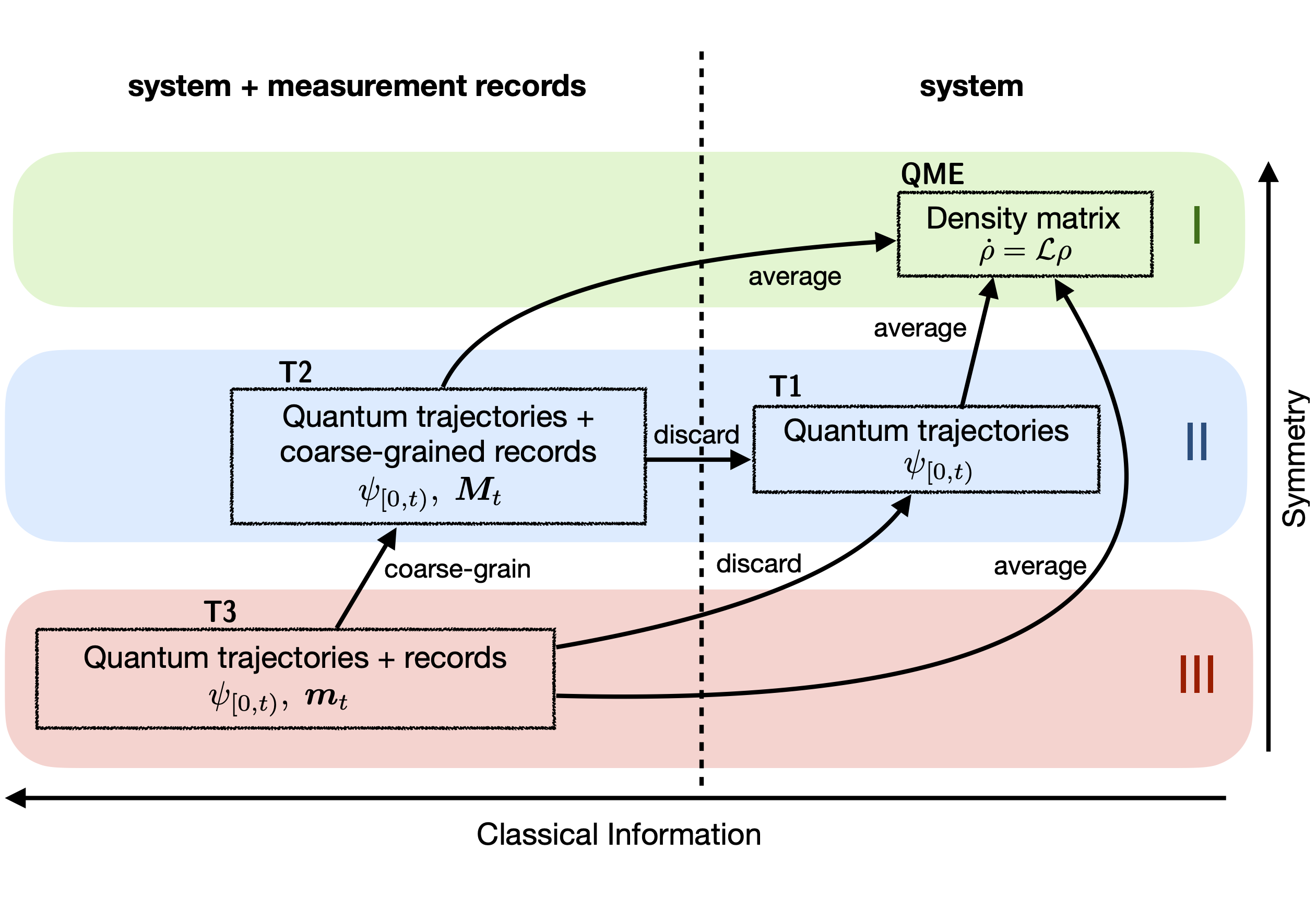}
    \vspace{-14pt}
    \caption{Detailed view of the second and third columns of Fig.~\ref{fig:outline}.   Sec.~\ref{sec:traj} gives the meanings of the boxes, the relationships between them (illustrated by arrows), and the conditions under which they are symmetric.}
    \label{fig:records}
\end{figure}

\subsubsection{Unravelled {quantum} dynamics}

We analyse quantum trajectories by focussing solely on the system {and its conditional states},  \emph{discarding} measurement records. This can be achieved in two ways: either by considering the distribution of conditional states at time~$t$, 
\begin{equation}
	P_t(\psi) = \int d\bm{m}_t\, p(\bm{m}_t | \psi_0) \,\delta\!\left[ \psi - \psi_{t}(\bm{m}_t) \right],
	\label{eq:P-t-m}
\end{equation}
 or by considering quantum trajectories with their associated probability density,
\begin{equation}
	p(\psi_{[0,t)}|\psi_0) = \int d\bm{m}_t\, p(\bm{m}_t | \psi_0)\, \delta\!\left[ \psi_{[0,t)} - \psi_{[0,t)}(\bm{m}_t) \right]\!.
	\label{eq:p-traj-m}
\end{equation} 
The resulting descriptions are equivalent, as we explain below; {they are both} represented by Box \Btraj\,in Figs.~\ref{fig:outline} and~\ref{fig:records}.

The probability density\footnote{%
In the following we sometimes refer to probability densities simply as probabilities, where there is no ambiguity.} 
$P_t(\psi)$ for the conditional state $\psi_t$ evolves according to the Fokker-Planck like equation,
\begin{equation}\label{eq:P_dot}
    \frac{\partial}{\partial t}{P}_t = {\cal W}^\dag (P_t),
\end{equation}
where  ${\cal W}^\dag$ is the \emph{generator of the unravelled quantum dynamics} %\footnote{\rlj{breuer eq (6.10)}}, 
which acts as
\begin{multline}\label{eq:generator_trajectories}
	[\mathcal{W}^\dagger (P_t)](\psi) = -\nabla \cdot \left[P_t(\psi) \mathcal{B}(\psi)\right]
	\\	% \nonumber
	+\int \!d\psi' \left[P_t(\psi')W(\psi',\psi)-P_t(\psi)W(\psi,\psi') \right].
\end{multline}
In the first line,%the (super)-operator
\begin{equation}\label{eq:B}
	\mathcal{B}(\psi) = -i\He\psi+i\psi\He^\dag - \psi\Tr(-i\He\psi+i\psi\He^\dag)
\end{equation}
encodes the smooth deterministic {flow} of conditional states between transitions\footnote{To see this, write $\hat{\cal G}(\varphi)={\cal G}_t(\varphi)/\Tr[ {\cal G}_t(\varphi)]$ and note that $d\hat{\cal G}(\varphi)/dt= \mathcal{B}[\hat{\cal G}(\varphi)]$.} {and we will refer to it as the \emph{drift}}.
The second line describes {stochastic transitions between conditional states, which leads to discontinuities in quantum trajectories. These transitions are facilitated by the jump operators,}  the rate of stochastic transitions from $\psi$ to $\psi'$ is
\begin{equation}\label{eq:w}
	W(\psi,\psi') = \sum_{j=1}^d \delta\!\left\{ \psi' - \frac{\mathcal{J}_j(\psi)}{\Tr[\mathcal{J}_j(\psi)]}\right\} \Tr[\mathcal{J}_j (\psi)] .
\end{equation}
where the sum over all jump operators appears as a consequence of discarding measurement records. {We will refer to these (stochastic) transitions as \emph{jumps} and their rates as \emph{jump rates}.}

 The probability density
$P_t$ follows a \emph{deterministic path} in the space of probability distributions,
\begin{equation}
	P_{[0,t)}=(P_\tau)_{\tau\in[0,t)},
	\label{eq:path-P}
\end{equation}
with the initial condition $P_0(\psi)=\delta ( \psi-\psi_0)$.  This smooth path is analogous to the path $\rho_{[0,t)}$ obtained from the QME. {In fact}, the quantum master {dynamics} is again recovered on \emph{average} [cf.~Eqs.~\eqref{eq:rho_t_p} and~\eqref{eq:P-t-m}], so the solution of the QME is
\begin{equation}\label{eq:rho_t_P}
	\rho_t  = \int d\psi P_t(\psi) \,\psi.
\end{equation}

The generator in Eq.~\eqref{eq:generator_trajectories} uniquely specifies a piecewise deterministic Markov process for the conditional state $\psi_t$. Its \emph{stochastic paths} (sample paths),
\begin{equation}
	\psi_{[0,t)}=(\psi_\tau)_{\tau\in[0,t)},
	\label{eq:path-psi}
\end{equation}
are the quantum trajectories of the system, whose probability density is  {given by Eq.}~\eqref{eq:p-traj-m}.
We refer to this collection of stochastic paths with their associated probabilities as an \emph{ensemble of quantum trajectories}.

\subsection{Symmetries of quantum trajectories}
\label{sec:generator_symm}

{We now describe} the symmetry properties of unravelled quantum dynamics.
In particular, we {derive} Symmetry Condition~II, which determines whether the corresponding ensemble of quantum trajectories is symmetric (Box \Btraj\ in Figs.~\ref{fig:outline} and~\ref{fig:records}).
This condition is the first main result of this paper.

\subsubsection{Weak symmetries of unravelled quantum dynamics}

{We begin by explaining what it means for the unravelled quantum dynamics to retain the weak symmetry of the quantum master dynamics. 
To this end, we introduce a unitary operation
$\Upsilon$ acting 
on probability distributions as
\begin{equation}\label{eq:Upsilon}
	[\Upsilon (P)](\psi) = P\!\left[\mathcal{U}^\dag(\psi)\right],
\end{equation}
where ${\cal U}$ is a given unitary super-operator.
The unitarity 
$\Upsilon^\dag=\Upsilon^{-1}$ holds as a dual pairing\footnote{{Specifically, the adjoint $\Upsilon^\dag$ is defined via $\langle f,\Upsilon P\rangle = \langle \Upsilon^\dag f,P\rangle$ where $f$ is a real function acting on (pure) conditional states and the inner product is $\langle f,P\rangle = \int d\psi  f(\psi)P(\psi)$.} 
}
on the space of functions of $\psi$; this is the most general form of a unitary operation that acts linearly on conditional states.   
Then, we say that the unravelled quantum dynamics {\emph{retains}} the weak symmetry {of the quantum master dynamics when}
\begin{equation}\label{eq:weakU_W}
    \Upsilon {\cal W}^\dag \Upsilon^\dag = {\cal W}^\dag.
\end{equation}
{Indeed, as we show below,  Eq.~\eqref{eq:weakU_W} implies that the weak symmetry of the quantum master dynamics [Eq.~\eqref{eq:weakU_QME}] holds, for the same $\mathcal{U}$ that appears in Eq.~\eqref{eq:Upsilon}. 
That is, weak symmetry of the QME is a necessary condition for this  symmetry of the unravelled dynamics, although it is not sufficient (see below).  The order of the symmetry group for the unravelled quantum dynamics is the same as that of the quantum master dynamics.

To understand the physical meaning of Eq.~\eqref{eq:weakU_W}, 
suppose that the path $P_{[0,t)}$ in Eq.~\eqref{eq:path-P} solves Eq.~\eqref{eq:generator_trajectories} for {$P_0(\psi) = \delta\left(\psi-\psi_0\right)$, where  $\psi_0$} is a given (pure) initial state. Then, {Eq.~\eqref{eq:weakU_W} implies} that 
the symmetry-transformed path 
\begin{equation}
	\sytr P_{[0,t)}=[ \Upsilon (P_\tau) ]_{\tau\in[0,t)}  
	\label{eq:tilde-path-P}
\end{equation}
also solves Eq.~\eqref{eq:generator_trajectories}, {with the symmetry-transformed initial distribution, $\sytr P_0=  \Upsilon (P_0)$, which is simply a delta function at $\sytr\psi_0 = {\cal U}(\psi_0)$.}
If the initial state is symmetric, ${\cal U}(\psi_0)=\psi_0$, then so is the initial distribution, $\Upsilon(P_0)=P_0$, {and}  the symmetry-transformed path coincides with the original one, $	\sytr P_{[0,t)} = P_{[0,t)}$. {That is, the path is invariant under the symmetry operation $\Upsilon$: it is symmetric.} %., so that the whole path is symmetric.
This also means that the distribution of conditional states remains symmetric for all times,
\begin{equation}
	\label{eq:P-psi-symm}
	\Upsilon (P_0)= P_0 \quad \Rightarrow \quad\Upsilon (P_t) =  P_t .
\end{equation}

{Moreover, if an asymptotic distribution $P_\infty$ of the unravelled quantum dynamics is unique}~\cite{Benoist2019}, it {must be symmetric, with $\Upsilon (P_\infty) = P_\infty$}. 
This situation would be different if the symmetry in Eq.~\eqref{eq:weakU_W} was associated with a conservation law, but this is not the case here, so we identify Eq.~\eqref{eq:weakU_W} as a weak symmetry.  Examples are given in 
Sec.~\ref{sec:examples}.
 {These same results can also be derived from the spectral properties of ${\cal W}$ together with the weak symmetry, see Sec.~\ref{sec:implications}.  

The properties in Eq.~\eqref{eq:tilde-path-P} and~\eqref{eq:P-psi-symm} are analogous to the properties of the QME given in Eqs.~\eqref{eq:tilde-path-rho} and~\eqref{eq:rho-symm}. In fact, the latter follow from  Eqs.~\eqref{eq:tilde-path-P} and~\eqref{eq:P-psi-symm}, via cf.~Eq.~\eqref{eq:rho_t_P}.  This means that symmetry of the QME is necessary for the symmetry in Eq.~\eqref{eq:weakU_W}, as already noted above.

The symmetry in Eq.~\eqref{eq:weakU_W}  can also be characterised in terms of ensembles quantum trajectories.  (This is analogous to the symmetry of the QME being expressed either in terms of operators [Eq.~\eqref{eq:weakU_QME}] or in terms of the paths $\rho_{[0,t)},\sytr\rho_{[0,t)}$.)
Define a symmetry-transformed quantum trajectory as  [cf.~Eq.~\eqref{eq:path-psi}]
\begin{equation}\label{eq:tilde-psi}
	\sytr{\psi}_{[0,t)}=\left[{\cal U}(\psi_\tau)\right]_{\tau\in[0,t)}.
\end{equation}
We explain in Sec.~\ref{sec:symm-II-deriv} below that Eq.~\eqref{eq:weakU_W} implies that 
 any {quantum} trajectory has the same probability density as its symmetry-transformed counterpart,
\begin{equation}
	p(\psi_{[0,t)}|\psi_0) = p\big(\sytr\psi_{[0,t)}|\sytr\psi_0\big)\,.
	\label{eq:tilde-path-psi}
\end{equation}
Moreover, if Eq.~\eqref{eq:tilde-path-psi} holds for all paths then Eq.~\eqref{eq:weakU_W} also holds.
That is, the symmetry of the ensemble of quantum trajectories [Eq.~\eqref{eq:tilde-path-psi}]} is equivalent to the weak symmetry of the generator {in Eq.~\eqref{eq:weakU_W}.  
When the initial state is symmetric ($\sytr\psi_0 = \psi_0$), {it follows that} the ensemble of {quantum} trajectories is invariant under the symmetry, {i.e., symmetric}. We emphasize however that in general individual quantum trajectories do not have any generic symmetry properties.

\subsubsection{{Symmetry Condition~II}}\label{sec:Wsymm_cond}
\label{sec:symm-II-deriv}

We now derive sufficient and necessary conditions for the weak symmetry of Eq.~\eqref{eq:weakU_W} to be present in the unravelled quantum dynamics.  In contrast to the quantum master operator ${\cal L}$, we note that the unravelled generator ${\cal W}^\dag$ depends on the representation of the QME, so the presence of symmetry can depend on the representation.  This is indeed the case.

In Appendix~\ref{app:UWU}, we show that $\Upsilon \mathcal{W}^\dagger\Upsilon^\dagger$ in Eq.~\eqref{eq:weakU_W} is obtained from ${\cal W}^\dagger$ in Eq.~\eqref{eq:generator_trajectories} by the replacements
\begin{equation}
\He \mapsto {\cal U}(\He), \qquad \mathcal{J}_j \mapsto {\cal U}\mathcal{J}_j {\cal U}^\dag \quad\forall j \;,
\label{eq:UJH}
\end{equation}
which must be applied in the drift ${\cal B}(\psi)$ of Eq.~\eqref{eq:B}, and the jump rates $W(\psi,\psi')$ of Eq.~\eqref{eq:w}. %one replaces

This has two important consequences. First, if $\psi_{[0,t)}$ is a sample path generated by ${\cal W}$ with probability $ p(\psi_{[0,t)}|\psi_0)$ then [recalling Eqs.~(\ref{eq:path_prob_m},\ref{eq:varphi},\ref{eq:p-traj-m})] the transformed path in Eq.~\eqref{eq:tilde-psi} is generated by $\Upsilon \mathcal{W}^\dagger\Upsilon^\dagger$, with the same probability.  Hence Eq.~\eqref{eq:weakU_W} implies that  Eq.~\eqref{eq:tilde-path-psi} holds for all paths.

Second, the replacement in Eq.~\eqref{eq:UJH} is shown in Appendix~\ref{app:UWU} to be equivalent to that of Eq.~\eqref{eq:UJH0}. 
Therefore, the weak symmetry of the unravelled {quantum} dynamics in Eq.~\eqref{eq:weakU_W} holds if and only if the two representations $\mathcal{U}(H)$,$\mathcal{U}(J_1),\dots,\mathcal{U}(J_d)$ and $H$, $J_1,\dots,J_d$ both share the same generator ${\cal W}^\dag$.  This last observation is crucial because means that we can exploit a recent result that characterise all the different representations that yield the same generator~\cite{Generators}.

The result is given as Theorem 2 below and it is framed
%\km{Such representations are characterised 
in terms of \emph{sets of jumps with equal destinations} (SJEDs), which are equivalence classes for jump operators  whose action, as given in Eq.~\eqref{eq:superJ}, is proportional for any pure state. 
Despite this abstract definition, the SJEDs have a straightforward construction, given in~\cite{Generators}.  We briefly summarise it here.  First, any sets of reset jump operators~\cite{Carollo2019} which share the same destination are identified as SJEDs.  Among the remaining jump operators, any sets of proportional operators (those with $J_i\propto J_j$) are also identified as SJEDs.  Any remaining jump operators form SJEDs with one element only.  The resulting SJEDs are disjoint and contain all jump operators (it is assumed that all $J_j\neq 0$).  We denote the $\alpha$th SJED by $S_\alpha$ with $\alpha=1,...,\dC$.  For further details of SJEDs, see Ref.~\cite{Generators} and also Sec.~\ref{sec:implications}.

The composite action of the jump operators in SJED $\alpha$ is defined as
\begin{equation}\label{eq:superA}
	\cgo_\alpha = \sum_{j\in \SJED} \mathcal{J}_j \quad\forall \alpha .%\sum_{k\in \SJED} J_k\psi J_k^\dag
\end{equation}
In particular, if $\psi$ is pure then so is ${\cal A}_\alpha(\psi)$.
{The definition of the SJEDs then ensures that 
%Due to their equal destinations, 
only these composite actions are relevant for the unravelled quantum dynamics in Eqs.~\eqref{eq:generator_trajectories}.
Moreover,
the jump rates in Eq.~\eqref{eq:w} can be expressed as
\begin{equation}\label{eq:w_alpha}
	W(\psi,\psi') = \sum_{\alpha=1}^{\dC} \delta\!\left\{ \psi' - \frac{\mathcal{A}_\alpha(\psi)}{\Tr[\mathcal{A}_\alpha(\psi)]}\right\} \Tr[\mathcal{A}_\alpha(\psi)] .
\end{equation}
This means that if two representations have the same composite action operators and the same Hamiltonian, they have the same unravelled quantum dynamics.  
It is shown in~\cite{Generators} that the converse of this statement also holds: to see the reason,
recall Eq.~\eqref{eq:generator_trajectories}, and note
%where a pure conditional state $\psi$ remains pure under the action of $\cgo_\alpha(\psi)$. 
that the drift (determined by the effective Hamiltonian) and the jumps (determined by the  SJED actions) give rise to different characteristics of unravelled quantum dynamics (smooth deterministic flow versus discontinuous stochastic transitions).
Given all these considerations, we arrive at Theorem 3 of Ref.~\cite{Generators} which in the current context becomes:\\
\par\smallskip \noindent $\bullet\; $
\textbf{Theorem~II} 
\par\noindent
\emph{For two representations of a given master operator,
$H, J_1, \dots, J_d$ %$H,\{J_k\}_{k=1}^d$
and
$\tilde H, \tilde J_1, \dots, \tilde J_{\tilde d}$,
the corresponding unravelled generators coincide, {$\tilde{\mathcal{W}}^\dag=\mathcal{W}^\dag$}, if and only if
\begin{subequations}\label{eq:theorem II_rep}
	\begin{align}
        \Tilde{H} & = H + r\mathbb{1}, \quad r\in\mathbb{R}, 
	%\end{equation}
	\\
	%\begin{align}
        \dCt&=\dC \quad \text{and} \quad \Tilde{\cgo}_{\alpha} = \cgo_{\pic(\alpha)} \quad \forall \alpha,% \; \cp{\emph{option 2}}
       %\label{eq:theorem II_repb}
	\end{align}
\end{subequations}
where  $\pic$ is a permutation of $\{1,...,\dC\}$.}% and $\dC=\tilde \dC$.}\\
\smallskip\\

{We now apply} Theorem II to
the representations $H,J_1,\dots,J_d$ and $\mathcal{U}(H),\mathcal{U}(J_1),\dots,\mathcal{U}(J_d)$, {which give rise to the generators ${\cal W}^\dag $ and $\Upsilon {\cal W}^\dag \Upsilon^\dag$ of the unravelled quantum dynamics, respectively. The two representations have the same number of SJEDs, as from} 
the definition of the SJEDs one sees that if $J_j,J_k$ are in the same SJED for the {former} representation then ${\cal U}(J_j),{\cal U}(J_k)$ are in the same SJED for the {latter} one.  Hence we obtain the following necessary and sufficient {conditions} for the weak symmetry in the {quantum} unravelled dynamics:\\
 
\smallskip \noindent $\bullet\; $
\textbf{Symmetry Condition~II} 
\par\noindent
\emph{For a representation $H,J_1,\dots,J_d$ of a quantum master operator with $\dC$ SJEDs, the corresponding unravelled dynamics features the weak symmetry $   \Upsilon {\cal W}^\dag \Upsilon^\dag = {\cal W}^\dag$ if and only if
\begin{subequations}\label{eq:symm_cond_II}
\begin{equation}\label{eq:symm_cond_IIa}
	 \mathcal{U}(H)= H \qquad
\end{equation}
and  there exists a permutation $\pic$ of $\{1,2,\dots,\dC\}$ such that
\begin{equation}\label{eq:symm_cond_IIb}
	 \mathcal{U}\cgo_\alpha\mathcal{U}^\dag = \cgo_{\pic(\alpha)}  \quad \forall \alpha.
\end{equation}
\end{subequations}	
}
\smallskip\\

From the definition of SJEDs, the composite actions ${\cal A}_\alpha$ and ${\cal A}_\beta$ cannot be equal unless $\alpha=\beta$, which means that the permutation $\pic$ in Eq.~\eqref{eq:symm_cond_IIb} is \emph{unique}~\cite{Generators}.  The physical meaning of this permutation will be discussed further when considering symmetries associated with measurement records and quantum trajectories in Sec.~\ref{sec:measurement_records}.  

The permutation also induces a discrete symmetry group whose order is the smallest integer $\Nc$ such that $\pic^{\Nc}$ is the identity. 
However, the weak symmetry associated with $\Upsilon^{\Nc}$ is not trivial unless $\mathcal{U}^{\Nc}=\mathcal{I}$, which does not hold in general (as $N_c \leq N$). {For example, weakly symmetric representations [Eq.~\eqref{eq:weakR}] have $\Nc=1$ but this still leaves considerable freedom in both $\Upsilon$ and ${\cal U}$.}} 

As already noted above, symmetric unravelled quantum dynamics [Eq.~\eqref{eq:weakU_W}] is sufficient for symmetric quantum master dynamics [Eq.~\eqref{eq:weakU_QME}].  
To show this explicitly, note that Eq.~\eqref{eq:symm_cond_II} implies\footnote{Summing Eq.~\eqref{eq:symm_cond_IIb} over $\alpha=1,...,\dC$ gives Eq.~\eqref{eq:symm_cond_II_to_Ib}. 
Then multiplying Eq.~\eqref{eq:symm_cond_II_to_Ib} by an arbitrary state $\psi$ and taking the trace gives $\sum_{j=1}^d 	\mathcal{U} (J_j^\dagger J_j)=\sum_{k=1}^d J_k^\dagger J_k$.   Using this with  Eq.~\eqref{eq:symm_cond_IIa} gives Eq.~\eqref{eq:symm_cond_II_to_Ia}
}
 both
\begin{subequations}\label{eq:symm_cond_II_to_I}
	\begin{equation}\label{eq:symm_cond_II_to_Ia}
		\mathcal{U}(\He)= \He \qquad
	\end{equation}
	and
	\begin{equation}\label{eq:symm_cond_II_to_Ib}
	 \sum_{j=1}^d	\mathcal{U}\mathcal{J}_j\mathcal{U}^\dag = \sum_{j=1}^d \mathcal{J}_k.
	\end{equation}
	\end{subequations}
%[To see this: .] 
From Eq.~\eqref{eq:rho_dot} we have also $\mathcal{L}(\rho)=-i \He \rho + i\rho \He^\dag+  \sum_{j=1}^d	\mathcal{J}_j(\rho),$  so that $\mathcal{U}\mathcal{L}\mathcal{U}^\dag (\rho)=-i \mathcal{U}(\He) \rho + i\rho\,\mathcal{U}(\He^\dag)+ 	 \sum_{j=1}^d	\mathcal{U}\mathcal{J}_j\mathcal{U}^\dag(\rho)$.  Therefore Eq.~\eqref{eq:symm_cond_II_to_I} implies Eq.~\eqref{eq:weakU_QME}, which also means that Symmetry Condition~II implies Symmetry Condition~I.  

Physically, this means that
 \emph{weak symmetries of unravelled quantum dynamics are inherited by the  quantum master dynamics}:
\beq\nonumber
\Upsilon\mathcal{W}^\dag\Upsilon^\dag = \mathcal{W}^\dag  \quad \Longrightarrow \quad \mathcal{U}\mathcal{L}\mathcal{U}^\dag = \mathcal{L} .
\eeq
The reverse implication does not hold which means that Symmetry Condition II is more restrictive than Symmetry Condition I. For example, Eq.~\eqref{eq:symm_cond_I} does not imply the symmetry in Eq.~\eqref{eq:symm_cond_II} of the Hamiltonian.
In Sec.~\ref{sec:examples}, we present examples illustrating theses cases, including when Symmetry Condition~II  is violated so that the weak symmetry of the unravelled quantum dynamics is absent, while Symmetry Condition~I still holds. 

Symmetry Condition~II above is phrased in terms of the {super}operators ${\cal A}_\alpha$, which {were defined in Eq.~\eqref{eq:superA} in terms of the actions of the jump operators $J_j$ from the corresponding SJED $S_\alpha$}.  We now {present} Symmetry Condition~\IIb, which is equivalent, but expressed directly in terms of the jump operators:  \\

\smallskip \noindent $\bullet\; $
\textbf{Symmetry Condition~\IIb} 
\par\noindent
\emph{For a representation $H,J_1,\dots,J_d$ of a quantum master operator with $\dC$ SJEDs, the corresponding unravelled dynamics features the weak symmetry $\Upsilon\mathcal{W}^\dag\Upsilon^\dag=\mathcal{W}^\dag$ if and only if 
\begin{subequations}\label{eq:UIIR}
\begin{equation}\label{eq:UIIRa}
\mathcal{U}(H) = H\qquad 
\end{equation}
and there exists a permutation $\pic$ of $\{1,2,\dots,\dC\}$ and a $d\times d$ matrix $\mathbf{X}$  such that
\begin{equation} \label{eq:X}
	\mathcal{U}(J_j) = \sum_{k=1}^{d} \mathbf{X}_{jk}J_k \quad \forall{j} 
\end{equation}
\end{subequations}
{and for $i,j\in S_\alpha$, $k,l\in S_{\pic(\alpha)}$ with $\alpha=1,...,\dC$
\begin{subequations}\label{eq:X_all}
\begin{align}\label{eq:Va2}
	\sum_{k\in S_{\pic(\alpha)}}  \mathbf{X}_{ik} \mathbf{X}_{jk}^*  & = \delta_{ij} \qquad \text{if} \quad |S_\alpha| \leq |S_{\pic(\alpha)}|,  \\
	\label{eq:Va3}
\sum_{j\in S_{\alpha}} \mathbf{X}_{jk}^*  \mathbf{X}_{jl}  & = \delta_{kl}  \qquad \text{if} \quad |S_\alpha| \geq |S_{\pic(\alpha)}|,
\end{align}
while
\begin{align}
	\mathbf{X}_{jk} = 0 \quad \text{unless} \quad j\in S_\alpha, \,\, k\in S_{\pic(\alpha)} \;  \text{\,for\, some\,\,} \alpha.
	\label{eq:Va1}
\end{align}
\end{subequations}
}}
\smallskip

To obtain this condition, note that
the properties of the matrix $\textbf X$ given in Eq.~\eqref{eq:X_all} follow from  
Eq.~\eqref{eq:symm_cond_IIb}, which requires (for each $\alpha$) the equality between two completely positive superoperators.  This means in turn that the underlying sets of operators must be related by an isometry~\cite{Wolf2012}.  Hence the elements of $S_\alpha$ and $S_{\pic(\alpha)}$ are related by isometric mixing, which induces the block structure in Eq.~\eqref{eq:Va1}.  The non-zero elements of $\mathbf{X}$ are elements of the relevant isometric matrices and Eqs.~(\ref{eq:Va2},\ref{eq:Va3}) are the corresponding isometric condition.

We also note that Eq.~\eqref{eq:symm_cond_II_to_Ib} is again an equality of two completely positive superoperators; 
{using this with  Eq.~\eqref{eq:symm_cond_II_to_Ia} we obtain
\begin{subequations}
\label{eq:not_cond_II}
\begin{align}
    {\cal U}(H)  & =H, 
    \\ 
    %\qquad 
    \mathcal{U}(J_j) & = \sum_{k=1}^{d} \mathbf{U}_{jk} J_k \quad \forall j,
    \label{eq:not_cond_IIb}
 \end{align}
\end{subequations}
for some unitary $d\times d$ matrix $\mathbf{U}$.  The converse also holds, so overall Eq.~\eqref{eq:symm_cond_II} $\Rightarrow$ Eq.~\eqref{eq:symm_cond_II_to_I}  $\Leftrightarrow$  Eq.~\eqref{eq:not_cond_II}.}
Noting that $\Tr[\mathcal{U}(J_j)]=\Tr(J_j)$, we also have
\begin{equation}\label{eq:U-Tr}
			\Tr(J_j) = \sum_k\mathbf{U}_{jk}\Tr(J_k) \quad \forall j,
\end{equation}
which combined with Eqs.~\eqref{eq:not_cond_IIb} and \eqref{eq:traceless_rep_b} recovers \eqref{eq:symm_cond_Ib}.
A similar argument can be used to recover~\eqref{eq:symm_cond_Ia}, which again confirms that Symmetry Condition II [or equivalently Condition \IIb] implies Symmetry Condition I.

Despite the similarity of Eqs.~(\ref{eq:X},\ref{eq:not_cond_IIb}), the matrix $\mathbf{X}$ is not unitary in general.  If the jump operators $J_k$ are linearly independent then $\mathbf{X},\mathbf{U}$ are uniquely determined by their respective equations, leading to unitary $\mathbf{X}=\mathbf{U}$.   In the general case however $\mathbf{X}$ is not unique, and it even can be that no unitary choice exists, an example is given in Sec.~\ref{sec:examples}.   Under Symmetry Condition II, we nevertheless show in  Appendix~\ref{app:boldU} that there always exists a unitary $\mathbf{U}$ that obeys Eq.~\eqref{eq:not_cond_IIb}, with the additional property
\beq
\label{eq:UU-partial-lemma}
\sum_{j \in S_{\alpha}} \mathbf{U}^*_{jk} \,{\cal U}(J_j) = 
\begin{cases}  J_k , \quad &  k \in S_{\pic(\alpha)}
	\\
	0 ,  &  \text{otherwise} \; .
\end{cases}
\eeq
The existence of this matrix will be important in later Sections.

\subsection{Symmetries of quantum trajectories and measurement records}\label{sec:measurement_records}

{We now consider the symmetry properties of quantum trajectories and measurement records together} (see {Box} \Bftraj\, in Figs.~\ref{fig:outline} and~\ref{fig:records}).  In particular, we present Symmetry Condition~III, which determines whether their dynamics are symmetric, and is more restrictive than Symmetry Condition II.  We explain this physically by observing that the mapping from measurement records to quantum trajectories reduces (classical) information.

\subsubsection{Measurement records and their symmetries}\label{sec:full_time_recs}

The combined description of the system and {the} measurement record at time $t$ is 
\begin{equation} 
	 \left(\psi_{t},\bm{m}_t\right),
	\label{eq:psi-m}
\end{equation}
where $\psi_{t}$ is a conditional state and $\bm{m}_t$ is the measurement record (recall Sec.~\ref{sec:measure-intro}).
{The conditional state in Eq.~\eqref{eq:psi_t} is fully determined
by the measurement record in Eq.~\eqref{eq:full_record} and the initial state $\psi_0$.  Thus,} the joint probability density for $(\psi_{t},\bm{m}_t)$ is
\begin{equation}
	p\!\left(\psi_t,\bm{m}_t | \psi_0\right) =	p(\bm{m}_t | \psi_0) \,\delta\!\left[ \psi_t - \psi_{t}(\bm{m}_t) \right].
	\label{eq:p-m-psi}
\end{equation}
Then $P_t(\psi)$ of Eq.~\eqref{eq:P-t-m} is the associated marginal density for $\psi_t$.

We may also consider a measurement record combined with its quantum trajectory:
\begin{equation}	\label{eq:path_psi_m}
	\left(\psi_{[0,t)},\bm{m}_t\right).
\end{equation} 
Again, the {quantum trajectory in Eq.~\eqref{eq:path-psi} is fully determined by the measurement record and} the initial condition, so the corresponding probability density is
\begin{equation}
	p\!\left(\psi_{[0,t)},\bm{m}_t|\psi_0\right)	=p(\bm{m}_t | \psi_0) \,\delta\!\left[ \psi_{[0,t)} - \psi_{[0,t)}(\bm{m}_t) \right],
	\label{eq:p-m-traj}
\end{equation}
and the probability $p(\psi_{[0,t)}|\psi_0)$  in Eq.~\eqref{eq:p-traj-m} is the corresponding marginal density for $\psi_{[0,t)}$.

Recalling the definition Eq.~\eqref{eq:tilde-psi} in terms of ${\cal U}$ of the symmetry-transformed quantum trajectory $\sytr\psi_{[0,t)}$, we identify a symmetry-transformed counterpart of $\bm{m}_t$ as 
\begin{equation}\label{eq:tilde-m}
	\sytr{\bm{m}}_t=\left[t_1,\pi(j_1);\dots;t_n,\pi(j_n)\right],
\end{equation} 
for a permutation $\pi$ of $\{1,\dots,d\}$.  The dynamics considered in this subsection feature a symmetry (dependent on ${\cal U},\pi$) if and only if  
\begin{equation}
	p\big(\psi_{[0,t)},\bm{m}_t\big|\psi_0\big) = p\big(\sytr\psi_{[0,t)},\sytr{\bm m}_t\big|\sytr\psi_0\big) \; ,
	\label{eq:path-psi-m-symm}
\end{equation}
for all $\left(\psi_{[0,t)},\bm{m}_t\right)$.  
{This result is analogous to the symmetry property Eq.~\eqref{eq:tilde-path-psi}, which may be recovered by integrating over measurement records.  By averaging the conditional state with respect to this distribution we also obtain the symmetry property Eq.~\eqref{eq:tilde-path-rho} for the QME.  That is, the symmetry of trajectories and measurement records implies symmetry of trajectories alone, as well as the symmetry of the QME.}

{The mapping from $\left(\psi_{t},\bm{m}_t\right)$ to $(\sytr\psi_{t},\sytr{\bm{m}}_t)$ is local in time; it is also separable (it acts independently on the conditional state and the measurement record); and it is unitary [see Eq.~\eqref{eq:Pi}, below].  In fact, this is the most general mapping that respects these three properties.  In principle one might consider some generalised version of the symmetry property \eqref{eq:path-psi-m-symm}, in which the underlying mapping does not respect these properties.  However, our focus is on unitary symmetries that imply the QME symmetry when discarding measurement records (hence we assume separability); and which hold for trajectories of all length $t$ (hence we assume local in time).  Therefore we focus on symmetry properties of the form \eqref{eq:path-psi-m-symm}.}

Considering the marginal probability of the final conditional state and the measurement records, {Eq.~\eqref{eq:tilde-m}}  implies 
\begin{equation}
	p(\psi_t,\bm{m}_t\big|\psi_0) = p(\sytr\psi_t,\sytr{\bm m}_t\big|\sytr\psi_0)  ,
	\label{eq:psi-m-symm}
\end{equation}
{while} for the measurement record alone [cf.~Eq.~\eqref{eq:p-m-traj}], we obtain 
\begin{equation}
	p(\bm{m}_t\big|\psi_0) = p(\sytr{\bm m}_t\big|\sytr\psi_0)  .
	\label{eq:m-symm}
\end{equation}
{In contrast to the symmetry property for quantum trajectories in Eq.~\eqref{eq:tilde-path-psi}, the symmetry property for measurement records in Eq.~\eqref{eq:m-symm} does imply both Eqs.~\eqref{eq:psi-m-symm} and~\eqref{eq:path-psi-m-symm}, as quantum trajectories are uniquely determined by measurement records and initial states. Hence, all these characterisations of the symmetry [Eqs.~(\ref{eq:path-psi-m-symm}-\ref{eq:m-symm})] are equivalent.}

\subsubsection{Labelled quantum dynamics}

To characterise the above symmetry in terms of the generator of a stochastic process,
we introduce a vector $\bm{q}_t$ whose $j$th element $(\bm{q}_t)_j$ {(with $j=1,...,d$)} is the number of times the {outcome} $j$ appears in the {measurement} record $\bm{m}_t$.  Physically, this is the  number of quanta of type $j$ emitted between times $0$ and $t$.  {We will refer to $\bm{q}_t$ as the vector of \emph{full counts}, to distinguish it from situations (considered below) where only some types of outcome are distinguished.}
% \footnote{'km{KM: I think we needs some name for it, I added 'full' to distinguish from the total counts which would be over everything.}}.
Crucially, the (stochastic) path 
\begin{equation}\label{eq:path-q}
	\bm{q}_{[0,t)}=(\bm{q}_\tau)_{\tau\in[0,t)}
\end{equation} 
contains the same (classical) information as the measurement record $\bm{m}_t$.  

We then consider the {(stochastic) combined} state
\begin{equation}\label{eq:psi-q}
	\left(\psi_t,\bm{q}_t\right)
\end{equation}
{of the (quantum) system and a (classical) counter, for which sample paths are}
\begin{equation}\label{eq:path-psi-q}
	\left(\psi_{[0,t)},\bm{q}_{[0,t)}\right).
\end{equation} 
We will refer to these paths as \emph{labelled quantum trajectories}.
They carry exactly the same information as the combined quantum trajectories and measurement records in Eq.~\eqref{eq:p-m-traj}. Indeed, each jump for the conditional state $\psi_t$ is accompanied by a jump in {$(\bm{q}_t)_j$ for some $j=1,...,d$, which allows for the label of corresponding outcome $j$ to be inferred. 

It follows that}  the probability density for labelled quantum trajectories is directly related to the joint probability density for {quantum} trajectories and measurement records:
\begin{equation}\label{eq:p-path-q-traj}
	p\!\left(\psi_{[0,t)},\bm{q}_{[0,t)}|\psi_0\right)=	p\!\left(\psi_{[0,t)},\bm{m}_t|\psi_0\right).
\end{equation}
This induces an \emph{ensemble of labelled quantum trajectories} which is equivalent to the ensemble of paths described by Eq.~\eqref{eq:p-m-traj}.

{The generator for the stochastic process of Eq.~\eqref{eq:psi-q}  is denoted by $\WF^\dag$, where the subscript  ${\textsc f}$ indicates that we consider full counts [cf.~Eq.~\eqref{eq:P_dot}].  We refer to $\WF^\dag$ as the \emph{generator of the labelled quantum dynamics}. It acts on the probability $P_t(\psi,\bm{q})$ for the conditional state to be $\psi_t$ and the full counts to be $\bm{q}_t$ (that is, the combined state of the system and the counter at $t$): specifically,
\begin{multline}\label{eq:Wf}
	\left[\WF ^\dag
	(P_t)\right](\psi,\bm{q}) =-\nabla\cdot\left[ P_t(\psi,\bm{q}) \,{\cal B}(\psi)\right]  \\ % \nonumber
	+  \sum_{j=1}^d\int \!d\psi' \Big[ P_t(\psi',\bm{q}\!-\!\bm{e}_j) \,w_j(\psi',\psi) 
	\\
	- P_t(\psi,\bm{q}) \, w_j(\psi,\psi') \Big]
\end{multline}
{[cf.~Eq.~\eqref{eq:generator_trajectories}].} Here, $(\bm{e}_j)_k=\delta_{jk}$ for $k=1,...d$, so that the number of detected quanta of type $j$  increases by $1$ when the {jump} from $\psi$ to $\psi'$ {is facilitated by the} jump operator $J_j$; this {occurs with} rate
\begin{equation}
\label{eq:w_j}
w_j(\psi,\psi') = \delta\!\left\{\psi' - \frac{\mathcal{J}_j(\psi)}{\Tr[\mathcal{J}_j(\psi)]}\right\}\Tr[\mathcal{J}_j (\psi)].
\end{equation}
Note that  $\sum_{j=1}^d w_j(\psi,\psi')=W(\psi,\psi')$ as defined {in}  Eq.~\eqref{eq:w}.

\subsubsection{{Weak} symmetries of labelled {quantum} dynamics}

The symmetry transformed measurement record of Eq.~\eqref{eq:tilde-m} corresponds to a symmetry-transformed path for the counter, defined via the action of the same permutation $\pi$.   
We define
\begin{equation}\label{eq:tilde-path-psi-q}
	\left(\sytr\psi_{[0,t)},\sytr{\bm{q}}_{[0,t)}\right)  =\left[{\cal U}(\psi_\tau),\bm{\pi}(\bm{q}_\tau)\right]_{\tau\in[0,t)},
%     \left(\sytr\psi_{[0,t)},\sytr{\bm{q}}_{[0,t)}\right)  =\left({\cal U}(\psi_\tau),\bm{\pi}(\bm{q}_\tau)\right)_{\tau\in[0,t)},
\end{equation}
where $\bm{\pi}(\bm{q})_j = (\bm{q})_{\pi(j)}$ {for $j=1,...,d$}.  That is, when {$\bm{\pi}$} acts on a vector, it permutes its entries as {the permutation $\pi$} would their indices. 
Then, the symmetry property in Eq.~\eqref{eq:path-psi-m-symm} is equivalent to
\begin{equation}
	p\!\left(\psi_{[0,t)},\bm{q}_{[0,t)}|\psi_0\right)=	p\!\left(\sytr\psi_{[0,t)},\sytr{\bm{q}}_{[0,t)}|\tilde\psi_0\right).
	\label{eq:path-psi-q-symm}
\end{equation}
{Hence also $p(\bm{q}_{[0,t)}|\psi_0)=p(\sytr{\bm{q}}_{[0,t)}|\sytr\psi_0)$, cf. Eq.~\eqref{eq:m-symm}.}
{Analogously to  Eq.~\eqref{eq:tilde-path-psi}, this property also means if the initial state is symmetric, then the whole ensemble of labelled quantum trajectories in invariant under the symmetry. 

Equations~\eqref{eq:tilde-path-psi-q} and \eqref{eq:path-psi-q-symm} show that the symmetry property of quantum trajectories and measurement records in Eqs.~\eqref{eq:path-psi-m-symm} can be equivalently formulated in terms of the symmetry transformations of the combined state in Eq.~\eqref{eq:psi-q}.  This depends on the system superoperator $\mathcal{U}$ and the permutation $\bm{\pi}$ for the state of the counter.
This motivates the symmetry operation in Eq.~\eqref{eq:path-psi-m-symm} to be local in time and also independent of time.

By analogy with the unravelled quantum dynamics, symmetry-transformed labelled quantum trajectories in Eq.~\eqref{eq:tilde-path-psi-q} can be viewed as (sample) paths generated by $\Yf\WF^\dag\Yf^\dag$, where $\Yf$ is a unitary operation acting on  joint probability distributions as
\begin{equation}
	[\Yf (P)](\psi, \bm{q}) = P\!\left[\mathcal{U}^\dag(\psi),  \bm{\pi}^{-1}(\bm{q} )\right],
	\label{eq:UpsF}
\end{equation}
 This operation can be related to $\Upsilon$ from Eq.~\eqref{eq:Upsilon} by introducing 
\begin{equation}\label{eq:Pi}
	{[\Pi(P)]}(\psi,\bm{q}) = P[\psi,\bm{\pi}^{-1}(\bm{q})],
\end{equation}
so that 
\begin{equation}\label{eq:UpsF_Pi}
	\Yf = \Upsilon\,\Pi,
\end{equation}
where $\Upsilon$ is understood to act as $ [\Upsilon (P)](\psi,\bm{q}) = P(\mathcal{U}^\dag(\psi),\bm{q})$.  The physical idea is that the symmetry operation $\Yf $ acts  \emph{independently} on the system and counter [cf.~Eq.~\eqref{eq:psi-q}].
Then, the symmetry property of the ensembles in Eq.~\eqref{eq:path-psi-q-symm} is equivalent to the equality of their generators,
\begin{equation}\label{eq:Wf_symm}
	\Yf\WF^\dag\Yf^\dag = \WF^\dag \; .
\end{equation}
{Equation~\eqref{eq:Wf_symm} is a \emph{weak symmetry of the labelled quantum dynamics} and properties analogous to  Eqs.~\eqref{eq:tilde-path-P} and~\eqref{eq:P-psi-symm} follow for the evolution of the associated probability distribution $P_t(\psi,\bm{q})$.}

\subsubsection{Symmetry Condition~III}\label{sec:symm_Wf}

We now identify {necessary and sufficient} conditions for the {weak symmetry [Eq.~\eqref{eq:Wf_symm}] of the labelled quantum dynamics}.  For this, we
use Theorem 2 of \cite{Generators}, {which can be rephrased using the definitions above as}:\\
\par\smallskip \noindent $\bullet\; $
\textbf{Theorem III}
\par\noindent
\emph{
Consider two representations of a given master operator, $H, J_1, \dots, J_d$ %$H,\{J_k\}_{k}$
and $\tilde H, \tilde J_1, \dots, \tilde J_{{d}}$, %$\tilde{H},\{\tilde{J}_k\}_{k}$,
both of which have $d$ jump operators.  Given a permutation $\pi$ of $\{1,2,\dots,d\}$, %and defining $\Pi$ \km{as in Eq.~\eqref{eq:Pi},}
the corresponding generators for the labelled {quantum dynamics  obey
\begin{equation}\label{eq:PiWf}
    \Pi \tilde{\mathcal{W}}^\dag_F \Pi^\dag = \mathcal{W}_F^\dag
\end{equation}}
%with $\Pi$ as in \eqref{eq:Pi}, 
if and only if
{\begin{subequations}\label{eq:WF_cond}
\begin{equation}%\label{eq:WF_cond}
\label{eq:WF_cond_H}
    \Tilde{H}=H+r\mathbb{1}, \quad r\in\mathbb{R},
\end{equation}
and
\begin{equation}%\label{eq:WF_cond}
\label{eq:WF_cond_jump}
\Tilde{J}_j = J_{\pi(j)}e^{i\phi_j}, \quad \phi_j\in\mathbb{R}\quad \forall j.
\end{equation}
\end{subequations}
}
\smallskip
}\\

Analogous to Sec.~\ref{sec:Wsymm_cond}, we {now} apply this Theorem to the two representations $H,J_1,\dots,J_d$ and ${\cal U}(H),{\cal U}(J_1),\dots,{\cal U}(J_d)$.  
{To do so, we observe that $\Yf\WF^\dag\Yf^\dag = \Pi\WFt^\dag\Pi^\dag$, where $\WFt^\dag=\Upsilon \WF^\dag \Upsilon^\dag$  from Eq.~\eqref{eq:UpsF_Pi}. Then, we note that  $\WFt^\dag$  is obtained from $\WF^\dag$ by the change of representation in Eq.~\eqref{eq:UJH0}. [A proof analogous to that for $\tilde{\mathcal{W}}^\dag$ in Appendix~\ref{app:UWU} shows that $\WFt^\dag$ is obtained from $\WF^\dag$ by the replacement in Eq.~\eqref{eq:UJH}.]
With these definitions, the symmetry Eq.~\eqref{eq:Wf_symm} is equivalent to Eq.~\eqref{eq:PiWf}, so we have:\\
\par\smallskip \noindent $\bullet\; $
\textbf{Symmetry Condition~III}
\par\noindent
\emph{For a representation $H, J_1,\dots, J_d$ %\{J_j\}_{j=1}^d$
of a quantum master operator, the corresponding labelled quantum dynamics features {the} weak symmetry $   \Yf \WF^\dag \Yf^\dag = \WF^\dag$ if and only if
\begin{subequations}	\label{eq:symm_cond_III}
\begin{equation}
	\mathcal{U}(H)= H
	\label{eq:symm_cond_III_H}
\end{equation}
and 
	\begin{equation}
	\mathcal{U}(J_j) = J_{\pi(j)}e^{i\phi_j}, \quad\phi_j\in\mathbb{R}\quad \forall j \; ,
			\label{eq:symm_cond_III_jump}
	\end{equation}
	where $\pi$ is the permutation which appears in  $\Upsilon_F$.
\end{subequations}
	\smallskip}

This condition shows that the presence of a weak symmetry in the labelled unravelled dynamics restricts the action of the symmetry superoperator $\mathcal{U}$ on the jump operators  $J_1,...,J_d$ only to their permutation [up to multiplication by a phase, which does not affect quantum trajectories, cf.~Eqs.~\eqref{eq:path_prob_m}-\eqref{eq:psi_t}]. Furthermore, this result clarifies that it is this action that determines the permutation $\pi$ of measurement records in Eq.~\eqref{eq:tilde-m}, which characterises their symmetry properties.
A special case of Symmetry Condition~III is that of the trivial permutation $\pi(j)=j$, which means %[by  Eq.~\eqref{eq:weakU_QME}] 
that the considered representation is weakly symmetric and [see Eq.~\eqref{eq:weakR}, the phase $\delta_j$ in that equation corresponds to the phase $\phi_j$  in  Eq.~\eqref{eq:symm_cond_III_jump}.]

Note also that Eq.~\eqref{eq:symm_cond_III_jump} is equivalent to Eq.~\eqref{eq:not_cond_IIb} holding with
\begin{equation} \label{eq:UIIIRb_U}
\mathbf{U}_{jk} = e^{i\phi_j} \delta_{\pi(j),k},
\end{equation}
and $\phi_j\in \mathbb{R}$.  One sees that each row of $\mathbf{U}$ contains a single non-zero element, this is a much more detailed constraint than the block structure of $\mathbf{X}$ prescribed by Symmetry Condition \IIb. 
Recalling also the discussion of Appendix~\ref{app:boldU}, there may be several different (unitary) matrices $\mathbf{U}$ such that Eq.~\eqref{eq:UIIIRb_U} holds.  In some cases it can be that Eq.~\eqref{eq:UIIIRb_U} holds simultaneously for more than one permutation $\pi$.  Examples are given in Sec.~\ref{sec:examples}.

It is immediate that 
\beq\nonumber
\text{Eq.}~\eqref{eq:symm_cond_III} \quad \Longrightarrow \quad \text{Eq.}~\eqref{eq:symm_cond_II},
\eeq
so that Symmetry Condition~III implies Symmetry Condition II, hence it also implies Condition I.  
Equivalently, \emph{the unravelled quantum dynamics inherits the weak symmetry from the labelled quantum dynamics}:
\beq\nonumber
\Yf\WF^\dag\Yf^\dag = \WF^\dag \quad \Longrightarrow \quad \Upsilon
\mathcal{W}^\dag\Upsilon^\dag = \mathcal{W}^\dag. 
\eeq
 This holds because of the symmetry transformation of quantum trajectories and measurement records acting separately on these parts [cf.~Eqs.~\eqref{eq:path-psi-m-symm},~\eqref{eq:path-psi-q-symm}, and~\eqref{eq:UpsF_Pi}], so when full counts are discarded from the labelled quantum dynamics, the weak symmetry is inherited by the resulting unravelled quantum dynamics.
One also sees easily that the reverse implication does not hold, and Symmetry Condition III may be violated while Condition II holds.  For example, Eq.~\eqref{eq:UIIIRb_U} is much more restrictive than the conditions on $\mathbf{X}$ in Symmetry Condition IIb. 

The above discussion establishes the promised hierarchy of Symmetry Conditions, that III is the most restrictive and I the least.  In cases where Symmetry Condition II holds but Symmetry Condition III is violated, the physical interpretation is that the adding labels to the quantum trajectories can break the symmetry of the dynamics.  
The next subsection shows how this symmetry-breaking can be avoided by using coarse-grained labels.
{Sec.~\ref{sec:partial-symm-further} below includes additional comments on the interpretation of symmetry-breaking via labelling.}

\subsection{Symmetries of quantum trajectories and coarse-grained measurement records} \label{sec:CG-records}

We now discuss coarse-grained measurement records and the symmetries of their dynamics (together with quantum trajectories), this is the content of Box \Bcgtraj~in Figs.~\ref{fig:outline} and \ref{fig:records}.  One sees from Fig.~\ref{fig:records} that these measurement records are obtained from full {measurement} records (Box \Bftraj) by coarse-graining. One may recover quantum trajectories from this description (Box \Btraj) by discarding the coarse-grained measurement records.

\subsubsection{Coarse-grained {measurement} records\\ and their symmetries}

For any given measurement record $\bm{m}_t$ [cf. Eq.~\eqref{eq:full_record}], the corresponding coarse-grained record 
 is obtained by classifying {its outcomes} according to the SJEDs of the associated jump operators.
{Physically, this corresponds to a quantum-optical  {erasure superoperator acting on} emitted quanta (before or after their detection), replacing them by new quanta of type associated with the corresponding  SJED. See also Sec.~\ref{sec:CG_cMPS}.}

For the measurement record $\bm{m}_t$ in Eq.~\eqref{eq:full_record}, the corresponding coarse-grained record is
\begin{equation}\label{eq:partial_record}
    \bm{M}_t =(t_1,\alpha_1;\, \dots; t_n,\alpha_n),
\end{equation}
where $\alpha_i$ is the SJED of {jump operator} $J_{j_i}$ ({i.e.}, $j_i\in S_{\alpha_i}$).
The probability of such a coarse-grained record is the sum of the {probabilities} $p(\bm{m}_t|\psi_0)$ over {all} full measurement records $\bm{m}_t$ which yield $\bm{M}_t$ under coarse-graining [{i.e.}, records with $j_i\in S_{\alpha_i}$ for $i=1,...,n$]. 
Hence 
[cf.~Eq.~\eqref{eq:path_prob_m}]
\begin{align}
	p(\bm{M}_t|\psi_0) & = \Tr[\varphi_t(\bm{M}_t)], 
	\label{eq:path_prob_M}
\end{align}
where  [cf.~{Eqs.~\eqref{eq:varphi} and~\eqref{eq:superA}}]
\begin{equation} \label{eq:coarse_traj}
	\varphi_t(\bm{M}_t)  
	=  
	\mathcal{G}_{t-t_n}  \cgo_{\alpha_n}  \dots  \mathcal{G}_{t_2-t_1}  \cgo_{\alpha_1}  \mathcal{G}_{t_1}(\psi_0).
\end{equation}
The system state conditioned on this record is  [cf.~Eq.~\eqref{eq:psi_t}]
\begin{equation}
	\psi_t(\bm{M}_t) 
	=      \frac{	\varphi_t(\bm{M}_t) }{\Tr[\varphi_t(\bm{M}_t)] }.
	\label{eq:psi_t_M}
\end{equation}
{This state is pure, as by the definition of SJEDs~\cite{Generators}, conditional states $\psi_t(\bm{m}_t) $ coincide for any full measurement records $\bm{m}_t$ that yield $\bm{M}_t$ under coarse-graining.} 
Thus, also its (stochastic) path [cf.~Eq.~\eqref{eq:m-to-psi}]
\begin{equation}
	\psi_{[0,t)}{(\bm{M}_t)} = \left[ \psi_\tau(\bm{M}_\tau) \right]_{\tau\in|0,t)}
	\label{eq:M-to-psi}
\end{equation}
is the same as {the quantum trajectory $\psi_{[0,t)}{(\bm{m}_t)}$, for any such $\bm{m}_t$}.

The combined {quantum} trajectory and  coarse-grained measurement record up to time $t$ are [cf.~Eq.~\eqref{eq:path_psi_m}]
\begin{equation}
 (\psi_{[0,t)},\bm{M}_t), 
\label{eq:path_psi_M}
\end{equation}
with the associated probability density [cf.~Eq.~\eqref{eq:p-m-traj}]
\begin{equation}
	p\!\left(\psi_{[0,t)},\bm{M}_t|\psi_0\right) = p(\bm{M}_t | \psi_0) \,\delta\!\left[ \psi_{[0,t)} - \psi_{[0,t)}(\bm{M}_t) \right]. 
	\label{eq:p-M-traj}
\end{equation}
By analogy with full measurement records, we consider the following symmetry property for the coarse-grained measurement records
[cf.~Eq.~\eqref{eq:path-psi-m-symm}]
\begin{equation}
	p\!\left(\psi_{[0,t)},\bm{M}_t|\psi_0\right)	=p\!\left(\sytr\psi_{[0,t)},\sytr{\bm{M}}_t|\sytr\psi_0\right),
	\label{eq:path-psi-M-symm}
\end{equation}
where the symmetry-transformed coarse-grained record is [cf.~Eq.~\eqref{eq:tilde-m}] 
\begin{equation}\label{eq:tilde-M}
	\sytr{\bm{M}}_t=\left[t_1,\pic(\alpha_1);\dots;t_n,\pic(\alpha_n)\right]
\end{equation}	
{for} a permutation $\pic$ {of $\{1,...,\dC\}$}. %such that $\mathcal{U} \mathcal{A}_\alpha\mathcal{U}^\dag=\mathcal{A}_{\pic(\alpha)}$.
 {This property is equivalent to} [cf.~Eq.~\eqref{eq:m-symm}]
\begin{equation}
	p(\bm{M}_t\big|\psi_0) = p(\sytr{\bm M}_t\big|\sytr\psi_0)   \; .
	\label{eq:M-symm}
\end{equation}
{Our results below imply that when this symmetry property of coarse-grained measurements holds true, the permutation $\pic$ in  Eq.~\eqref{eq:tilde-M} is determined by how the composite actions of SJEDs in Eq.~\eqref{eq:superA} are transformed under the symmetry, recall Eq.~\eqref{eq:symm_cond_IIb}. This permutation is always \emph{unique}.}

When discarding coarse-grained measurement records while keeping quantum trajectories, the symmetry property of the former in Eq.~\eqref{eq:path-psi-M-symm} guarantees the symmetry property of the latter in Eqs.~\eqref{eq:tilde-path-psi}, as was the case of for labelled quantum dynamics. Hence, Eq.~\eqref{eq:path-psi-M-symm} implies Symmetry Condition~II, with the same permutation $\pic$.  Crucially, the coarse-grained record $\bm{M}_t$ is constructed such that the reverse implication also holds: we show below that Symmetry Condition~II also implies Eq.~\eqref{eq:path-psi-M-symm}.
{That is, if an ensemble of quantum trajectories features the symmetry property in Eq.~\eqref{eq:tilde-path-psi}, then the corresponding
% joint ensemble of trajectories and coarse-grained measurement records is also symmetric.
combined ensemble of quantum trajectories with associated coarse-grained measurement records has the corresponding symmetry property in Eq.~\eqref{eq:p-M-traj}.
The physical interpretation of this equivalence is discussed in Sec.~\ref{sec:partial-symm-further}.
}

\subsubsection{Partially-labelled {quantum} dynamics}

The description in Eq.~\eqref{eq:path_psi_M} can be equivalently replaced by a \emph{partially-labelled quantum trajectory}, that is the (stochastic) path [cf.~Eq.~\eqref{eq:path-psi-q}]
\begin{equation}\label{eq:path-psi-Q}
	\left(\psi_{[0,t)},\bm{Q}_{[0,t)}\right)=(\psi_\tau, \bm{Q}_\tau)_{\tau\in[0,t)},
\end{equation} 
with a vector $\bm{Q}_t$  in which $(\bm{Q}_t)_\alpha$ {for $\alpha=1,...,\dC$ is} the total number of {outcomes with a label $j$ from $S_\alpha$ in $\bm{m}_t$} [i.e., the number of times $\alpha$ appears in $\bm{M}_t$].  {Therefore, we can view $(\psi_t, \bm{Q}_t)$ as a combined stochastic state of the (quantum) system and a coarse-grained (classical) counter,  and we will refer to $\bm{Q}_t$ as \emph{coarse-grained counts} (obtained by coarse-graining of full counts). The path probabilities obey [cf.~Eq.~\eqref{eq:p-path-q-traj}] 
\begin{equation}\label{eq:p-path-Q-traj}
	p\!	\left(\psi_{[0,t)},\bm{Q}_{[0,t)}|\psi_0\right)=	p\!\left(\psi_{[0,t)},\bm{M}_t|\psi_0\right),
\end{equation}
 %\footnote{** this sentence is not nice}
so the \emph{ensemble of partially-labelled quantum trajectories} is equivalent to the collection of quantum trajectories with associated coarse-grained measurement records in Eq.~\eqref{eq:path_psi_M}.
%and their joint probability density in Eq.~\eqref{eq:p-M-traj}.

Partially-labelled quantum trajectories are sample paths of {a} stochastic process with {the} generator {$\WC^\dag$ below. The generator acts on probability distributions $P_t(\psi,\bm{Q})$
% of $(\psi_t,\bm{Q}_t)$
as  [cf.~Eq.~\eqref{eq:Wf}] }
\begin{multline}\label{eq:Wc}
	\left[\WC ^\dag
	(P_t)\right]\!(\psi,\bm{Q}) =-\nabla\cdot\left[ P_t(\psi,\bm{Q}) \,{\cal B}(\psi)\right]  \\ % \nonumber
	+  \sum_{\alpha=1}^{\dC}\int \!d\psi' \Big[ P_t(\psi'\!,\bm{Q}\!-\!\bm{E}_\alpha) \,W_\alpha(\psi',\psi) 
	\\ \!-\! P_t(\psi,\bm{Q}) \, W_\alpha(\psi,\psi') \Big],
\end{multline}
where  $(\bm{E}_\alpha)_\beta=\delta_{\alpha\beta}$ {for $\beta=1,...,\dC$}, and
\begin{equation}
	\label{eq:W_alpha}
	W_\alpha(\psi,\psi') = \delta\!\left\{ \psi' - \frac{\mathcal{A}_\alpha(\psi)}{\Tr[\mathcal{A}_\alpha(\psi)]}\right\}\Tr[\mathcal{A}_\alpha (\psi)]
\end{equation}
is the {total jump rate from $\psi$ to $\psi'$, as facilitated by jump operators from the SJED $S_\alpha$, i.e.,} $W_\alpha(\psi,\psi')=\sum_{j\in S_\alpha} w_j(\psi,\psi')$ [cf.~Eq.~\eqref{eq:w_j}]. {We then have $\sum_{\alpha=1}^{\dC}W_\alpha(\psi,\psi')=W(\psi,\psi')$ [cf.~Eq.~\eqref{eq:w_alpha}]. The subscript ${\textsc c}$ in $\WC $ refers to the coarse-grained counts.}

\subsubsection{{Weak symmetries of partially-labelled quantum dynamics}}
\label{sec:partial-symm}

To derive {necessary and sufficient} conditions for the {symmetry property} in Eq.~\eqref{eq:M-symm}, we consider 
the symmetry operation $\Yc$ defined  [cf.~Eq.~\eqref{eq:UpsF}] as
\begin{equation}
	[\Yc (P)](\psi, \bm{Q}) = P\!\left[\mathcal{U}^\dag(\psi), \Pic^{-1}(\bm{Q} )\right],
	\label{eq:UpsC}
\end{equation}
where $\Pic(\bm{Q})_\alpha= (\bm{Q})_{\pic(\alpha)}$ with $\pic$ being a permutation of $\{1,\dots,\dC\}$. Then, a \emph{weak symmetry in the partially-labelled {quantum} dynamics} is [cf.~Eq.~\eqref{eq:Wf_symm}]
\begin{equation}\label{eq:Wc_symm}
	\Yc\WC^\dag\Yc^\dag = \WC^\dag.
\end{equation}
Following the same arguments as Sec.~\ref{sec:generator_symm}, one finds that this symmetry is equivalent to Eq.~\eqref{eq:M-symm}. 

{We now establish the \emph{weak symmetry equivalence}:} 
\beq\nonumber
\Yc\WC^\dag\Yc^\dag = \WC^\dag 
\quad \Longleftrightarrow \quad 
\Upsilon
\mathcal{W}^\dag\Upsilon^\dag = \mathcal{W}^\dag. 
\eeq
Here, the implication from left to right states that the unravelled quantum dynamics inherits symmetries of the partially-labelled quantum dynamics, as expected since removing labels cannot break symmetries.  It follows
 directly from the definitions of $\WC^\dag$ and $\Yc$, by summing over all possible states of the counter.
 The implication from right to left means that Symmetry Condition~II is sufficient for Eq.~\eqref{eq:Wc_symm}, which may be verified directly by constructing $\Yc\WC^\dag\Yc^\dag$ and applying  Eq.~\eqref{eq:symm_cond_II}.

\subsubsection{Effects of labelling on Symmetry Conditions}
\label{sec:partial-symm-further}

To end this Section, we provide a physical interpretation of the fact that the partially-labelled quantum dynamics is symmetric under exactly the same conditions as the unravelled quantum trajectory dynamics, in which labels are absent.  
We emphasize that it is not possible in general to reconstruct measurement records from quantum trajectories: two different records
% $\bm{M}_t,\bm{M}_t'$
may correspond to the same trajectory, via Eq.~\eqref{eq:varphi} or \eqref{eq:psi_t_M}.
% The same also holds for full measurement records.

Let us write
the symmetry property  of fully-labelled trajectories [Eq.~\eqref{eq:path-psi-m-symm}] in terms of the conditional distribution of the measurement record, given the trajectory:
\begin{equation}
	p\big(\psi_{[0,t)}\big) p\big(\bm{m}_t|\psi_{[0,t)}\big)	=
	p\big(\sytr\psi_{[0,t)}\big) p\big(\sytr{\bm{m}}_t|\sytr\psi_{[0,t)}\big)
	\label{eq:path-psi-m-cond}
\end{equation}
If the trajectory ensemble features the same symmetry property in Eq.~\eqref{eq:tilde-path-psi}, then the labelling will break the symmetry unless 
\beq
\label{eq:record-m-cond}
p\big(\bm{m}_t|\psi_{[0,t)}\big)=p\big(\sytr{\bm{m}}_t|\sytr\psi_{[0,t)}\big) \quad \forall \, \psi_{[0,t)}.
\eeq

Loosely speaking, the conditional distribution $p(\bm{m}_t|\psi_{[0,t)})$ can be obtained by identifying the jumps in the quantum trajectory $\psi_{[0,t)}$ and associating to each jump a distribution of possible outcomes (labels).\footnote{If there are jump operators such that $J\psi J^\dag \propto \psi$ then it may not be trivial to identify the jumps in $\psi_{[0,t)}$, so the conditional distribution of $\bm{M}_t$ may contain a random number of emitted quanta, as well as random labels.  This does not affect the arguments given here.}  The structure of the labelled quantum dynamics means that for any jump from state $\psi$ to $\psi'$, the probability to assign label $j$ is zero unless ${\cal J}_j(\psi) = r_j \psi'$; otherwise it is  $%P(j|\psi,\psi')=
(r_j/r)$ where $r$ ensures the normalisation.\footnote{Note both $r_j$ and $r$ depend on $\psi$, this structure is familiar from quantum jump Monte Carlo, where $r_j$ is the rate associated with the jump operator $J_j$ acting on $\psi$, i.e. $r_j=\Tr[\mathcal{J}_j(\psi)]$~\cite{Carmichael1993,Dalibard1992,Dum1992,Molmer93,Plenio1998,Daley2014}. In the current context, we only consider jump operators that are consistent with the given quantum trajectory.}
Nothing in this construction ensures Eq.~\eqref{eq:record-m-cond} so this labelling generally breaks the symmetry property of the quantum trajectory ensemble.  However, if Symmetry Condition III holds then ${\cal U}{\cal J}_j(\psi)={\cal J}_{\pi(j)}{\cal U}(\psi)$ which is enough to ensure Eq.~\eqref{eq:record-m-cond}, and hence that the quantum trajectory ensemble is symmetric.

Repeating this argument for the partially-labelled quantum dynamics, the relevant condition is
% Eq.~\eqref{eq:record-m-cond} but with $\bm{m}_t$ replaced by $\bm{M}_t$.
\begin{equation}\label{eq:record-M-cond}
    p\big(\bm{M}_t|\psi_{[0,t)}\big)=p\big(\sytr{\bm{M}}_t|\sytr\psi_{[0,t)}\big) \quad \forall \, \psi_{[0,t)}
\end{equation}
To construct $p(\bm{M}_t|\psi_{[0,t)})$, the probability to assign coarse-grained label $\alpha$ is zero unless ${\cal A}_\alpha(\psi) = r^{(\alpha)} \psi'$; otherwise it is  $%P(\alpha|\psi,\psi')=
(r^{(\alpha)}/r)$ where $r$ is the same normalisation constant as above.  By analogy with the fully-labelled case, Symmetry Condition II is then sufficient to ensure symmetry of the partially-labelled trajectory ensembles, via Eq.~\eqref{eq:record-M-cond}.

Combining this analysis with properties of quantum trajectories from Sec.~\ref{sec:generator_symm}, one sees that if two representations of the QME both obey Symmetry Condition II with the same composite actions ${\cal A}_1,{\cal A}_2\dots$ then they generate the {same ensemble of partially-labelled quantum trajectories}.\footnote{We specifically assume that the two representations have $\tilde{\cal A}_j={\cal A}_j \forall j$.} This property can be traced back to the definition of the SJEDs~\cite{Generators}.  Moreover, by isometric mixing of the jumps in each SJED, it is always possible to construct another (third) representation that again generates the same ensemble, but also respects Symmetry Condition III, and hence generates a symmetric ensemble of fully-labelled quantum trajectories.  (See Sec.~\ref{sec:SJED_reps}.)   
Then the interpretation of the difference between Symmetry Conditions II and III is that Symmetry Condition II remains valid on any isometric mixing that preserves the composite actions: this generally breaks the symmetry of the fully-labelled quantum trajectory ensemble, but the coarse-graining step in Fig.~\ref{fig:records} means that the partially-labelled quantum trajectories are unaffected by the mixing and hence retain the symmetry.

\section{Symmetries of joint quantum dynamics}\label{sec:cMPS}

In Sec.~\ref{sec:traj}, the system was treated quantum-mechanically through its density matrix, but its environment was {described} classically, through  measurement {records}.
This Section considers {a} joint quantum state for the system and its environment, which {in the absence of measurements} remains pure under {the} joint evolution.

An overview of the resulting scenario is {shown} in Fig.~\ref{fig:states}.  The joint quantum state for system and environment is a continuous matrix product state (cMPS), {which} evolves unitarily. We show that this evolution is symmetric whenever the QME is symmetric, i.e., whenever Symmetry Condition~I holds.  Projective measurements on the environment allow the evolution of the {resulting average} state {of the system and the environment} to be described by a continuous matrix product operator (cMPO).  {As we explain below, } depending on the {structure} of the measurement, the (non-unitary) evolution of this cMPO may be symmetric under either Symmetry Condition~II or Symmetry Condition~III.

We make two observations, before starting the analysis.  First, while the joint dynamics of system and environment is naturally described in the formalism of cMPOs, the results can be equivalently stated in the language of quantum stochastic calculus.  Indeed, most of the results are presented in a way that does not rely on the calculus of matrix product states and operators.  Second, we aim to keep this Section self-contained so we present the joint dynamics without explicit reference to the measurement records and quantum trajectories discussed in  Sec.~\ref{sec:traj}.  The connections between these two formalisms will be presented in the following Sec.~\ref{sec:discussion1}.

\subsection{The joint state and its dynamics}
\label{sec:stoch-ham}

As in previous Sections, we consider a system interacting with a surrounding field (environment) through the emission and absorption of quanta.  The description of the evolution can be equivalently phrased in terms of the quantum stochastic calculus~\cite{Hudson1984,Parthasarathy1992}, the input-output formalism~\cite{Gardiner2004,Wiseman2010}, as well as continuous matrix product states~\cite{Verstraete2010,Osborne2010}. {Here,} we choose the last framework {while} borrowing  notation from quantum stochastic calculus (cf.~\cite{Kiukas2015,Garrahan2016}).

We assume that the system and the field are initially uncorrelated and described by a pure state\footnote{For simplicity of notation, we write state vectors rather than density matrices when considering pure states of system and environment.} 
\begin{equation}\label{eq:cMPS_t0}
	|\Psi_0\rangle=|\psi_0\rangle \otimes |\text{vac}\rangle,
\end{equation}
where the environment is initialised in the vacuum state $|\vac\rangle$.  (It is possible to consider other initial states for the environment, see Appendix~\ref{app:non-vac}, but we consider a vacuum initial state throughout the main text.)
The joint dynamics of the system and the environment is unitary {so their} joint state {can be written} as 
\begin{equation}\label{eq:cMPS_dH}
	|\Psi_t\rangle= %\mathcal{T} e^{-i\int_0^t d H_{\tau} } \big( |\psi_0\rangle\otimes |\text{vac}\rangle \big) ,
	\mathcal{T} e^{-i\int_{t'}^t d H_{\tau} } 	|\Psi_{t'}\rangle ,
	%|\Psi_0\rangle, %|\psi_0\rangle\otimes |\text{vac}\rangle,
\end{equation}
where $\mathcal{T}$ denotes time ordering and the \emph{stochastic Hamiltonian} \cite{Hudson1984,Parthasarathy1992} is
\begin{equation}\label{eq:dH_tau}
	d H_{t}=  H \otimes \IE\,dt+ i \sum_{j=1}^{d} \left(J_j \otimes dB^\dagger_{j,t} -J_j^\dagger\otimes dB_{j,t}\right)\!.
	%d H_{\tau}\equiv  H \otimes \IE\,d\tau+ i \sum_j \left(J_j \otimes dB^\dagger_{j,\tau} -J_j^\dagger\otimes dB_{j,\tau}\right)\! .
\end{equation}
Here, $H,J_1,\dots,J_d$ is a representation of the system QME, see Eq.~\eqref{eq:LL-outline}, while
{$dB_{j,t}$ and} $dB^\dag_{j,t}$ denote the (unnormalised) {annihilation and} creation operators, respectively, for a quantum of type $j$ emitted at time $t$. 
{Environmental basis states are of the form $dB_{j_n,t_n}^\dag \dots dB_{j_2,t_2}^\dag dB_{j_1,t_1}^\dag|{\rm vac}\rangle$ with $0<t_1<t_2<\dots<t_n$.  The resulting Ito table~\cite{Gardiner2004} is}
%These together give rise to an Ito table:
\begin{subequations}\label{eq:Ito}
	\begin{align}
		dB_{j,t} dB^\dagger_{k,t}&=  \delta_{jk} \IE\,dt , \\
		dB^\dagger_{j,t} dB_{k,t}&=  0, \\
		dB_{j,t} dB_{k,t}&=  0.
	\end{align}
\end{subequations}

This joint state in Eq.~\eqref{eq:cMPS_dH} is a cMPS~\cite{Verstraete2010,Osborne2010}.  
The stochastic differential equation in Eq.~\eqref{eq:Psi_dot} is to be understood as a shorthand notation for Eq.~\eqref{eq:cMPS_dH}.\footnote{Eq.~\eqref{eq:Psi_dot} is expressed in Stratonovich convention, although it is also possible to use Ito.  Eq.~(\ref{eq:cMPS_dH}) is true in both conventions, as are the other results of this Section.}
By tracing out the environment, the density matrix of the system in Eq.~\eqref{eq:rho_t} is recovered,
\begin{equation}\label{eq:cMPS_trace}
	\rho_t=% \Tr_E(\Psi_t), %|\Psi_0\rangle, %|\psi_0\rangle\otimes |\text{vac}\rangle,
	\Tr_E\left(|\Psi_t\rangle\!\langle 	\Psi_t|\right).
\end{equation}
Note that the left hand side solves the QME and is independent of its representation.  The stochastic Hamiltonian and $|\Psi_t\rangle$ both depend on the representation but tracing over the environment eliminates this dependence.
We next clarify how cMPSs for different representations are related, before analysing symmetries of the joint dynamics.

\begin{figure}
    \centering
    \includegraphics[width=\linewidth]{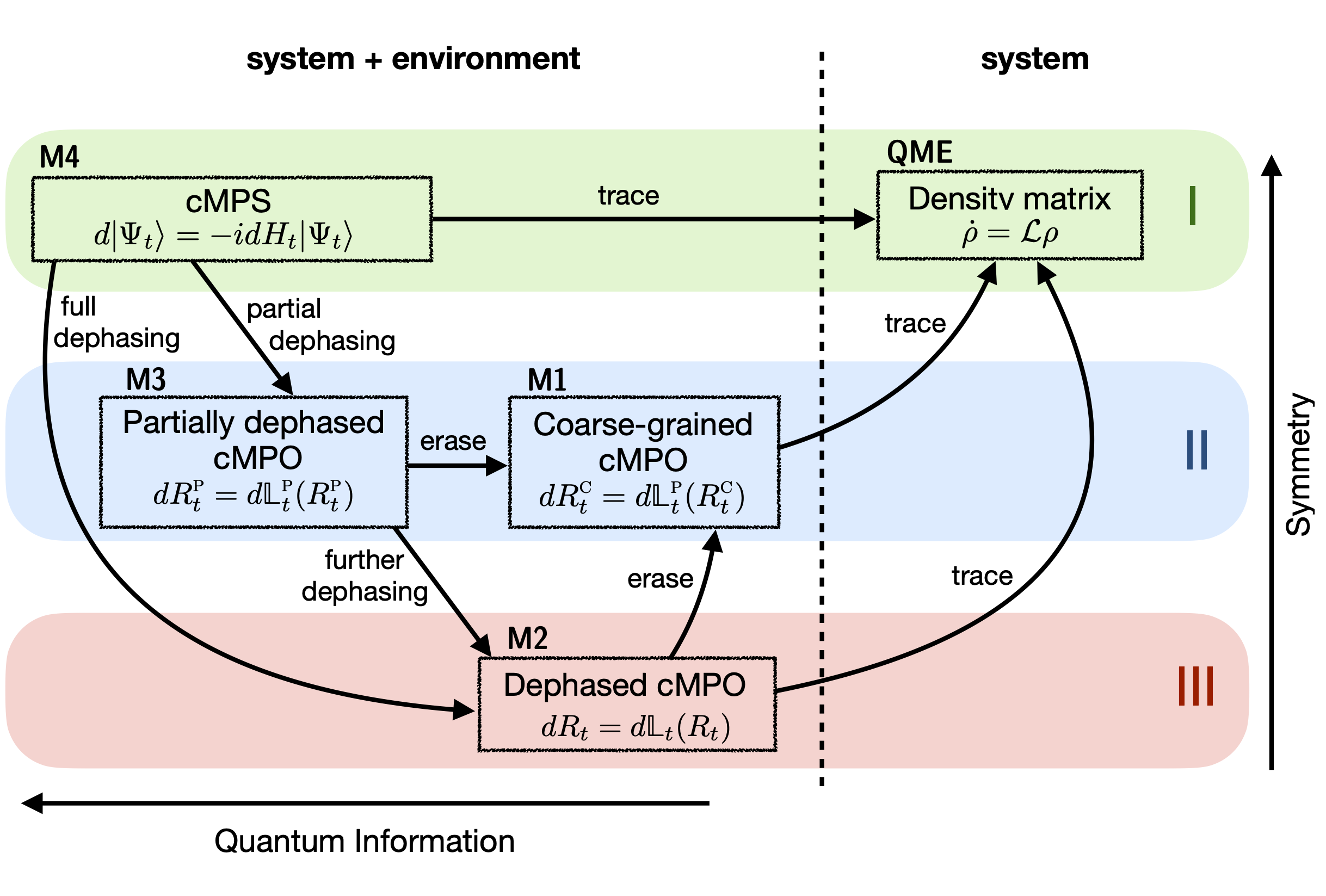}
    \caption{Detailed first and third columns of Fig.~\ref{fig:outline}.   Sec.~\ref{sec:cMPS} gives the meanings of the boxes, the relationships between them (illustrated by arrows), and the conditions under which they are symmetric.} 
    \label{fig:states}
\end{figure}

\subsection{{Representations and measurement basis}}
\label{sec:measure-basis}

{Physically, each representation of the QME corresponds to a measurement procedure in the environment, which also determines the ensemble of quantum trajectories, via the system's conditional state.  Indeed, we explain in Sec.~\ref{sec:dephased_symm} that the counting measurement corresponds to dephasing in a particular basis of quanta, which is determined by the choice of representation and to which we refer as the measurement basis (of the environment).  Here, we explain that transformations between different representations of the QME can be related to transformations of stochastic Hamiltonians (basis changes) for the cMPS.}

{Two types of transformations are relevant.
First, we consider a unitary transformation on the environment}
\begin{equation}\label{eq:rotating_frame}
	|\Psi_t'\rangle = \left(\mathbb{1}\otimes D_t \right)|\Psi_t\rangle , \qquad D_t = e^{-i\int_0^t dQ_\tau} ,
\end{equation}
with 
\begin{equation}\label{eq:Qt}
	dQ_t =  \frac{i}{\dS}\sum_{j=1}^d\left[dB_{j,t}\Tr(J_j^\dag)-dB^\dag_{j,t}\Tr(J_j)\right] .
\end{equation}
In the context of linear optics, this transformation corresponds to a coherent displacement of the field.\footnote{In contrast to Eq.~\eqref{eq:cMPS_dH}, no time-ordering operator is needed in Eq.~\eqref{eq:rotating_frame} because $dQ_t$ operators at different times commute with each other.}  
{Such a time-dependent unitary transformation takes the environment into a rotating frame of reference.
The transformed stochastic Hamiltonian is denoted $d H_{\tau}'$ and defined as
\beq
\label{eq:dH-dH'}
\mathcal{T} e^{-i\int_{t'}^t d H_{\tau}' } = \left(\mathbb{1}\otimes D_t \right) \big( \mathcal{T} e^{-i\int_{t'}^t d H_{\tau} } \big) \left(\mathbb{1}\otimes D_{t'} \right)^\dag .
\eeq  
This shows explicitly the connection to the interaction picture of quantum mechanics.
From Eqs.~(\ref{eq:cMPS_dH},\ref{eq:rotating_frame}) the transformed joint state evolves unitarily as
\begin{equation}
	\label{eq:def_dH'}
	|\Psi_t'\rangle = \mathcal{T}e^{-i\int^t_{t'} dH'_\tau} |\Psi'_{t'}\rangle \; .
\end{equation}
%Using this with Eqs.~(\ref{eq:cMPS_dH},\ref{eq:rotating_frame}), 
We show
in Appendix~\ref{app:rotate} that the transformed stochastic Hamiltonian is given by
 \begin{equation}\label{eq:dH'}
 	dH'_t = H' \! \otimes \IE \,dt
 	+ i \sum_{j=1}^d \left[J_j' \otimes dB^\dagger_{j,t} -\left(J_j'\right)^\dagger\otimes dB_{j,t}\right],
 \end{equation}
where $H',J_1',\dots,J_d'$ are related to $H,J_1,\dots,J_d$ by Eq.~\eqref{eq:traceless_rep}.   That is,  $H',J_1',\dots,J_d'$ is a traceless representation of the system QME.  Since the transformation in Eq.~\eqref{eq:rotating_frame} acts only on the environment both cMPSs recover the same system (QME) dynamics, see Eq.~\eqref{eq:cMPS_trace}.}

{The second type of transformation corresponds in linear optics to isometric mixing of the emitted quanta.  For this we consider another set of creation operators, denoted by $d\tilde{B}_{j,t}^\dag$ with $j=1,2,\dots,\tilde{d}$; annihilation operators are defined similarly, and the corresponding vacuum is denoted $|\tilde{\rm vac}\rangle$.  We assume $\tilde{d}\geq d$, the other case is analogous.  Then we introduce an isometric operator $V_E$
% (acting on the environment)
that transforms between the two sets of quanta as:
\begin{equation}\label{eq:VE}
	%d\tilde{B}^\dag_{j,t} = \sum_{k=1}^d \mathbf{V}_{jk}^*dB_{k,t}^\dag, \quad |\tilde{\vac}\rangle = |\vac\rangle,
	V_E \, d {B}_{j,t}^\dag \, V_E^\dag = \sum_{k=1}^{\tilde d} \mathbf{V}_{kj}\, d\tilde B_{k,t}^\dag, \quad V_E |{\vac}\rangle = |\tilde{\vac}\rangle,
\end{equation}
where $\mathbf{V}$ is  a $\tilde{d}\times d$ isometric matrix.\footnote{The isometry ensures that creation/annihilation operators of both representations respect the Ito table Eq.~\eqref{eq:Ito}; we also have $V_E^\dag V_E=\mathbb{1}_E$.}}

We consider the action of this operator on the stochastic Hamiltonian for the traceless representation $H',J_1',\dots,J_d'$:
\begin{equation}
\label{eq:dH-dual}
 d\tilde{H}'_t  = 
(\mathbb{1} \otimes V_E) \, d{H}'_t \, (\mathbb{1} \otimes V_E)^\dag   .
\end{equation}
This (isometric) transformation is analogous to the unitary transformation Eq.~\eqref{eq:dH-dH'}, 
so $d\tilde{H}'_t$ corresponds to the same QME as $d{H}'_t$, because of Eq.~\eqref{eq:cMPS_trace}.
One finds that
 \beq\label{eq:dH'tilde}
 	d\tilde{H}'_t%&= H' dt\otimes \IE \\
 	 =\tilde{H}' \! \otimes \IE \,dt+ i\sum_{k=1}^{\tilde{d}}\left[\tilde{J}_k'\otimes d\tilde{B}^\dag_{k,t}
	 -\!(\tilde{J}_k')^\dag\!\!\otimes d\tilde{B}_{k,t}\right] ,
 \eeq
where $\tilde{H}',\tilde{J}_1',\dots,\tilde{J}_{\tilde d}'$ are related to $H',J_1',\dots,J_d'$ according to Eq.~\eqref{eq:theorem_I}.

From Theorem I, we conclude that by combining suitable environmental transformations $D_t$ and $V_E$, one may transform between the stochastic Hamiltonians (and hence also the cMPSs) for any two representations of the QME.  Specifically, if $dH_t,d\tilde{H}_t$ are two such stochastic Hamiltonians (with $\tilde d\geq d)$ then 
\beq
\label{eq:general_rep_change}
\mathcal{T} e^{-i\int_{t'}^t d  \tilde H_{\tau} } = \left(\mathbb{1}\otimes  V_{E,t}\right) \big( \mathcal{T} e^{-i\int_{t'}^t d H_{\tau} } \big) \big(\mathbb{1}\otimes V_{E,t} \big)^\dag ,
\eeq
where $V_{E,t} =  \tilde{D}_t^\dag V_{E} D_t$, and the matrix $\mathbf{V}$ in Eq.~\eqref{eq:dH-dual} must be compatible with Eq.~\eqref{eq:theorem_I};\footnote{We consider two representations of the same QME so Theorem I ensures that such a matrix always exists, although it is not unique in general.  In this case Eq.~\eqref{eq:general_rep_change} holds simultaneously for more than one choice of $V_{E,t}$.} 
also $\tilde{D}_t$ is defined similarly to \eqref{eq:rotating_frame}, but now for the representation $\tilde{H},\tilde{J}_1,\dots,\tilde{J}_{\tilde d}$.
This construction shows explicitly how changes of measurement basis correspond to different representations of a given QME. 
The corresponding cMPS is
\beq
%|\tilde\Psi_t\rangle = 
(\mathbb{1}\otimes V_{E,t}) |\Psi_t\rangle \; .
\label{eq:kasia-tilde}
\eeq
These results have implications for symmetries of the stochastic Hamiltonian, as we discuss in the next Subsection.

\subsection{Symmetries of the joint state}\label{sec:cmps_symm}

{We now explain that the joint dynamics of the} cMPS (Box \Bcmps~of Fig.~\ref{fig:states}) is symmetric under Symmetry Condition~I, that is whenever {the quantum master dynamics features a weak symmetry. We also discuss the resulting properties of the cMPS}.

\subsubsection{{Stationary and dynamical symmetries}}

The joint dynamics has a \emph{joint stationary symmetry} associated with the system superoperator $\mathcal{U}$ if 
\begin{equation}
	\label{eq:weakU_dH}
	\left (U\otimes \UE \right)dH_t \left(U\otimes \UE\right)^\dag = dH_t \;,
\end{equation}
where $\UE$ is a ({time-independent}) unitary operator on the environment.
{This definition is equivalent to
\begin{equation}
\label{eq:weakU_dH_exp}
   \left(U\otimes U_{E}\right) \mathcal{T} e^{-i\int_{t'}^t dH_\tau} \left(U\otimes U_{E}\right)^\dag = \mathcal{T} e^{-i\int_{t'}^t dH_\tau}. 
\end{equation}
holding for all $t,t'$.
We show below that in order for this symmetry to hold, $\UE$ must correspond to unitary mixing of emitted quanta, specifically
\begin{equation}\label{eq:UE}
	%\superUE \big( dB_{j,t}^\dag \big) = \sum_{k=1}^d \mathbf{U}_{jk}^* dB_{k,t}^\dag, \quad \UE|\vac\rangle = |\vac\rangle,
		\UE \,  dB_{j,t}  \,\UE^\dag = \sum_{k=1}^d \mathbf{U}_{jk}\, dB_{k,t}, \quad \UE|\vac\rangle = |\vac\rangle,
	%	\UE dB_{j,t}^\dag \UE^\dag = \sum_j \mathbf{V}_{jk}^* dB_{k,t}^\dag, \quad \UE|\vac\rangle = |\vac\rangle.
\end{equation}
for a unitary $d \times d$ matrix $\mathbf{U}$.
We also consider a \emph{joint dynamical symmetry} of $d H_t$~\cite{Buca2022}
defined as  
\beq
\label{eq:symm_dH_Ut}
    \left(U\otimes \UEt\right) \mathcal{T} e^{-i\int_{t'}^t dH_\tau} \left(U\otimes \UEtp\right)^\dag = \mathcal{T} e^{-i\int_{t'}^t dH_\tau} ,
\eeq
where $\UEt$ is a time-dependent unitary operator.\footnote{As above, we write Eq.~\eqref{eq:symm_dH_Ut} in integral/exponential form to avoid potential confusion with Ito/Stratonovich calculus.}  
If $\UEt$ is independent of $t$, Eq.~\eqref{eq:symm_dH_Ut} {reduces} to Eq.~\eqref{eq:weakU_dH_exp}.
In practice, we always take 
\begin{align}\label{eq:UEt}
	\UEt = 
	{e^{i\int_0^t dQ_\tau}} \UE \,{e^{-i\int_0^t dQ_\tau}} \; .
\end{align}
which is the transformation of $\UE$ into the rotating frame of reference of Eq.~\eqref{eq:rotating_frame}, see also below.
%The physical meaning of this choice is reladiscussed below.

The symmetry {in Eq.~\eqref{eq:symm_dH_Ut}} can be expressed in terms of the (determinstic) path that is followed by the symmetry-transformed cMPS
\begin{equation}\label{eq:tilde-path-Psi}
    |\sytr\Psi_t\rangle = (U\otimes \UEt) |\Psi_t\rangle. 
\end{equation}
Equation~\eqref{eq:symm_dH_Ut} implies that the path followed by this $|\sytr\Psi_t\rangle$ can also be constructed as in Eq.~\eqref{eq:weakU_dH_exp}, by taking initial condition $|\sytr\Psi_0\rangle=(U\otimes\UE)|\Psi_0\rangle$ at $t'=0$, and the same stochastic Hamiltonian.
This situation is analogous to Eq.}~(\ref{eq:tilde-path-rho}) for the QME.
For a symmetric initial state, we then have for {all times}
\begin{equation}\label{eq:symm_Psi0_evolution}
    (U\otimes \UE)|\Psi_{0}\rangle = e^{i\phi} |\Psi_{0}\rangle \implies (U\otimes \UEt)|\Psi_{t}\rangle = e^{i\phi}|\Psi_{t}\rangle.
\end{equation}
That is, an initial state prepared in an eigenstate of $U\otimes \UE$ evolves at time $t$ into an eigenstate of $U\otimes \UEt$ with the same eigenvalue.  This is analogous to Eq.~\eqref{eq:rho-symm} for the QME.

When tracing out the environment as in Eq.~\eqref{eq:cMPS_trace}, the symmetry property for the cMPS in Eq.~\eqref{eq:symm_dH_Ut} implies the symmetry property for the density matrix in Eq.~\eqref{eq:tilde-path-rho}, and thus the weak symmetry of the quantum master dynamics in Eq.~\eqref{eq:weakU_QME}.  In order for this to be the case, the symmetry of the cMPS must be separable, as assumed in Eq.~\eqref{eq:symm_dH_Ut}.  This observation justifies our restriction to symmetries of this form.
%We show next that the weak symmetry of the quantum master dynamics is always associated with the presence of a joint separable symmetry.

\subsubsection{Necessary and sufficient conditions\\ (Symmetry Condition~I)}
\label{sec:symm-joint-uni-conds}

{Consider a traceless representation of the QME and its corresponding stochastic Hamiltonian $dH'_t$ [cf. Eq.~\eqref{eq:dH'}].  A direct calculation shows that the symmetry}  
\begin{equation}
\label{eq:UUE-dH}
\left(U\otimes \UE\right)dH'_t \left(U\otimes \UE\right)^\dag = dH'_t,
\end{equation}
{holds if and only if  
 $\mathcal{U}(H') = H'$ and $\mathcal{U}(J_j') = \sum_{k=1}^{d} \mathbf{U}_{jk}J_k'$, where $\mathbf{U}$ is the matrix appearing in Eq.~\eqref{eq:UE} [cf. Eqs.~\eqref{eq:VE} and \eqref{eq:dH'tilde}]. Recalling Sec.~\ref{sec:qme_symm}, one sees that such a symmetry exists if and only if Symmetry Condition I holds.
That condition does not uniquely fix $\mathbf{U}$ in general (recall Appendix~\ref{app:boldU}), in which case Eq.~\eqref{eq:UUE-dH} holds simultaneously for all the corresponding choices of $U_E$.
}

{The above result applies to traceless representations.  However, Eq.~\eqref{eq:dH-dH'} and \eqref{eq:dH'} show how the stochastic Hamiltonian for a general representation can be transformed to a traceless one by using a rotating reference frame.}  Together with Eqs.~(\ref{eq:UEt},\ref{eq:UUE-dH}) this implies
\begin{equation}
	\label{eq:symm_dH_Ut_expanded-v2}
	\left(U\otimes \UEt\right)  \mathcal{T} e^{-i\int_{t'}^t dH_\tau}  \left(U\otimes \UEtp\right)^\dag = \mathcal{T} e^{-i\int_{t'}^t dH_\tau}\,.
\end{equation}
This time-dependent symmetry holds for any stochastic Hamiltonian as long as (i) the underlying QME is symmetric and (ii) the matrix $\mathbf{U}$ in Eq.~\eqref{eq:UE} is consistent with Eq.~\eqref{eq:symm_cond_Ib} in Symmetry Condition I.
We also show in Appendix~\ref{app:UEt=UE} that the joint symmetry is stationary if and only if Eq.~\eqref{eq:not_cond_IIb} is satisfied, that is
	\beq\nonumber
\UEt=\UE \quad \Longleftrightarrow \quad \text{Eq.}~\eqref{eq:not_cond_II}.
	\eeq
We return to this point below when considering how the joint symmetry can persist despite performing measurements.}

Recall that Symmetry Condition I holds for all representations of a weakly-symmetric QME.  Since this condition is sufficient for a symmetry of the stochastic Hamiltonian, it follows that all representations of such a QME must have symmetric stochastic Hamiltonians.  This result is natural because stochastic Hamiltonians for different representations are related by
Eq.~\eqref{eq:general_rep_change}.
Specifically, consider two stochastic Hamiltonians $dH_t,d\tilde{H}_t$ (with $\tilde{d}\geq d$), that correspond to different representations of a weakly-symmetric QME.  
{Using the symmetry of $dH_t$ in Eq.~\eqref{eq:symm_dH_Ut} together with Eq.~\eqref{eq:general_rep_change}, one sees that 
\beq\label{eq:Yt}
( U \otimes \tilde Y_{E,t} ) \, {\cal T} e^{-i \int_{t'}^t d\tilde{H}_\tau } \,  ( U \otimes \tilde Y_{E,t'} )^\dag = {\cal T}  e^{-i \int_{t'}^t d\tilde{H}_\tau } .
\eeq
with $\tilde Y_{E,t}=\tilde D_t^\dag V_{E} U_{E} V_{E}^\dag \tilde D_t$.
If $\tilde{d}=d$ then  $V_E$ is unitary, so this is a unitary symmetry of $d\tilde{H}_t$. In this case the underlying environmental symmetry operator is $\tilde{U}_{E}=V_{E} U_{E} V_{E}^\dag$, which one 
 recognises as the symmetry operator $U_E$ of the original representation, transformed into the new measurement basis.
For $\tilde d>d$, the operator $\tilde Y_{E,t}$ is not unitary, and an additional step is required to relate the  unitary symmetry of $d \tilde{H}_\tau$ to the change of basis, details are given in 
Appendix~\ref{app:mpo-UCB}.}
The essential feature is that the symmetries of different representations are related through Eq.~\eqref{eq:general_rep_change}.
Physically, this means that all representations of a weakly symmetric QME share the same underlying symmetry, with corresponding symmetry operators related by changes of basis.

{To conclude this analysis of the stochastic Hamiltonian, recall that we exploited Symmetry Condition~I to ensure Eq.~\eqref{eq:UUE-dH}, which is also equivalent to Eq.~\eqref{eq:symm_dH_Ut_expanded-v2}. Before, we also argued that such a joint separable symmetry implies the weak symmetry of quantum master dynamics. Thus, we conclude that} Symmetry Condition~I is necessary and sufficient for Eq.~\eqref{eq:UUE-dH} {to hold, so that}:	
\smallskip
\\
{\emph{The joint unitary dynamics features {a separable} symmetry if and only if Symmetry Condition~I holds.}}
\vspace{4mm}
\\
This result corresponds to the first row of Figs.~\ref{fig:outline} and~\ref{fig:states}}. Appendix~\ref{app:non-vac} shows that an analogous result applies if {the field is not initially in the vacuum [cf.~Eq.~\eqref{eq:cMPS_t0}].

\subsubsection{{From strong to weak joint symmetries}}

The joint stationary symmetry in Eq.~\eqref{eq:weakU_dH} is similar to the weak symmetry of the quantum master dynamics in Eq.~\eqref{eq:weakU_QME}, but the dynamics is unitary so this is a strong symmetry~\cite{Buca2012,Albert2014}. This also the case for a general time-dependent symmetry in Eq.~\eqref{eq:symm_dH_Ut}, as evidenced by the corresponding conservation laws in Eq.~\eqref{eq:symm_Psi0_evolution} (cf.~Ref.~\cite{Gough2015}). 

To obtain corresponding weak symmetries, we consider a joint density matrix of the system and the environment, which is a cMPO (see also~\cite{Verstraete2004}).  For the unitary evolution considered so far,  
this cMPO is pure and equal to $\Psi_t = |\Psi_t\rangle\!\langle\Psi_t|$.
From Eq.~\eqref{eq:cMPO_dH}}, it evolves as 
\begin{equation}\label{eq:cMPO_dH}
\Psi_t = {\cal T} {\rm e}^{{-i}\int_{t'}^t d{\mathbb{H}}_\tau}(\Psi_{t'})
\end{equation}
 with
 	\begin{equation}\label{eq:Ht-double}
 		-i \,d\mathbb{H}_t(\Psi_t) =  %-i [ dH_t, \Psi_t]. 
 		-i dH_t \Psi_t + i \Psi_t dH_t.
 		\end{equation}
The symmetry in Eq.~\eqref{eq:symm_dH_Ut_expanded-v2} becomes 
\begin{equation}
\label{eq:symm-double}
\mathbb{U}_t \,\mathcal{T} e^{-i\int_{t'}^t d{\mathbb{H}}_\tau} \, \mathbb{U}_{t'}^\dag= 
 \mathcal{T} e^{-i\int_{t'}^t d{\mathbb{H}}_\tau}
\end{equation}
with
\begin{equation}\label{eq:Ut-double}
  \mathbb{U}_t = {\cal U} \otimes \superUEt,
\end{equation}
{where} $\superUEt(\cdot) = \UEt(\cdot)\UEt^\dag$.

Note that $d\mathbb{H}_t$ and $\mathbb{U}_t$ are superoperators acting {jointly on the system and environment, which are} analogous to  ${\cal L}$ and ${\cal U}$ for the system.  Thus, Eq.~\eqref{eq:symm-double} describes a {weak joint symmetry} in analogy to Eq.~\eqref{eq:weakU_QME}.  In particular, the results analogous to Eqs.~\eqref{eq:tilde-path-rho} and~\eqref{eq:rho-symm} follow for the pure cMPO. 
That is, 
\begin{equation}\label{eq:tilde-path-Psi_double}
	\sytr \Psi_t =   \mathbb{U}_t (\Psi_t)
\end{equation}
is the solution of the dynamics in Eq.~\eqref{eq:cMPO_dH} for the symmetry transformed initial joint state $\sytr{\Psi}_0=\mathbb{U}(\Psi_0)$, which we refer to as its (weak)  symmetry property [cf.~Eqs.~\eqref{eq:tilde-path-Psi}]. When the initial joint state is symmetric, it remains so at all times [cf. Eq.~\eqref{eq:symm_Psi0_evolution}]
\begin{equation}\label{eq:symm_Psi0_evolution_double}
	\mathbb{U}(\Psi_0)=(\Psi_0)	\implies   \mathbb{U}_t(\Psi_0)=(\Psi_0).
	%  \sytr{\Psi}_0= \Psi_0 \implies    \sytr{\Psi}_t= \Psi_t .
\end{equation}
Recall that such symmetries hold whenever Symmetry Condition I applies, so whenever the QME is symmetric.  The following subsections analyse how counting measurements can break this symmetry.

\renewcommand{\l}{\langle}
\renewcommand{\r}{\rangle}
\subsection{Symmetries of {fully-dephased} cMPO}\label{sec:dephased_symm}

This subsection discusses the effect of counting measurements on the {joint state of system and enviroment, which results in a fully-dephased c}MPO (Box \Bdcmps \, in Fig.~\ref{fig:states}).  We explain that the dynamics of this {mixed joint state} is symmetric under Symmetry Condition~III.

\subsubsection{{Joint dynamics under counting} measurement}

As in Sec.~\ref{sec:generator_symm}, we  assume that  quanta emitted into the environment are detected by projective measurement, through their (time-dependent) number operators. This can be done either at a final time or continuously as quanta are emitted resulting in the same joint state. We refer to those as \emph{full} measurements, to  distinguish from other measurements considered later. 

The joint state after measurement is obtained from  $\Psi_t$ by removing all coherences between different numbers (or types) of quanta. It is in general a mixed cMPO that we denote by $R_t$.  We refer to it as a \emph{dephased cMPO}\footnote{We sometimes also write ``fully-dephased cMPO'', for the avoidance of possible confusion with other types of cMPO discussed below}. Its evolution is
\begin{equation}
	R_t = \mathcal{T}e^{\int_{t'}^t \dL_\tau} (R_{t'}) \, ,
	\label{eq:QME-MPO}
\end{equation}	
with a superoperator 
\begin{multline}
    \dL_t(\deph_t) =  \left[-i(H_\text{eff}\otimes\IE)\deph_t+i\deph_t(H_\text{eff}^\dag\otimes\IE)\right] dt \\
    +\sum_{j=1}^d (J_j\otimes dB^\dag_{j,t})\deph_t(J_j^\dag\otimes dB_{j,t}) \,,
    \label{eq:dLf}
\end{multline}
which is analogous to the {quantum master} operator ${\cal L}$, but now acting on the {joint} cMPO instead of a {system} density matrix.  Further details of the cMPO $R_t$ are given in Sec.~\ref{sec:discussion1}, including its relationship to counting measurements.

Comparing with the system master operator, one identifies $J_j\otimes dB^\dag_{j,t}$ for $j=1,\dots,d$ as {joint} jump operators that act {separably} on the system and the environment. The joint effective Hamiltonian $H_\text{eff}\otimes\IE$ has a Hermitian part $H\otimes\IE$ that originates from the system Hamiltonian; its non-Hermitian contribution is related to the joint jump operators $J_j\otimes dB_{j,t}^\dag$, since by the Ito table {in Eq.~\eqref{eq:Ito} $ (J_j^\dag\otimes dB_{j,t}) (J_j\otimes dB^\dag_{j,t})=J_j^\dag J_j \otimes\IE dt $.
In the outline, we wrote the equation of motion \eqref{eq:QME-MPO} in shortand notation as Eq.~\eqref{eq:dL_intro}.\footnote{{Eq.~\eqref{eq:dL_intro} corresponds to Stratonovich convention.}}

{As in the case of a pure joint state, the density matrix of the system in Eq.~\eqref{eq:rho_t} is recovered by tracing out the environment
\begin{equation}\label{eq:cMPO_trace}
	\rho_t=
	\Tr_E\left(R_t\right).
\end{equation}
This follows from Eq.~\eqref{eq:cMPS_trace} by observing that the measurement is performed on the environment only.}

\subsubsection{{Weak joint symmetries}}
\label{sec:weak-joint-symm-def}

{If the measurement does not break the joint symmetry in the unitary dynamics, the dynamics of $R_t$ \emph{retains} the weak dynamical symmetry of Eq.~\eqref{eq:symm-double}. That is,} 
\begin{equation}
\label{eq:symm-deph-rot}
\mathbb{U}_t \,  \mathcal{T} e^{\int_{t'}^t d\mathbb{L}_\tau}  \, \mathbb{U}_{t'}^\dag= 
 \mathcal{T} e^{\int_{t'}^t d\mathbb{L}_\tau} \; .
\end{equation}  
{For a stationary symmetry, $\mathbb{U}_t$ in Eq.~\eqref{eq:Ut-double} becomes}
	\begin{equation}\label{eq:U-double}
		\mathbb{U} = {\cal U} \otimes \superUE \; 
	\end{equation}
{and Eq.~\eqref{eq:symm-deph-rot} can be expressed in terms of the generator in Eq.~\eqref{eq:dLf} as}
\begin{equation}\label{eq:dLsymm}
   \mathbb{U}\, \dL_t\, \mathbb{U}^\dag = \dL_t.
\end{equation}
We show below that this is always the case when Eq.~\eqref{eq:symm-deph-rot} holds.

{Another way to understand this symmetry is to describe the measurement process with dephasing, which occurs at a final time $t$, or continuously.  Specifically $R_t = \mathbb{F}(\Psi_t)$ where $\mathbb{F}$ is a (full) dephasing 
% (projection)
superoperator, independent of time.  Then a stationary symmetry of the unitary dynamics [Eq.~\eqref{eq:weakU_dH_exp}] also exists as a symmetry of dephased joint dynamics [Eq.~\eqref{eq:dLsymm}], if the dephasing operator is symmetric: $\mathbb{U}^\dag \mathbb{F} \mathbb{U} = \mathbb{F}$.  We return to this point in Sec.~\ref{sec:discussion1} below, where we discuss how the symmetries of different dynamics are related to each other.}

Analogously  to Eq.~\eqref{eq:tilde-path-Psi_double}, the weak symmetry in Eq.~\eqref{eq:dLsymm} can be {equivalently formulated} by considering the symmetry-transformed {fully-dephased cMPO}
\begin{equation}\label{eq:tilde-path-R}
	\sytr R_t = \UUE {(R_t)}, 
\end{equation}
If the symmetry holds then this
solves $\sytr R_t = \mathcal{T}e^{\int_{0}^t \dL_\tau} (\sytr R_{0})$ with $\sytr R_0 = \UUE (R_0)$.  That is, the symmetry-transformed cMPO $\sytr R_t$ is the solution of the joint dynamics (under the counting measurement) for the symmetry transformed initial condition $\sytr R_0$. In {this case,} an initially symmetric {joint state remains so at all times,}
\begin{equation}
\label{eq:URR}
    \mathbb{U}{(R_0)}= R_0 \implies \mathbb{U}{(R_t)} = R_t \,,
\end{equation}
{and the symmetry of the joint state is unaffected by the measurement process.}

\subsubsection{{Necessary and sufficient conditions\\(Symmetry Condition~III)}}
\label{sec:cond-deph}

We show in Appendix~\ref{app:full_deph} that: 
\smallskip\par\noindent
\emph{The {joint dynamics under a counting measurement features {a} weak symmetry} Eq.~\eqref{eq:symm-deph-rot} if and only if Symmetry Condition~III holds for some {permutation} $\pi$.}
\smallskip

 To {understand the implications of this statement,} recall that the {environmental part of the symmetry operator in Eq.~\eqref{eq:U-double} is parametrised in Eq.~\eqref{eq:UE} by the matrix $\mathbf{U}$. Under Symmetry Condition III, there exists a matrix $\mathbf{U}$ of the form given in Eq.~\eqref{eq:UIIIRb_U}, such that \eqref{eq:not_cond_IIb} also holds.  {Then take $\mathbf{U}$ in Eq.~\eqref{eq:UE} such that
\begin{equation}\label{eq:UE_III}
	\superUE (dB_{j,t}) = dB_{\pi(j),t} e^{-i\phi_j} %\, e^{i\delta_j} \; ,
\end{equation}
with $\pi,\phi_j$ as in Eq.~\eqref{eq:UIIIRb_U}:
this gives the $\UUE$ that yields the specific symmetry of Eq.~\eqref{eq:symm-deph-rot}.  (The permutation $\pi$ is not unique in general but Symmetry Condition III guarantees that a suitable choice exists.)
With this choice, the joint symmetry simply permutes the joint jump operators according to $\pi$:
\begin{equation}
\label{eq:UE_III_joint}
	 \UUE (J_j\otimes dB^\dag_{j,t}) = J_{\pi(j)}\otimes dB^\dag_{\pi(j),t}.
\end{equation}
This result is valuable when establishing Eq.~\eqref{eq:symm-deph-rot}, and for the symmetry of the stochastic Hamiltonian in Eq.~\eqref{eq:weakU_dH}.\footnote{{Note that taking different phases in Eq.~\eqref{eq:UE_III} would not affect Eq.~\eqref{eq:symm-deph-rot} although it does affect the stochastic Hamiltonian.  The choice of Eq.~\eqref{eq:UE_III} is always sufficient to construct a symmetry of either object, if it exists.}}
}

{Finally, recalling that Symmetry Condition~III implies {Symmetry Condition~II, which in turn implies Eq.~\eqref{eq:not_cond_II}, it follows that \emph{any weak symmetry of $d\mathbb{L}_\tau$ is always stationary},} 
\begin{equation}
\mathbb{U}_t=\mathbb{U}.
\end{equation}
{That is, Eq.~\eqref{eq:dLsymm} holds only if Eq.~\eqref{eq:symm-deph-rot} does too. This can be explained by the fact that the fully-dephased cMPO features only definite number of quanta [with their (classical) randomness associated with probabilities of different measurement outcomes]. This property must be preserved under the action of the symmetry, see~Eqs.~\eqref{eq:UE} and~\eqref{eq:UEt}. It then follows that the action of $\superUE$ must correspond to permutation of quanta, cf.~Eq.~\eqref{eq:UE_III}.}

\subsection{Symmetries of partially {dephased} cMPO} \label{sec:partial_dephasing}

This subsection introduces a partially-dephased cMPO, {whose dynamics features the weak joint symmetry} whenever Symmetry Condition~II holds (Box \Bpdcmps~in Fig.~\ref{fig:states}).  The motivation is similar to the coarse-graining procedure in Sec.~\ref{sec:CG-records}: we identify a measurement in the environment such that the resulting joint state has the same symmetry properties as the associated quantum trajectory ensemble.

\subsubsection{{Joint dynamics under partial counting measurement}}

{In the fully-dephased cMPO, all coherences between different numbers of quanta at any time have been removed, due to measurement of number operators for all types $j=1,...,d$ of quanta}.  The construction {here} retains some of these coherences, leading to a \emph{partially-dephased cMPO}. {This is achieved by defining a \emph{partial measurement} which considers the total number of quanta associated with each SJED ($\alpha=1,...,\dC$). This measurement is still projective, but it only leads to partial dephasing of the joint pure state $\Psi_t$, because of the degeneracy of measured number operators.}

The partially-dephased cMPO is $R^\textsc{p}_t$.  It evolves as
\begin{equation}\label{eq:Rp_time_evolution}
    R^\textsc{p}_t = \mathcal{T}e^{\int_{t'}^t d\mathbb{L}^\textsc{p}_t} \big( R^\textsc{p}_{t'} \big),
\end{equation}
with superoperator 
\begin{multline}\label{eq:dLp}
	\dL^\textsc{p}_t(\deph^\textsc{p}_t) =  \left[-i(H_\text{eff}\otimes\IE)\deph^\textsc{p}_t+i \deph^\textsc{p}_t(H_\text{eff}^\dag\otimes\IE)\right] dt  \\
	+\sum_{\alpha=1}^{\dC} \bigg(\sum_{j\in S_\alpha}  J_j\otimes dB^\dag_{j,t}\bigg)\deph^\textsc{p}_t\bigg(\sum_{j'\in S_\alpha} J_{j'}^\dag\otimes dB_{j',t}\bigg).
\end{multline}
{Similarly} to Eq.~\eqref{eq:dLf}, {$\dL^\textsc{p}_t$ is a super}operator acting jointly on the system and environment.  The joint jump operators are $\sum_{j\in S_\alpha}  J_j\otimes dB^\dag_{j,t}$ for $\alpha=1,\dots,\dC$, they are not separable in general, so their action on the vacuum of the environment creates superpositions of the emitted quanta.  Nevertheless, the action of a joint jump operator on a separable state yields a separable state, because a pure system state remains pure under the composite SJED actions ${\cal A}_\alpha$.
The joint effective Hamiltonian is again $H_\text{eff}\otimes\IE$, where the Hermitian part $H\otimes\IE$ comes from the system Hamiltonian and the non-Hermitian part originates from the joint jump operators via} $ \sum_{\alpha=1}^{\dC} \sum_{j,j'\in S_\alpha} (J_j^\dag\otimes dB_{j,t})(J_{j'}\otimes dB_{j',t}^\dag) = \sum_{j=1}^d J_j^\dag J_j \otimes \IE dt$ [recall Eq.~\eqref{eq:Ito}].

\subsubsection{{Weak joint symmetries}}

A weak dynamical symmetry of the {joint dynamics under the partial measurement} is 
\begin{equation}
    \UUE_t(\mathcal{T}e^{\int_{t'}^t d\mathbb{L}_\tau^\textsc{p}}) \UUE_{t'}^\dag = \mathcal{T}e^{\int_{t'}^t d\mathbb{L}_\tau^\textsc{p}}
\label{eq:LP-symm_t}
\end{equation}
which is the analogue of {Eq.~\eqref{eq:symm-deph-rot} for this partially-dephased cMPO. 
This holds if the partially-dephased (measured) cMPO retains the weak symmetry of the joint unitary dynamics in Eq.~\eqref{eq:symm_dH_Ut_expanded-v2}.

As in the fully-dephased case, the all symmetries turn to be stationary in which case Eq.~\eqref{eq:LP-symm_t} is equivalent to 
\begin{equation}\label{eq:LP-symm}
	\mathbb{U} \,d\mathbb{L}^{\textsc p}_t \,\mathbb{U}^\dag = d\mathbb{L}^{\textsc p}_t.
\end{equation}
[cf.~Eq.~\eqref{eq:dLsymm}].
In this case, for a general initial state, the symmetry can be {also} expressed by considering the symmetry-transformed {partially-dephased cMPO}, by direct analogy with Eq.~(\ref{eq:tilde-path-R},\ref{eq:URR}).
In the symmetric case, the symmetry transformed state $\sytr R_t^\textsc{p} = \UUE {(R_t^\textsc{p})}$ solves
\begin{equation}
    \sytr R_t^\textsc{p} = \mathcal{T}e^{\int_0^t \dL^\textsc{p}_\tau} (\sytr R_{0}^\textsc{p})
\end{equation}
with 
$\sytr R_{0}^\textsc{p} = \UUE(R_{0}^\textsc{p})$.
Also, an initially symmetric {joint} state
$
%\begin{equation}
    \UUE {(R_0^\textsc{p})}= R_0^\textsc{p}$ implies $\UUE  {(R_t^\textsc{p})}= R_t^\textsc{p}
$ for all $t$.
%\end{equation}

\subsubsection{{Necessary and sufficient conditions\\(Symmetry Condition~II)}}\label{sec:cond-partial}

In Appendix \ref{app:partial_deph}, we derive a result for the partially-dephased cMPO, analogous to that {for full dephasing} in Sec.~\ref{sec:dephased_symm}:

\smallskip\par\noindent
\emph{The {joint dynamics under a partial counting measurement features {a} weak} symmetry  Eq.~\eqref{eq:LP-symm_t} if and only if Symmetry Condition~II holds.}
\smallskip

As in Sec.~\ref{sec:cond-deph}, the {environmental part of the symmetry operator in Eq.~\eqref{eq:LP-symm_t} is parametrised  by the matrix $\mathbf{U}$, via Eq.~\eqref{eq:UE}.  Under Symmetry Condition II, there exists a matrix of the form of Eq.~\eqref{eq:UU-partial-lemma} such that Eq.~\eqref{eq:not_cond_IIb} also holds.  Using this $\mathbf{U}$ in Eq.~\eqref{eq:UE} means that the symmetry permutes the joint jump operators as
	\begin{equation}\label{eq:UE_II}
		\UUE \bigg( \! \sum_{j\in S_\alpha}J_j\otimes dB^\dag_{j,t}\bigg) = \!\!\!\sum_{k\in S_{\pic(\alpha)}}  \!\!\!J_k\otimes dB_{k,t}^\dag \; ,
	\end{equation}
where $\pic$ is the (unique) permutation that appears in both Eqs.~(\ref{eq:UU-partial-lemma},\ref{eq:symm_cond_IIb}).\footnote{Similarly to the case of full dephasing with Eq.~\eqref{eq:UE_III}, additional phase factors could be added to Eq.~\eqref{eq:UE_II} without affecting the (equivalent) weak joint symmetry of the dynamics in Eq.~\eqref{eq:LP-symm}, but the stochastic Hamiltonian in Eq.~\eqref{eq:symm-deph-rot} would no longer respect that joint symmetry.}  This result is valuable when establishing Eq.~\eqref{eq:LP-symm_t}.
Such a $\mathbf{U}$ exists if and only if Symmetry Condition~II holds, see Appendix \ref{app:UEt=UE}.

{As} Symmetry Condition~II additionally implies Eq.~\eqref{eq:not_cond_IIb}, \emph{{any weak} symmetry is always
stationary}, 
 \begin{equation}
 	\mathbb{U}_t =\mathbb{U},
 \end{equation}
 and Eq.~\eqref {eq:LP-symm} holds whenever Eq.~\eqref{eq:LP-symm_t} does. This is related to the fact that the partially-dephased cMPO at any time features only definite number of quanta associated with each SJED, and this property needs to be preserved under the symmetry transformation. As in the case of the fully-dephased cMPO, this forbids symmetries that involve a coherent displacement [cf.~Eq.~\eqref{eq:UEt}], but unitary mixing of quanta within each SJED still allowed in Eq.~\eqref{eq:UE_II}}, contrary to Eq.~\eqref{eq:UE_III_joint}. As we discuss next, this explains why weak symmetry of the joint dynamics under the partial counting measurement relies on Symmetry Condition~II, rather than the stricter Symmetry Condition~III.
 
We also state one more result that will be useful in the following, and is obtained by considering different representations of the ${\cal A}_\alpha$, see~\cite{Generators}. {If Symmetry Condition II holds then 
we can define the family of isometric environmental operators $V_E$ that only mix quanta from within the same SJED.  The resulting transformations [cf Eqs.~(\ref{eq:dH'tilde},\ref{eq:general_rep_change})]
\beq
\label{eq:rep-ii}
d  \tilde H_{\tau}  = \left(\mathbb{1}\otimes  V_{E}\right)  d H_{\tau} \big(\mathbb{1}\otimes V_{E} \big)^\dag 
\eeq
lead to new representations $\tilde{H},\tilde{J_1},\dots,\tilde{J}_{\tilde d}$,
whose composite action operators are invariant under the transformation, that is $\tilde{\cal A}_\alpha={\cal A}_\alpha$ for all $\alpha$.}
Since the composite actions are the same then these new representations all obey Symmetry Condition II. Moreover, there always exists at least one such transformation such that the new representation obeys Symmetry Condition III, see Sec.~\ref{sec:SJED_reps}.  
[It is not possible to obtain a weakly symmetric representation in this way, unless $\pic(k)$ in Eq.~\eqref{eq:symm_cond_II} is the trivial permutation.]

Physically, this means that there is some freedom to vary the measurement basis without affecting system conditional states (composite actions are preserved); it also provides a judicious choice of measurement basis in which the (fully-)dephased cMPO is symmetric, as is the ensemble of (fully)-labelled trajectories. In fact, these freedoms are the same as those discussed in Sec.~\ref{sec:partial-symm-further}, in the context of trajectory labelling.
We also define a super-operator 
\beq
\label{eq:superV-ii}
\mathbb{V}(\cdot)= \left(\mathbb{1}\otimes  V_{E}\right) (\cdot) \big(\mathbb{1}\otimes V_{E} \big)^\dag 
\eeq
such that Eq.~\eqref{eq:rep-ii} becomes $d  \tilde H_{\tau}=\mathbb{V}( d H_{\tau})$; we reserve this notation specifically for transformations $V_E$ that mix quanta within SJEDs.

The partially dephased cMPO is discussed further in Sec.~\ref{sec:discussion1} below, including its physical connection to Symmetry Condition~II. Before doing this, we introduce a coarse-grained cMPO which helps to elucidate the connection of these joint states with ensemble of quantum trajectories and coarse-grained measurement records.

\subsection{Symmetries of coarse-grained cMPO}\label{sec:CG_cMPS}

This subsection {defines} a coarse-grained cMPO (Box \Bcgcmps~in Figs.~\ref{fig:outline} and~\ref{fig:states}).  The construction of this state provides a connection to partially-labelled quantum trajectories of Sec.~\ref{sec:traj}.  As a result, {the dynamics of this joint state features a weak symmetry} under Symmetry Condition~II.  We {also} discuss the connection of this cMPO to the other cMPOs considered so far.

\subsubsection{{Joint dynamics under coarse-grained counting measurement}}

The {coarse-grained cMPO} $ R_t^\textsc{c}$  is obtained  from the joint state of the system and environment by replacing any emitted quanta (of types $j=1,...d$) by quanta of new types labelled by $\alpha=1,...,\dC$, according the SJED of the corresponding jump operator \cite{Kabernik2018}. Let  $dC_{\alpha,t}$ and $dC_{\alpha,t}^\dag$ denote (unnormalised) annihilation and creation operators, respectively, with Ito table [cf.~Eq.~\eqref{eq:Ito}]
\begin{subequations}\label{eq:Ito_alpha}
	\begin{align}
		dC_{\alpha,t} dC^\dagger_{\beta,t}&=  \delta_{\alpha\beta} \IE  \,dt, \\
		dC^\dagger_{\alpha,t} dC_{\beta,t}&=  0, \\
		dC_{\alpha,t} dC_{\beta,t}&=  0.
	\end{align}
\end{subequations} 
The evolution of $ R_t^\textsc{c}$ is
\begin{equation}\label{eq:Rc_time_evolution}
    R^\textsc{c}_t = \mathcal{T}e^{\int_{t'}^t d\mathbb{L}^\textsc{c}_\tau} \big( R^\textsc{c}_{t'} \big),
\end{equation}
with the superoperator
\begin{multline}\label{eq:dLc}
	d\mathbb{L}^\textsc{c}_t(R^\textsc{c}_t) =  \left[-i(H_\text{eff}\otimes\IE)\deph^\textsc{c}_t+i\deph^\textsc{c}_t(H_\text{eff}^\dag\otimes\IE)\right] dt \\
	+\sum_{\alpha=1}^{\dC}\sum_{j\in S_\alpha}(J_j\otimes dC_{\alpha,t}^\dag) R_t^\textsc{c} (J_j^\dag\otimes dC_{\alpha,t}).
\end{multline}
The action of $d\mathbb{L}^\textsc{c}_t$ on the coarse-grained cMPO $ R_t^\textsc{c}$ is similar to the action of $d\mathbb{L}_t$ on the fully-dephased cMPO $R_t$ in  Eq.~\eqref{eq:dLf}.  The jump operators $J_j\otimes dC_{\alpha,t}^\dag$ are  \emph{separable}, and $H_\text{eff}\otimes\IE$ is the effective Hamiltonian whose non-Hermitian part arises via $(J_j^\dag\otimes dC_{\alpha,t})(J_j\otimes dC_{\alpha,t}^\dag)=J_j^\dag J_j\otimes\IE\,dt $ from Eq.~\eqref{eq:Ito_alpha}. Comparing with  Eq.~\eqref{eq:dLf}, the coarse-graining effect is that different system jump operators in the same SJED are all associated with the same creation operator.  The second line of Eq.~\eqref{eq:dLc} can also be represented as 
$\sum_\alpha (\mathcal{A}_\alpha\otimes d\mathcal{C}_{\alpha,t}^\dagger)(R_t^\textsc{c} ) $, where $d\mathcal{C}_{\alpha,t}^\dagger(\cdot)= dC_{\alpha,t}^\dag(\cdot)dC_{\alpha,t}$, which shows the role of the composite action, see Eq.~\eqref{eq:superA}.

{When $\dC< d$, the above construction corresponds to an \emph{erasure} operation in the context of quantum optics, which can be performed either continuously as the quanta are emitted by the system, or  at the final time $t$. After measuring the number operators for different types of quanta $j$, they are replaced by the same numbers of new quanta of type $\alpha$ such that $j\in S_\alpha$.  The same operation can also be performed if one only measures the total number of quanta emitted from each SJED.  Therefore, the corresponding joint state $R^\textsc{c}_t$ is a cMPO that can be obtained either from $R_t$ or $R^{\textsc p}_t$ by such an erasure operation, see Fig.~\ref{fig:states}. The corresponding composed measurement from the joint pure state $\Psi_t$ is no longer projective, and we refer to it as \emph{coarse-grained counting measurement}. As its action is restricted to the environment we can again recover the system density matrix as [cf.~Eq.~\eqref{eq:cMPO_trace}]
\begin{equation}\label{eq:cMPO_trace_cg}
	\rho_t=
	\Tr_E\left(R^{\textsc c}_t\right).
\end{equation}}

\begin{table*}
\begin{center}
{\def\arraystretch{1.6}
\begin{tabular}{p{3.82cm} | c | c | c | c } %| c | c | c | c | c }
& \quad {\bf Generator} \quad & \quad {\bf Symmetry} \quad & \quad {\bf Transformation} \quad  & %{\bf Box}
\\ \hline 
Density matrix $\rho$ 
& $\frac{d}{dt}\rho_t = 
\mathcal{L}\,\rho_t$ 
& $\mathcal{U\,L\,U}^\dag=\mathcal{L}$ 
&
$\mathcal{U}\rho=U\rho U^\dag$ 
%&  \Bqme
& \textbf{\lvla{I}}
\\ \hline
Quantum trajectories &  $\frac{\partial}{\partial t}P_t(\psi) = \mathcal{W}^\dag\! P_t(\psi)$ & $\Upsilon\mathcal{W}^\dag\Upsilon^\dag = \mathcal{W}^\dag$ 
& $\Upsilon {P}(\psi) = {P}({\cal U}^\dag\psi) $
%& \Btraj
& \textbf{\lvlb{II}}
\\ \hline
Partially-labelled  q.~traj. &  $\frac{\partial}{\partial t}P_t(\psi,\bm{Q}) = \WC^\dag P_t(\psi,\bm{Q})$ 
& \quad $\Yc\WC^\dag\Yc^\dag = \WC^\dag$ \quad
& \ $\Yc {P}(\psi,\bm{Q}) = {P}({\cal U}^\dag\psi,\Pic{^{-1}}\bm{Q}) $ \ 
%& \Bcgtraj
& \textbf{\lvlb{II}}
\\ \hline
Labelled q.~traj.&  $\frac{\partial }{\partial t}P_t(\psi,\bm{q}) = \WF^\dag P_t(\psi,\bm{q})$ 
& \quad $\Yf\WF^\dag\Yf^\dag = \WF^\dag$ \quad
& $\Yf{ P}(\psi,\bm{q}) = {P}({\cal U}^\dag\psi,\bm{\pi}{^{-1}}\bm{q}) $
%& \Bftraj
& \ \textbf{\lvlc{III}} \
\\ \hline
Fully-dephased cMPO 
& $ dR_t = d\mathbb{L}_t R_t$ 
& $\mathbb{U} \,d\mathbb{L}_t\,\mathbb{U}^\dag =d\mathbb{L}_t $
& \; $\mathbb{U}R = ({\cal U} \!\otimes \superUE)R$
%& \Bdcmps
& \ \textbf{\lvlc{III}} \
\\ \hline
Coarse-grained cMPO
& $ dR_t^{\textsc c} = d\mathbb{L}_t^{\textsc c} R_t^{\textsc c} $
%& $\mathbb{U}^{\textsc c} d\mathbb{L}_t^{\textsc c}{\mathbb{U}^{\textsc c}}^\dag =d\mathbb{L}_t^{\textsc c} $
& $\mathbb{U}^{\textsc c} \,d\mathbb{L}_t^{\textsc c}\,{\mathbb{U}^{\textsc c}}^\dag =d\mathbb{L}_t^{\textsc c} $
 & $\mathbb{U^{\textsc c}}R^{\textsc c} = ({\cal U} \!\otimes\superUE^{\textsc c})R^{\textsc c}$
%& \Bcgcmps
& \textbf{\lvlb{II}}
\\ \hline
% Partially dephased cMPO 
% & $ dR_t^{\textsc p} = d\mathbb{L}_t^{\textsc p}(R_t^{\textsc p}) $
% %& $\mathbb{U}\, d\mathbb{L}_t^{\textsc p}\, \mathbb{U}^\dag =d\mathbb{L}_t^{\textsc p} $
% & $ \mathbb{U}_t ({\cal T}{\rm e}^{\int_{t'}^t d\mathbb{L}^{\textsc p}_\tau}) 
%  \mathbb{U}_{t'}^\dag =  {\cal T}{\rm e}^{\int_{t'}^t d\mathbb{L}^{\textsc p}_\tau} $
% & $\mathbb{U}_tR = ({\cal U} \otimes {\cal U}_{{\rm E},t})R$
% & \Bpdcmps
% & \textbf{\lvlb{II}}
% \\ \hline 
Partially-dephased cMPO %now without t-dependence on symm
& $ dR_t^{\textsc p} = d\mathbb{L}_t^{\textsc p} R_t^{\textsc p} $
%& $\mathbb{U}\, d\mathbb{L}_t^{\textsc p}\, \mathbb{U}^\dag =d\mathbb{L}_t^{\textsc p} $
&  $\mathbb{U} \,d\mathbb{L}_t^{\textsc p}\,\mathbb{U}^\dag =d\mathbb{L}_t^{\textsc p} $
% & $ \mathbb{U} ({\cal T}{\rm e}^{\int_{t'}^t d\mathbb{L}^{\textsc p}_\tau}) 
% \mathbb{U}^\dag =  {\cal T}{\rm e}^{\int_{t'}^t d\mathbb{L}^{\textsc p}_\tau} $
& $\mathbb{U} R{^{\textsc p}} = ({\cal U} \!\otimes \superUE)R{^{\textsc p}}$
%& \Bpdcmps
& \textbf{\lvlb{II}}
\\ \hline 
 cMPS 
& $d|\Psi_t \rangle = -i dH_t |\Psi_t \rangle$ 
& \ $(U\!\otimes \UE) dH'_t (U\!\otimes \UE)^\dag = dH_t'$  \ 
& $\superUE (dB_{j,t}^{\dag}) = \sum_k\! \mathbf{U}_{jk}^{*} dB_{k,t}^{\dag}$
%& \Bcmps
& \textbf{\lvla{I}}
\\ \hline
\end{tabular}

}
\end{center}
\caption{\textbf{Summary of main objects} for the various  {descriptions of open quantum dynamics} and environmental information {therein}. The generators of the dynamics {(in Stratonovich convention for cMPOs and cMPS), their} symmetries and the relevant symmetry {transformations} are all given. The final column {shows the relevant Symmetry Conditions}.}
\label{tab:summ}
\end{table*}

\subsubsection{{Weak joint symmetries}}

{We define a weak (coarse-grained)} symmetry of {the joint dynamics under the coarse-grained measurement as} 
\begin{equation}
	\mathbb{U}^{\textsc c} \,d\mathbb{L}_t^{\textsc c}\,{\mathbb{U}^{\textsc c}}^\dag =d\mathbb{L}_t^{\textsc c},
	\label{eq:symm-LC}
\end{equation}
where {the (coarse-grained) joint symmetry superoperator}
\begin{equation}\label{eq:UUEC}
\UUE^\textsc{c}=\mathcal{U}\otimes\superUE^\textsc{c}
\end{equation}	
{with $\superUE^\textsc{c}(\cdot)=\UE^\textsc{c}(\cdot)(\UE^\textsc{c})^\dag$ acting on the new type of quanta in the environment  as 
\begin{equation}
	\superUE^\textsc{c}\left(dC_{\alpha,t}\right) = \sum_{\beta=1}^{\dC} \mathbf{U}^{\textsc c}_{\alpha\beta} \,dC_{\beta,t}, %dC_{\pic(\alpha),t}
	\quad \UE^\textsc{c} |\vac\rangle=|\vac\rangle.
	\label{eq:UEC_V}
\end{equation}
%for a permutation $\pi$.
Here, $\mathbf{U}^{\textsc c} $} is a $\dC\times \dC$ unitary matrix. 
{In analogy to the cases of fully- and partially-dephased cMPOs,} it is sufficient to consider stationary symmetries here. {The types of quanta considered here are distinct from previous cases and constitute a different Fock space, but the coarse-grained cMPO similarly features only definite number of quanta, cf.~Appendix~\ref{app:coarse_deph}.}

As in previous cases, the symmetry can be {also} expressed in terms of the symmetry-transformed cMPO, by direct analogy with Eq.~(\ref{eq:tilde-path-R},\ref{eq:URR}).
In the symmetric case,
$
	\sytr R_t^\textsc{c} = \UUE^\textsc{c} (R_t^\textsc{c})
$
 solves 
\beq
\sytr R_t^\textsc{c} = \mathcal{T}e^{\int_0^t \dL_\tau} (\sytr R_{0}^\textsc{c})
\eeq
with 
$\sytr R_{0}^\textsc{c} = \UUE^\textsc{c}(R_{0}^\textsc{c})$.
Also, an initially symmetric {joint} state remains symmetric:
$
    \UUE^\textsc{c} {(R_0^\textsc{c})}= R_0^\textsc{c}$ implies $\UUE^\textsc{c}  {(R_t^\textsc{c})}= R_t^\textsc{c}
$ for all $t$.

\subsubsection{{Necessary and sufficient conditions\\(Symmetry Condition~II)}}
\label{sec:cond-cg}

We show in Appendix~\ref{app:coarse_deph} that:

\smallskip\par\noindent
\emph{The {joint dynamics under a coarse-grained counting measurement features the (coarse-grained) weak symmetry in Eq.~\eqref{eq:symm-LC}} if and only if Symmetry Condition~II holds.}
\smallskip\\

Similar to previous subsections, the {environmental part of the symmetry operator in Eq.~\eqref{eq:symm-LC} is parametrised  by the matrix $\mathbf{U}^\textsc{c}$, via Eq.~\eqref{eq:UEC_V}.  Under Symmetry Condition II, this matrix may be chosen such that 
\begin{equation}\label{eq:UEC}
   % \superUE^\textsc{c}(dC_{\alpha,t}^\km{\dag}) = dC^\km{\dag}_{\pic(\alpha),t} e^{\km{-}i\phi_\alpha} \; 
   \superUE^\textsc{c}(dC_{\alpha,t}) = dC_{\pic(\alpha),t} % \,e^{i\phi_\alpha^{\textsc c}} 
   \; ,
\end{equation}
where the permutation $\pic$ of $\{1,\dots,\dC\}$ is the one appearing in  Eq.~\eqref{eq:symm_cond_IIb}.\footnote{Similar to Eq.~\eqref{eq:UE_II}, arbitrary phase factors could also be included in Eq.~\eqref{eq:UEC}, but the specific choice given here is always sufficient.
%$\phi_1^{\textsc c},\dots,\phi^{\textsc c}_{\dC}\in \mathbb{R}$ are arbitrary.
}  
With this choice of $ \superUE^\textsc{c}$, we find
\begin{equation}
     \UUE^\textsc{c}(\mathcal{A}_\alpha\otimes d\mathcal{C}^\dag_{\alpha,t}) \UUE^\textsc{c}= \mathcal{A}_{\pic(\alpha)}\otimes d\mathcal{C}^\dag_{\pic(\alpha),t}
\end{equation}
which ensures Eq.~\eqref{eq:symm-LC}.

We have already emphasized that Eq.~\eqref{eq:cMPO_trace_cg} requires that the symmetry condition for $R^{\textsc c}_t$ should be at least as restrictive as Symmetry Condition I, which applies to the QME.  Noting that the mechanism for reduced symmetry is dephasing caused by the measurement, it is also not surprising that the relevant condition for these coarse-grained measurements should be weaker than Symmetry Condition III, which applies to the fully dephased cMPO $R_t$, whose measurements are more detailed.  However, it is non-trivial that the same Symmetry Condition II applies to the $R^{\textsc c}_t$ as well as to the partially-dephased cMPO $R^{\textsc p}_t$ and the ensemble of quantum trajectories considered in Sec.~\ref{sec:traj}.  The connections between the symmetries of these different objects is explored in the next Section.

\section{Unified hierarchy of symmetries}\label{sec:discussion1}

We have presented results for quantum trajectories, measurement records, and cMPOs; we explained how their symmetries are ruled by Symmetry Conditions I, II, and III.  This Section explains the relationships between these objects, and their symmetries.
 We also clarify the structure of the quantum and classical information that impact the symmetries shown in Fig.~\ref{fig:outline}; the symmetry properties of the relevant objects are also summarised in Table~\ref{tab:summ}.

\subsection{Symmetry Condition~I}

Using properties of the time-ordered exponential, 
the {pure} joint state of the system and {the field at time $t$ in Eq.~\eqref{eq:cMPS_dH} can be expressed as}
\begin{align}
	|\Psi_t\rangle
	\label{eq:cMPS}
	&= \int \sqrt{p(\bm{m}_t|\psi_0)} \,|\psi_t (\bm{m}_t)\rangle\otimes |d\bm{m}_t\rangle.
\end{align}
where the {measurement records of Eq.~(\ref{eq:full_record})
 are encoded in (unnormalised) states} of the field, as
\begin{equation}\label{eq:psi_E}
%|\varphi_E(\bm{m}_t)\rangle=  
|d\bm{m}_t\rangle = dB^\dag_{j_n,t_n}\dots dB^\dag_{j_1,t_1}|\vac\rangle.
%  |\bm{m}_t\rangle = b^\dag_{j_n,t_n}\dots b^\dag_{j_1,t_1}|\vac\rangle
\end{equation}
The (normalised) conditional {states} of the system  [cf.~Eq.~\eqref{eq:psi_t}] are
	\begin{equation}\label{eq:psi_t_vec}
		|\psi_t (\bm{m}_t) \rangle =   \frac{ |\varphi_t (\bm{m}_t)\rangle}{\sqrt{p(\bm{m}_t|\psi_0)}},
	\end{equation} 
	where [cf.~Eq.~\eqref{eq:varphi}]
		\begin{align}
			%&
			|\varphi_t (\bm{m}_t) %(t_1, j_1;\dots;t_n,j_n)
			\rangle & =         G_{t-t_n} J_{j_n}  \cdots G_{t_2-t_1}  J_{j_1} G_{t_1} |\psi_{0}\rangle,
			%\end{equation}
			\nonumber \\
			%\begin{equation}\label{eq:G}
			%&
			G_t & =         e^{-i\He t} .
		\end{align}
 The {probability density} $p(\bm{m}_t|\psi_0)$ for time record $\bm{m}_t$ (conditional on the initial system state $|\psi_0\rangle$) was given in Eq.~\eqref{eq:path_prob_m}, and it is $p(\bm{m}_t|\psi_0)=\langle \varphi_t (\bm{m}_t)|\varphi_t (\bm{m}_t)\rangle$ in the notation of this Section. 
 
 To see that the joint state is a cMPS~\cite{Verstraete2010,Osborne2010}, we write Eq.~\eqref{eq:cMPS} as
 	\begin{align}
 		|\Psi_t\rangle
 		\label{eq:cMPS1}
 		&=\int |\varphi_t (\bm{m}_t) \rangle \otimes |d\bm{m}_t\rangle.
 	\end{align}
From the Ito table in Eq.~\eqref{eq:Ito}}, $\langle d\bm{m}_t |d\bm{m}_t\rangle=dt_1\cdots dt_n$;  the integrals in {Eqs.~\eqref{eq:cMPS} and~\eqref{eq:cMPS1}} should be interpreted as $\sum_{n=0}^\infty \sum_{j_1,\dots,j_n=1}^d\int_{0\leq t_1<\dots<t_n < t}$  [cf.~Eqs.~(\ref{eq:rho_t_p}-\ref{eq:p-traj-m})]. 

This formulation of the joint state has been constructed in the (measurement) basis associated with a given representation of the QME.
% Its behaviour in different measurement bases
The joint states for different representations
can be obtained by isometric transformations,
recall Eq.~\eqref{eq:kasia-tilde}. The symmetry operators for their stochastic Hamiltonians are also related by the same isometric transformation, see Sec.~\ref{sec:symm-joint-uni-conds}. Hence, the symmetry properties of pure joint states are related analogously, and are therefore present either for all representations or none.
This situation is covered by Symmetry Condition~I, whose physical implication is that
% all representations of a symmetric QME are symmetric.
all joint dynamics of the system and environment as constructed are symmetric if and only if the quantum master dynamics features the weak symmetry.
See Boxes~\Bcmps~and~\Bqme~in Fig.~\ref{fig:outline} and the first and last row of Tab.~\ref{tab:summ}.

\subsection{Symmetry Condition~III} \label{sec:unif-iii}

The fully-dephased cMPO {in Eq.~\eqref{eq:QME-MPO} can be expressed as} 
\begin{align}\label{eq:def_deph}
	\deph_t &%= \dephop(\Psi_t) \nonumber \\
	=\int p(\bm{m}_t |\psi_0)\, \psi_t(\bm{m}_t) \otimes |d\bm{m}_t\rangle\!\langle d\bm{m}_t|.
\end{align}
Comparing with Eq.~\eqref{eq:cMPS}, one verifies that
 this  state is obtained from the pure joint state $\Psi_t$ by  (full) dephasing  in the basis of $|d\bm{m}_t\rangle$.  That is, coherences 
between different measurement records $|d\bm{m}_t\rangle$ and $|d\bm{m}'_t\rangle$ are  removed when the (full) projective measurement is performed at the final time $t$. Then, the structure of the resulting joint state as a cMPO~\cite{Verstraete2004} follows by writing it as 
\begin{align}\label{eq:def_deph2}
	\deph_t &%= \dephop(\Psi_t) \nonumber \\
	=\int  \varphi_t(\bm{m}_t) \otimes |d\bm{m}_t\rangle\!\langle d\bm{m}_t|.
\end{align}
From this form, it is natural that separable symmetries of the cMPO [$\mathbb{U}(R_t)=R_t$] are possible only if the environmental part of the joint symmetry acts as
% $U_E  |d\bm{m}_t\rangle \propto |d\tilde{\bm{m}}_t\rangle$
$U_E|d\bm{m}_t\rangle\langle d\bm{m}_t|U_E^\dagger=|d\tilde{\bm{m}}_t\rangle\langle d\tilde{\bm{m}}_t|$
where $\tilde{\bm{m}}_t$ is a new time record.  (No superpositions are created by this symmetry operator, in the measurement basis.)  The relevant cMPO symmetries do have this form, in fact Symmetry Condition III can be rationalised by this constraint on the environmental part of the joint symmetry.
 
This cMPO form of $R_t$ can be connected back to the 
labelled quantum trajectories in Sec.~\ref{sec:measurement_records} by noting that 
\begin{equation}
\label{eq:psi_mm}
	\psi_t(\bm{m}_t) \otimes |d\bm{m}_t\rangle\!\langle d\bm{m}_t|
\end{equation}
is the (unnormalised) conditional joint state for an outcome $\bm{m}_t$ of the (full) projective  measurement at time $t$.  It contains the full time record $\bm{m}_t$ via the environmental state $|d\bm{m}_t\rangle\!\langle d\bm{m}_t|$, so it has exactly the same information content as the object defined in Eq.~\eqref{eq:psi-m}, which is the conditional system state combined with the (full) measurement record.  This object also contains the same information as the labelled quantum trajectory of Eq.~\eqref{eq:path-psi-q}. Since the environmental states $|d\bm{m}_t\rangle,|d\bm{m}_t'\rangle$ are all orthogonal [see Eq.~\eqref{eq:Ito}], one sees that 
the fully-dephased cMPO $R_t$ contains exactly the same information as the whole ensemble of (fully) labelled quantum trajectories. This equivalence -- as indicated by a green arrow between Boxes \Bdcmps~and \Bftraj~in Fig.~\ref{fig:outline} -- is responsible for the symmetry properties of those objects being characterised by the same Symmetry Condition~III, see Tab.~\ref{tab:summ}.

\subsection{Symmetry Condition~II}
\label{sec:unif-ii}

The partially-dephased cMPO in Eq.~\eqref{eq:Rp_time_evolution} can be expressed as
\begin{align}\label{eq:Rp_integral}
	R^{\textsc p}_t 
	=\int \!\!\int |\varphi_t(\bm{m}_t) \rangle\langle \varphi_t(\bm{m}_t')|   \otimes |d\bm{m}_t\rangle\langle d\bm{m}_t'|\, \delta(\bm{M}_t-\bm{M}'_t),
\end{align}
where $\bm{M}_t$ denotes the coarse-grained measurement record [cf. Eq.~\eqref{eq:partial_record}] that is obtained from $\bm{m}_t$ by replacing the types of emitted photons by their SJED labels. 

Comparing with Eq.~\eqref{eq:def_deph}, one sees that this joint state retains coherences between environmental states  $|d\bm{m}_t\rangle$ and $|d\bm{m}'_t\rangle$ in situations where  $\bm{m}_t,\bm{m}_t'$ both correspond to the same coarse-grained measurement record $\bm{M}_t$; all other coherences in the pure state $\Psi_t$ have been removed.  This process of partial dephasing (of $\Psi_t$) is caused by a partial projective measurement at the final time $t$.
Thanks to $\delta(\bm{M}_t-\bm{M}'_t)$  the double integral in Eq.~\eqref{eq:Rp_integral} should be interpreted as 
$\sum_{n=0}^\infty \sum_{j_1,j_1',\dots,j_n,j_n'=1}^d \prod_{i=1}^n \int_{0\leq t_1<\dots<t_n < t} \delta_{\alpha(j_i),\alpha(j_i')}$.

In contrast to Eq.~\eqref{eq:psi_mm}, the conditional joint state for a (coarse-grained) measurement record $\bm{M}_t$ includes coherences between different environmental states  $|d\bm{m}_t\rangle$ and $|d\bm{m}'_t\rangle$.  However it has the same conditional system state for each record [Eq.~\eqref{eq:psi_t_M}], and is therefore separable.
Physically, this reflects that records for partial measurement can be alternatively interpreted as records of a counting measurement associated with joint jump operators $\sum_{j\in S_\alpha} J_j \otimes dB^\dag_{j,t}$, which appear in the dynamics for the partially-dephased cMPO in Eq.~\eqref{eq:dLp}.

{The presence of coherences in the partially-dephased cMPO in Eq.~\eqref{eq:Rp_integral} allows for it to be connected by transformations as in Eq.~\eqref{eq:rep-ii} to cMPOs of any other representation with the same composite actions of the SJEDs. The symmetry operators for their joint unitary dynamics before partial measurements are related by the same isometric transformation, so the corresponding symmetry properties are retained after the measurements by either all such cMPOs or none. This happens exactly when their dynamics features the weak symmetry in Eq.~\eqref{eq:LP-symm}, which is characterised by}
Symmetry Condition II.  That explains why this is the relevant condition for symmetry of the partially-dephased cMPO, see  Box \Bpdcmps\ in Fig.~\ref{fig:outline}.  The same property ensures gauge independence of the SJED jump action under such transformations~\cite{Generators}, which is also the relevant condition for quantum trajectories [Box~\Btraj], see also Tab.~\ref{tab:summ}.

Note that isometric transformations that permute the SJEDs also leave the conditional state invariant: they only change the labelling of the emitted quanta in the coarse-grained measurement record.  Similarly, permuting the jumps themselves changes the labels in the full measurement record, leaving the conditional states invariant.  We recall that such permutations appear in the definitions of the symmetry operators $\Yf,\Yc$; this reflects that if two ensembles of labelled quantum trajectories are the same up to permutation of the labels, they describe physically equivalent situations.

Finally, we consider the coarse-grained cMPO in {Eq.~\eqref{eq:Rc_time_evolution}.  The situation is similar to previous cases, so we will be brief.  We have
%\begin{align}\label{eq:def_cg}
$
	R^\textsc{c}_t % &
	= \int p(\bm{M}_t|\psi_0) \,\psi_t(\bm{M}_t) \otimes |d\bm{M}_t\rangle\langle d\bm{M}_t|,
$
where the coarse-grained record $\bm{M}_t$ is now {encoded} by 
\begin{equation}
	|d\bm{M}_t\rangle=dC^\dag_{\alpha_n,t_n}\ldots dC^\dag_{\alpha_1,t_1}|\vac\rangle.
\end{equation}
while  $\psi_t(\bm{M}_t)$ is {the corresponding conditional system state} in Eq.~\eqref{eq:psi_t_M} and $p(\bm{M}_t|\psi_0)$ is the probability density in Eq.~\eqref{eq:path_prob_M}.  Alternatively we may write
\begin{equation}\label{eq:def_cg1}
	R^\textsc{c}_t 	= \int  \,\varphi_t(\bm{M}_t) \otimes |d\bm{M}_t\rangle\langle d\bm{M}_t|,
\end{equation}
%from Eqs.~\eqref{eq:varphi} and~\eqref{eq:coarse_traj}, it is clear how} 
from which one sees that this cMPO can be obtained from either $R_t$ {in Eq.~\eqref{eq:def_deph2}}  or $R^{\textsc p}_t$ in {Eq.~\eqref{eq:Rp_integral} by erasing the emitted quanta according to  which SJED their type corresponds to}, recall Fig.~\ref{fig:states}. 

The (unnormalised) conditional state for the record $\bm{M}_t$ of the coarse-grained  counting measurement is
\begin{equation}
	\psi_t(\bm{M}_t) \otimes |d\bm{M}_t\rangle\!\langle d\bm{M}_t| \; ,
\end{equation}
cf. Eq.~\eqref{eq:psi_mm}.
This joint state contains exactly the same information as the conditional system state  in Eq.~\eqref{eq:psi_t_M} combined with the coarse-grained measurement record in Eq.~\eqref{eq:partial_record}.  By the same arguments as Sec.~\ref{sec:unif-iii},
%From Eq.~\eqref{eq:Ito_alpha}, 
it follows that the coarse-grained cMPO introduced in Sec.~\ref{sec:CG_cMPS} contains exactly the same information as the ensemble of partially-labelled quantum trajectories  in Sec.~\ref{sec:CG-records}. 
This equivalence is responsible for the analogous symmetry properties of those objects and the weak symmetries in their dynamics characterised by the same Symmetry Condition~II, see Boxes \Bcgcmps~and~\Bcgtraj~in Fig.~\ref{fig:outline} and cf.~Tab.~\ref{tab:summ}.

Overall, the middle row of Fig.~\ref{fig:outline} includes four different ways of describing {the open quantum dynamics of} the system, at different levels of detail: they all {can feature weak} symmetries {whose presence is determined by the same} Symmetry Condition~II.  The next subsection  discusses further the relationship between these objects.

\subsection{Hierarchy of Symmetry Conditions}

{We have discussed in detail the symmetry properties of the various objects in Fig.~\ref{fig:outline}, and the associated Symmetry Conditions.  This Section provides additional discussion of the relationships between these objects.}

{\emph{Averaging} --}
starting from any object in Fig.~\ref{fig:outline}, the QME can be obtained by a suitable trace or average, as discussed in the relevant Sections above.  Since we only consider separable symmetries, this means that symmetry of the more detailed description always implies symmetry of the QME.  In other words, symmetry of the quantum master dynamics (characterised by Symmetry Condition I) is necessary for all other symmetries.

{\emph{Quantum trajectories and measurement records} -- 
The unravelled quantum dynamics includes classical information via conditioning on measurement outcomes, and is symmetric only under {Symmetry Condition~II}.  This condition, which is more restrictive than Symmetry Condition I, appears because  conditioning on the measurement outcomes can break the symmetry of the quantum master dynamics.}
If one further includes the information about all measurement outcomes via labelled quantum trajectories, the relevant condition is the yet stronger {Symmetry Condition~III}.  As discussed in Sec.~\ref{sec:partial-symm-further}, we only consider symmetries that act separably on trajectories and measurement records so any symmetry of the more detailed description (labelled quantum trajectories) is always sufficient to ensure symmetry of the ensemble of quantum trajectories, where labels are \emph{discarded}.  This general principle also applies to partially-labelled trajectories, and explains why increasing classical information in Fig.~\ref{fig:outline} corresponds to  conditions that are either more restrictive, or equal.

\emph{Coarse-graining} -- The coarse-graining operation, where fully-labelled quantum trajectories are replaced by partially-labelled ones, is more subtle.  The information content of the ensemble of partially-labelled quantum trajectories is intermediate between the ensembles of fully-labelled and unlabelled quantum trajectories.  Using again the separable structure, one sees that the conditions for the partially-labelled case must be at least as restrictive as the unlabelled case, and no more restrictive than the fully-labelled case.  In fact, they exactly match the unlabelled case (Symmetry Condition II).  
{The underlying reason -- as discussed in Sec.~\ref{sec:partial-symm-further} -- is that 
for different representations of the QME to generate the same ensemble of partially-labelled quantum trajectories, it is necessary and sufficient for their SJEDs to have the same composite actions~\cite{Generators}.  Indeed, Symmetry Condition II holds either for all such representations, or for none of them.  Moreover, if Symmetry Condition II holds then there is at least one representation  that obeys Symmetry Condition III, while still generating the same ensemble of partially-labelled quantum trajectories.}

{\emph{Transforming cMPSs between representations} -- 
We have emphasized that the stochastic Hamiltonians for different representations of the same QME are related by isometric transformations on the environment [see Eq.~\eqref{eq:kasia-tilde}, also recall Sec.~\ref{sec:measure-basis}].  If no measurements are performed then the joint state evolves unitarily and the existence of a joint symmetry is independent of the representation,}
 although the relevant symmetry operator must be transformed accordingly [recall Sec.~\ref{sec:symm-joint-uni-conds}].  Symmetry Condition I determines when such symmetries exist.

{\emph{Dephasing} -- Measurements can break symmetries of unitary joint dynamics, so the Symmetry Condition of measured cMPOs are more restrictive. The relationships between counting measurements and dephasing are most easily appreciated by comparing Eqs.~(\ref{eq:cMPS1},\ref{eq:def_deph2},\ref{eq:Rp_integral}), which shows that quantum information is reduced by dephasing, as coherences are removed.  As explained in Sec.~\ref{sec:weak-joint-symm-def}, symmetries of the unitary joint dynamics are preserved by measurement if the corresponding dephasing operator is symmetric.  Our results imply that necessary conditions for symmetry of the dephasing operator are Symmetry Condition III (for full dephasing) and Symmetry Condition II (for partial dephasing); see also Secs.~\ref{sec:unif-iii} and~\ref{sec:unif-ii}. This hierarchy of conditions is expected because more dephasing (reduced quantum information) leads to increasingly restrictive conditions for  the associated dephasing operator to be symmetric. (Fig.~\ref{fig:outline} also shows how full dephasing can be achieved in two stages, with the partially-dephased cMPO as an intermediate state.)}

{\emph{Erasure} -- The erasure operations in Fig.~\ref{fig:outline} take Eqs.~(\ref{eq:def_deph2},\ref{eq:Rp_integral}) to Eq.~\eqref{eq:def_cg1}. In the transformation from the fully-dephased cMPO to the coarse-grained cMPO the simple structure of $R_t$ means that erasure cannot remove its symmetry properties, which hold under Symmetry Condition III.}  
However, if this condition is violated but Symmetry Condition II holds, then erasure restores the symmetry of the dynamics, just as discarding labels restores the symmetry of quantum trajectories (under the same conditions).  The transformation from the partially-dephased cMPO to the coarse-grained cMPO can be achieved by first performing full dephasing, and then erasure as above.%, corresponding to the operator $\mathbb{C}\mathbb{F}$.}

{\emph{Symmetry Condition II revisited}} -- We observe that four different descriptions in Fig.~\ref{fig:outline} all share the same Symmetry Condition II.  One may ask why the diagram does not feature other Symmetry Conditions, different from both II and III.  This structure all hinges on the definition of the SJEDs, which are used extensively in the derivations of the Symmetry Conditions.  Their essential property is that the conditional states of the system (quantum trajectories) are invariant under changes of representation that preserve the SJED action operators ${\cal A}_\alpha$, recall Sec.~\ref{sec:cond-partial}.  

{
This structure can be further elucidated by using the algebra of joint super-operators (for example, $\mathbb{U},\mathbb{V},\mathbb{F},d\mathbb{H}_\tau$) to establish relationships between symmetry properties of different cMPOs (further insight may be obtained by introducing additional operators for coarse-graining and partial dephasing, see also below).
To briefly sketch the idea, 
consider a representation that obeys Symmetry Condition II but not Symmetry Condition III.  The joint unitary dynamics has a time-independent symmetry ($\mathbb{U} \, d\mathbb{H}_\tau \, \mathbb{U}^\dag =  d\mathbb{H}_\tau$); the evolution operator of the fully dephased cMPO may be written as 
\beq
{\cal T} e^{\int_0^t d \mathbb{L}_\tau} = \mathbb{F} \Big( {\cal T} e^{-i\int_0^t d\mathbb{H}_\tau} \Big) \mathbb{F}
\eeq 
but this is not symmetric because $[\mathbb{U},\mathbb{F}]\neq 0$.  However, Sec.~\ref{sec:cond-partial} explains that the operator $\mathbb V$ in Eq.~\eqref{eq:superV-ii} may be chosen such that it only mixes quanta within their SJEDs, and Symmetry Condition III holds for the new representation.  Hence the corresponding operator $d \tilde{\mathbb{L}}_\tau = \mathbb{V} \,d \mathbb{L}_\tau \mathbb{V}^\dag $ is symmetric, that is 
$\tilde{\mathbb{U}} \, d\tilde{\mathbb{L}}_\tau \, \tilde{\mathbb{U}}^\dag =  d\tilde{\mathbb{L}}_\tau$.  The relationship between $\tilde{\mathbb{U}}$ and $\mathbb{U},\mathbb{V}$ is discussed in Sec.~\ref{sec:symm-joint-uni-conds}.   
 Such relationships -- and their interplay with partial dephasing, erasure, and coarse-graining -- might be usefully investigated in future work.}

This concludes our discussion of Symmetry Conditions and their applicability to the objects in Fig.~\ref{fig:outline}.
The following Sec.~\ref{sec:examples} gives examples to illustrate this general picture, before
Sec.~\ref{sec:implications} describes some further implications of the symmetries derived so far.

\section{Examples}
\label{sec:examples}

We present three examples to illustrate Symmetry Conditions I, II, and~III, focusing on the dynamics of quantum trajectories.  In Secs.~\ref{sec:example_qubit} and \ref{sec:example_two_qubit}, we present systems with one and two qubits respectively, and we show how different representations of the QME satisfy different symmetry conditions.  In Sec.~\ref{sec:example_translate}, we discuss an example of a qutrit chain with translation and rotation symmetries.

\subsection{Single Qubit}\label{sec:example_qubit}

Consider a single qubit system, with Hamiltonian and jump operators given by
\begin{equation}\label{eq:ex1:rep1}
    H=\omega\sigma^z, \quad J_1 = \sqrt{\gamma_z}\sigma^z, \quad J_2 = \sqrt{\gamma_x}\sigma^x,
\end{equation}
where $\sigma^x$, $\sigma^z$ are Pauli matrices. %\cred{[Remove the $\gamma$'s?}.
The stationary state of the corresponding quantum master dynamics in Eq.~\eqref{eq:LL-outline} is unique and given by the fully mixed state $\mathbb{1}/2$.
Moreover, the dynamics features the weak symmetry with respect to the unitary operator  [cf. Eq.~\eqref{eq:weakU_QME}] 
\begin{equation}\label{eq:ex1:U}
U=\sigma^z.
\end{equation}
Thus, Symmetry Condition~I is satisfied for any representation of the QME.
{This is a parity symmetry, as $U^2=\mathbf{1}$ (so that $N=2$).}

The representation given in Eq.~\eqref{eq:ex1:rep1} is weakly symmetric   [cf. Eq.~\eqref{eq:weakR}] with
\begin{equation}
    \mathcal{U}(H) = H,\quad \mathcal{U}(J_1) = J_1, \quad \mathcal{U}(J_2) = -J_2,
\end{equation}
so that Symmetry Condition~I,~II, and~III are all satisfied (with $\pic$ and $\pi$ both trivial). The quantum trajectories,  are not ergodic on the Bloch sphere. For a given initial state, $\psi_0$, the corresponding asymptotic probability distribution is nevertheless symmetric with respect to the transformation under the symmetry superoperator $\mathcal{U}$ [cf. Eq.~\eqref{eq:tilde-path-P}], as it is supported on the cross-section with two planes where  $|\Tr(\psi\sigma_z)|=|\Tr(\psi_0\sigma_z)|$, and is uniform there. Moreover, unless $\psi_0$ is an eigenstate of $\sigma_z$, conditional states are not symmetric at any time.
Next, we consider other representations of the QME, which exhibit less symmetry in their quantum trajectories. In what follows, we always keep the Hamiltonian the same and change only the jump operators.

\emph{Symmetry Condition~III}: Consider the jump operators % given by % jump operators
\begin{equation}\label{eq:ex1:rep2}
    J_1 = \frac{1}{\sqrt{2}}(\sqrt{\gamma_z}\sigma^z + \sqrt{\gamma_x}\sigma^x), \quad J_2 = \frac{1}{\sqrt{2}}(\sqrt{\gamma_z}\sigma^z - \sqrt{\gamma_x}\sigma^x) \; .
\end{equation}
The trajectories are now ergodic and the asymptotic probability distribution is uniformly distributed over the Bloch sphere.
This representation is no longer weakly symmetric but does still satisfy Symmetry Condition~III. Indeed,  the jump operators are swapped by the symmetry,
\begin{equation}\label{eq:ex1:rep2_U}
    \mathcal{U}(J_1) = J_2, \quad \mathcal{U}(J_2)=J_1\,,
\end{equation}
so that $\pi(1)=2$ and $\pi(2)=1$ and $\phi_1=\phi_2=0$ in Eq.~\eqref{eq:symm_cond_III}.

\emph{Symmetry Condition~II}: 
Consider the jump operators 
\begin{align}\label{eq:ex1:rep3}
    &J_1 = c_1(\sqrt{\gamma_z}\sigma^z+\sqrt{\gamma_x}\sigma^x), \quad J_2 = c_2(\sqrt{\gamma_z}\sigma^z+\sqrt{\gamma_x}\sigma^x) \nonumber \\
    &J_3 = \frac{1}{\sqrt{2}} ( \sqrt{\gamma_z}\sigma^z-\sqrt{\gamma_x}\sigma^x ) ,
\end{align}
parameterised
 by non-zero $c_1,c_2\in \mathbb{C}$, subject to 
$|c_1|^2+|c_2|^2 = 1/2$.  Since $J_1 \propto J_2$  there are two (non-reset) SJEDs with  $S_1 = \{1,2\}$ and $S_2 = \{3\}$.  
Under the action of the symmetry, the jump operators themselves are not related by a permutation, but the corresponding SJED actions transform as
\begin{equation}\label{eq:ex1:rep3_AU}
    \mathcal{UA}_1\mathcal{U}^\dag = \mathcal{A}_2, \quad \mathcal{UA}_2\mathcal{U}^\dag = \mathcal{A}_1,
\end{equation}
so that $\pic(1)=2$ and $\pic(2)=1$ in Eq.~\eqref{eq:symm_cond_II}.
Hence this representation obeys Symmetry Condition~II. 
Suitable choices for the matrices $\mathbf{X},\mathbf{U}$ of Eqs.~(\ref{eq:UIIR},\ref{eq:not_cond_IIb}) are
\begin{align}
\mathbf{X} & = \sqrt{2}
\begin{pmatrix} 
0 & 0 & c_1 \\
0 & 0 & c_2 \\
c_1^* & c_2^* & 0 
\end{pmatrix},
\nonumber\\
\mathbf{U} & =  \sqrt{2}
\begin{pmatrix} 
 \sqrt{2}|c_2|^2  & - \sqrt{2} c_1 c_2^* & c_1  \\
- \sqrt{2}c_1^* c_2  &  \sqrt{2}|c_1|^2  & c_2  \\
c_1^*  & c_2^*  & 0 
\end{pmatrix}.
\end{align}
Indeed, $\mathbf{X}$ has the block structure induced by Eq.~\eqref{eq:Va1} (but is not unitary) while
$\mathbf{U}$ is unitary but lacks the block structure.  This $\mathbf{U}$ is also consistent with Eq.~\eqref{eq:UU-partial-lemma}.

 Note that $\pic$ above coincides with $\pi$ before. In fact, the SJEDs for Eq.~\eqref{eq:ex1:rep2} are singletons, and their action is transformed exactly as in Eq.~\eqref{eq:ex1:rep3_AU}, and $\mathbf{X}=\mathbf{U}$ as they are both uniquely defined, cf. Sec.~\ref{sec:implications}.  Thus, the weak symmetry of the unravelled quantum dynamics that coincides for both representations in Eq.~\eqref{eq:ex1:rep2} and~\eqref{eq:ex1:rep3} can be seen as inherited from the weak symmetry of the labelled quantum dynamics for the representation in  Eq.~\eqref{eq:ex1:rep2}.

\emph{Symmetry Condition~I}: Now consider the jump operators % with jump operators
\begin{equation}\label{eq:ex1:rep4}
    J_1 = a\sqrt{\gamma_z}\sigma^z + b\sqrt{\gamma_x}\sigma^x, \quad J_2 = b^*\sqrt{\gamma_z}\sigma^z\ - a^*\sqrt{\gamma_x}\sigma^x
\end{equation}
parameterised by non-zero $a,b\in\mathbb{C}$, subject to
 $|a|^2+|b|^2=1$ and $|a|\neq |b|$ [which excludes, e.g., Eqs.~\eqref{eq:ex1:rep1} and~\eqref{eq:ex1:rep2})]. Then, any such representation no longer satisfies Symmetry Conditions~II or~III, but does obey Symmetry Condition~I.  Indeed, we have that Eq.~\eqref{eq:symm_cond_I} holds for any such $a,b$ and hence also Eq.~\eqref{eq:weakU_QME},  as it should.

Finally, we use this system to show an explicit example of the non-uniqueness of $\mathbf{U}$. Consider a representation with two sets of proportional jump operators
\begin{align}\label{eq:ex1:rep5}
    &J_1 = c_1(\sqrt{\gamma_z}\sigma^z+\sqrt{\gamma_x}\sigma^x), \quad J_2 = c_2(\sqrt{\gamma_z}\sigma^z+\sqrt{\gamma_x}\sigma^x) \nonumber\\
    &J_3 = c_1(\sqrt{\gamma_z}\sigma^z-\sqrt{\gamma_x}\sigma^x), \quad J_4 = c_2(\sqrt{\gamma_z}\sigma^z-\sqrt{\gamma_x}\sigma^x),
\end{align}
with $|c_1|^2+|c_2|^2 = 1/2$, as before. This representation obeys Symmetry Condition~III with $\mathcal{U}(J_{1,2})=J_{3,4}$ and $\mathcal{U}(J_{3,4})=J_{1,2}$ [cf. Eqs.~\eqref{eq:ex1:rep2} and~\eqref{eq:ex1:rep2_U}]. There are many unitary matrices $\mathbf{U}$ that give these transformations (see Appendix~\ref{app:boldU-gen}). For concreteness we take real $(c_1,c_2)=\frac{1}{2}(\cos\theta,\sin\theta)$ in which case two (real) possibilities for $\mathbf{U}$ are
\begin{equation}
    \mathbf{U} =
    \begin{pmatrix}
        0&0&1&0\\
        0&0&0&1\\
        1&0&0&0\\
        0&1&0&0\\
    \end{pmatrix}
   , \;\;
    \mathbf{U} =
    \begin{pmatrix}
        0&0&\cos2\theta&\sin2\theta\\
        0&0&\sin2\theta&-\cos2\theta\\
        1&0&0&0\\
        0&1&0&0\\
    \end{pmatrix}.
\end{equation}
Note that the first of these satisfies Eq.~\eqref{eq:UIIIRb_U}, the existence of such a $\mathbf{U}$
is required by 
Symmetry Condition~III. 
This is important when considering the symmetry of the dephased cMPO (see Sec.\ref{sec:dephased_symm} and Appendix \ref{app:full_deph}).

\newcommand{\0}{\downarrow\downarrow}
\newcommand{\1}{\uparrow\uparrow}
\newcommand{\2}{\uparrow\downarrow}
\newcommand{\3}{\downarrow\uparrow}
\subsection{Two Coupled Qubits}\label{sec:example_two_qubit}

We now consider a system of two coupled qubits and describe the symmetry properties of various representations. In particular, we illustrate that reset SJEDs can exhibit a global parity symmetry of the unravelled quantum dynamics, despite the reset jump operators not being individually symmetric. 

We take Hamiltonian and jump operators
\begin{align}
\label{eq:ex2:rep1}
    H 
     &= \omega_1\sigma_1^x + \omega_2\sigma_2^x,
    \nonumber\\
    J_1 &= \sigma_1^-\bar{n}_2 + \sigma_1^+n_2, \quad J_2 = \sigma_1^-\bar{n}_2 - \sigma_1^+n_2,
    \nonumber \\
    J_3 &= \bar{n}_1\sigma_2^- + n_1\sigma_2^+, \quad J_4 = \bar{n}_1\sigma_2^- - n_1\sigma_2^+,
\end{align}
where $\sigma^\pm = (\sigma^x\pm i\sigma^y)/2$ and $n,\bar{n}=(\mathbb{1}\pm\sigma^z)/2$, with the subscript denoting the qubit on which the operator acts. %We use the single qubit basis $\{|\!\uparrow\rangle, |\!\downarrow\rangle\}$.
The corresponding quantum master dynamics features the weak symmetry with respect to the unitary operator
\begin{align}\label{eq:ex2:U}
    U = \sigma_1^x\sigma_2^x,
\end{align}
which satisfies $U^2=\mathbb{1}$. % (i.e. a $\mathbb{Z}_2$ symmetry).
The representation given in Eq.~\eqref{eq:ex2:rep1} is weakly symmetric, with 
\begin{align}
\mathcal{U}(J_{1})&= J_{1}, \quad \mathcal{U}(J_{2})= -J_{2},\nonumber\\
\mathcal{U}(J_{3})&= J_{3}, \quad \mathcal{U}(J_{4})= -J_{4}.
\end{align}
With respect to Eq.~\eqref{eq:symm_cond_III}, the permutation $\pi$ is trivial, but the phase factors are not, $\phi_1=\phi_3=0$ and $\phi_2=\phi_4=180^\circ$.

\emph{Symmetry Condition~III}: Consider the same Hamiltonian as above, with the \emph{reset} jump operators
\begin{align}\label{eq:ex2:repIII}
    J_1 =& \sigma_1^-\bar{n}_2, \quad J_2 = \sigma_1^+n_2, \nonumber\\
    J_3 =& \bar{n}_1\sigma_2^-  , \quad J_4 = n_1\sigma_2^+
\end{align}
(with reset destinations 
	$|00\rangle$, $|11\rangle$, $|00\rangle$, and $|11\rangle$,
	respectively, and where the Pauli operators correspond to the single-qubit basis % $|\!\!\!\uparrow\rangle, |\!\!\!\downarrow\rangle $
	$|1\rangle, |0\rangle $).
This representation is no longer weakly symmetric, but it %gives canonical purifications and
satisfies Symmetry Condition~III with 
\begin{align}
	\mathcal{U}(J_{1})&= J_{2}, \quad \mathcal{U}(J_{2})= J_{1},\nonumber\\
	\mathcal{U}(J_{3})&= J_{4}, \quad \mathcal{U}(J_{4})= J_{3}.
\end{align}
The permutation $\pi$ now consists of 2 cycles, $\pi(1)=2$,  $\pi(2)=1$, $\pi(3)=4$, $\pi(4)=3$, while the phase factors are trivial, $\phi_1=\phi_2=\phi_3=\phi_4=0$.

\emph{Symmetry Condition~II}: Keeping the Hamiltonian, we now consider the {reset} jump operators 
\begin{align}\label{eq:ex2:repII}
    J_1 &= a_1\sigma_1^-\bar{n}_2+a_2\bar{n}_1\sigma_2^-, \quad J_2 = a_2^*\sigma_1^-\bar{n}_2-a_1^*\bar{n}_1\sigma_2^-,
    \nonumber \\
    J_3 &= b_1 n_1\sigma_2^+ + b_2\sigma_1^+n_2,  \,\quad J_4 = b_2^*n_1\sigma_2^+- b_1^*\sigma_1^+n_2,
\end{align}
with non-zero $a_1,a_2,b_1,b_2\in\mathbb{C}$ such that $|a_1|^2+|a_2|^2 = |b_1^2|+|b_2|^2=1$ and $|a_1|\neq|b_1|$.
The SJEDs are $S_1=\{1,2\}, \, S_2=\{3,4\}$, of reset type~\cite{Generators} (with the reset destinations 
$|00\rangle$ and $|11\rangle$, respectively).
While Symmetry Condition~III does not hold, the representation does obey Symmetry Condition~II with
\begin{equation}\label{eq:ex2:condII}
        \mathcal{UA}_{1}\mathcal{U}^\dag = \mathcal{A}_{2} \; ,\quad      \mathcal{UA}_{2}\mathcal{U}^\dag = \mathcal{A}_{1} \; ,  
\end{equation}
so that $\pic(1)=2$ and  $\pic(2)=1$.  The weak symmetry of the corresponding unravelled or partially-labelled quantum dynamics can be seen as inherited from the weak symmetry of the labelled quantum dynamics for any representation with $|a_1|=|b_1|$ [which all satisfy Symmetry Condition~III, cf. the special case in Eq.~\eqref{eq:ex2:repIII}].
Note that the jump operators in each of the SJEDs are not proportional to each other [contrary to Eq.~\eqref{eq:ex1:rep3}]. 
This illustrates the gauge freedom of the reset jumps for dynamics of quantum trajectories; for any valid choice of $a_1,a_2,b_1,b_2\in\mathbb{C}$  in Eq.~\eqref{eq:ex2:repII}, we have that Eq.~\eqref{eq:ex2:condII} holds [in contrast to Eq.~\eqref{eq:ex1:rep4}].

\emph{Symmetry Condition~I}: Finally, consider the jump operators
\begin{align} \label{eq:ex2:repI}
    J_1 &= a_1\sigma_1^-\bar{n}_2+b_1 n_1\sigma_2^+, \quad J_2 = b_1^*\sigma_1^-\bar{n}_2-a_1^*n_1\sigma_2^+,
    \nonumber\\
    J_3 &= a_2\bar{n}_1\sigma_2^+ + b_2\sigma_1^+n_2, \,\quad J_4 = b_2^*\bar{n}_1\sigma_2^+-a_2^*\sigma_1^+n_2,
\end{align}
where non-zero $a_1,a_2,b_1,b_2\in\mathbb{C}$ obey $|a_1|^2+|b_1|^2 = |a_2|^2+|b_2|^2=1$  together with  $|a_1|\neq|b_1|$ [this excludes, e.g., Eqs.~\eqref{eq:ex2:rep1} and~\eqref{eq:ex2:repIII}]. For all such representations neither Symmetry Condition~II or~III hold, but we retain Symmetry Condition~I as expected since the QME features the weak symmetry [cf. Eq.~\eqref{eq:ex1:rep4}].

\subsection{Qutrit Chain with Translation and Rotation Symmetries}
\label{sec:example_translate}

So far we considered the unitary symmetries of a parity type, i.e., corresponding to the  $\mathbb{Z}_2$ group [cf. Eqs.~\eqref{eq:ex1:U} and~\eqref{eq:ex2:U}].
A form of weak unitary symmetry which is prevalent in many-body systems with periodic boundary conditions is translational symmetry.  This relevant group is $\mathbb{Z}_N$, where $N$ now coincides with the system size. Another type of directional independence in the dynamics, for internal rather than external degrees of freedom, are rotation symmetries.

\emph{Translation symmetry}: For simplicity, we consider a chain of $L$ sites in 1 dimension. Each site has a local Hilbert space dimension 3, for which we use the basis $|0\rangle, |1\rangle, |2\rangle$. 
The translation symmetry is encoded by the unitary operator $\trans$ defined so that
\begin{multline}
	\trans \left( |\psi_1\rangle\otimes|\psi_2\rangle\otimes\dots\otimes|\psi_L\rangle\right) \\
	= |\psi_L\rangle\otimes|\psi_1\rangle\otimes\dots\otimes|\psi_{L-1}\rangle ,
\end{multline}
where $ |\psi_\alpha\rangle$ is a pure state on site $\alpha=1,...,L$ (we take periodic boundary conditions, $L+1=1$). We have $\trans^L=\mathbb{1}$, so that indeed $N=L$. We then consider quantum master dynamics which exhibits the  weak translation symmetry. We assume a translationally symmetric Hamiltonian $H$, so that $\trans H \trans^\dag=H$.  %% ** should keep this here despite KM comment
The (reset) jump operators are chosen as
\begin{align}\label{eq:trans_jumps}
	J_\alpha^{(1)} &= %\frac{1}{\sqrt{2}}
    |\Omega\rangle\!\langle\Omega|\left[c_{\alpha}^{(1)}\cos\theta_\alpha + c_{\alpha}^{(2)}\sin\theta_\alpha \right], \nonumber \\
	J_\alpha^{(2)} &= %\frac{1}{\sqrt{2}}
	|\Omega\rangle\!\langle\Omega|\left[c_{\alpha}^{(1)}\sin\theta_\alpha -c_{\alpha}^{(2)} \cos\theta_\alpha \right]
\end{align}
for  $\theta_\alpha\in\mathbb{R}$ and $\alpha=1,...,L$. Here, $|\Omega\rangle = \bigotimes_{\alpha=1}^L |0\rangle$ and $c_{\alpha}^{(1)}$ and $c_{\alpha}^{(2)}$ denote the operators $c^{(1)}=|0\rangle\!\langle 1|$ and $c^{(2)}=|0\rangle\!\langle 2|$ acting on site $\alpha$ (while acting as the identity on the other sites).
While reset operators are atypical in many-body systems, any linear combinations of creation/anihilation operators as jump operators will have the same symmetry properties as described below.

Each SJED consists of $J_\alpha^{(1)}$ and $J_\alpha^{(2)}$ and its composite action is independent of $\theta_\alpha$; 
$\mathcal{A}_\alpha(\psi) = |\Omega\rangle\!\langle\Omega|\,\{\langle \Omega|c_{\alpha}^{(1)}\psi [c_{\alpha}^{(1)}]^\dag |\Omega\rangle +\langle\Omega|c_{\alpha}^{(2)}\psi [c_{\alpha}^{(2)}]^\dag |\Omega\rangle\}$.
It follows that for $\Trans(\cdot) = \trans(\cdot) \trans^\dag$, 
 \begin{equation}\label{eq:trans_symm_cond_II}
 	\Trans(H)=H,\qquad   \Trans\mathcal{A}_\alpha\Trans^\dag = \mathcal{A}_{\alpha+1},
 \end{equation}
 where $ \alpha=1,...,L $, so that $\pic$ acts as a translation on the cycle of length $L$. Thus, Symmetry Condition~II is satisfied by such representations, and Symmetry Condition~I follows, so that the weak translation symmetry holds. Symmetry Condition~III, however, is not fulfilled with respect to that symmetry  by the jump operators in Eq.~\eqref{eq:trans_jumps}, unless  $\theta_\alpha=\theta $, $\theta+90^\circ$, $\theta+180^\circ$ or $\theta+270^\circ$ for each $ \alpha=1,...,L $ and some $\theta\in \mathbb{R}$.\footnote{Here,  $|\Omega\rangle$ can be viewed as the vacuum state while the jump operators can be interpreted as local anihilation of superpositions of particles of type $1$ and $2$ from states with only a single particle. The composite action of each SJED is removal of any particles from the corresponding site, so that Symmetry Condition~II is satisfied.	Then the Symmetry Condition~III is not obeyed unless effectively the same (coherent) types of particles are anihilated at every site.
}
For more details on the various representations of these SJEDs and their connections, see the example in \cite{Generators}, which discusses a similar model on a single site.

\emph{Rotation symmetry}: This quantum master dynamics can also feature a weak symmetry with respect to any rotation of the $\{|1\rangle,|2\rangle\}$ subspace of any site, which form the $SO(2)$ group. A rotation  by angle $\vartheta\in\mathbb{R}$ is described by the single-site unitary operator
\begin{multline}
U(\vartheta) = |0\rangle\!\langle 0| + \cos\vartheta\,(|1\rangle\!\langle1|+|2\rangle\!\langle2|) \\+\sin\vartheta\,(|2\rangle\!\langle1|-|1\rangle\!\langle2|),
\end{multline}
and the symmetry operator for the chain is then
\begin{equation}
	 \quad\urot = \bigotimes_{\alpha=1}^L U (\vartheta_\alpha) \; ,
\end{equation}
which depends implicitly on the angles $\{\vartheta_\alpha\}_{\alpha=1}^L$.
For $\Urot(\cdot) = \urot (\cdot) \urot^\dag $, 
the action of SJEDs themselves features a weak symmetry, with
\begin{equation}
	\Urot\mathcal{A}_\alpha\Urot^\dag = \mathcal{A}_\alpha,
\end{equation}
where $\alpha=1,...,L$. Thus, the permutation $\pic$ of the SJEDs is the trivial one (identity). Therefore, Symmetry Condition~II is satisfied if and only if $\Urot(H)=H$, i.e., the considered Hamiltonian is rotationally symmetric. By Eqs.~\eqref{eq:not_cond_IIb}, this is the only situation Symmetry Condition~I is satisfied as well, and we assume it below.

\emph{Combined symmetry}: Finally, we show that for any representation given by Eq.~\eqref{eq:trans_jumps}, there always exists a symmetry operator $\urot$, such that the considered representation satisfies Symmetry Condition~III with respect to the combined unitary symmetry
\begin{equation}\label{eq:ex3:U}
U=\trans\urot.
\end{equation}
For example, choosing $\vartheta_\alpha = \theta-\theta_\alpha$, for some $\theta\in \mathbb{R}$, we obtain
\begin{equation}
\mathcal{U}[J_\alpha^{(1)}] = J_{\alpha+1}^{(1)}, \quad \mathcal{U} [J_\alpha^{(2)}] = J_{\alpha+1}^{(2)}
\end{equation}
for $\alpha=1,...,L$. Indeed, note that [cf. Eq.~\eqref{eq:trans_jumps}]
\begin{align}\label{eq:UT_trans_jumps}
	\Urot(J_\alpha^{(1)}) =  
	|\Omega\rangle\!\langle\Omega|\left[c_{\alpha}^{(1)}\cos\tilde{\theta}_\alpha + c_{\alpha}^{(2)}\sin\tilde{\theta}_\alpha\right], \nonumber \\
	\Urot(J_\alpha^{(2)}) =  
	|\Omega\rangle\!\langle\Omega|\left[c_{\alpha}^{(1)}\sin\tilde{\theta}_\alpha -c_{\alpha}^{(2)}\cos\tilde{\theta}_\alpha\right],
\end{align}
where $\tilde{\theta}_\alpha=\theta_\alpha+\vartheta_\alpha$. These jump operators are exactly as in Eq.~\eqref{eq:trans_jumps} but with $\theta_\alpha$ replaced by $\tilde{\theta}_\alpha$.
Hence, for the choices of $\vartheta_\alpha$ above, we obtain the representation with
% $\theta_\alpha\rightarrow\theta \;\forall\alpha$.
$\tilde\theta_\alpha=\theta$, which satisfies the translation Symmetry Condition~III, as discussed earlier. Thus, the original representation obeys the Symmetry Condition~III for the combined symmetry operator in Eq.~\eqref{eq:ex3:U} (but, in general, not with respect to to $\trans$ or $\urot$ individually). This holds for any choice of $\theta$ (for which a generally different $\urot$ is determined via $\{\vartheta_\alpha\}_{\alpha=1}^L$).

\section{Further implications of symmetric unravelled dynamics}\label{sec:implications}

We discussed in Sec.~\ref{sec:traj} the conditions under which the unravelled quantum dynamics is symmetric (Symmetry Condition~II). 
This Section discusses some implications of this symmetry.

\subsection{Eigenfunctions of unravelled generators}\label{sec:eigenfunctions}

First we introduce the adjoint of the generator $\mathcal{W}^\dag$, which is denoted $\mathcal{W}$ and defined via $\langle f, \mathcal{W}^\dag P\rangle = \langle \mathcal{W} f,P\rangle$, recall Sec.~\ref{sec:generator_symm}. 
This ${\cal W}$ is the (backwards) generator on the space of functions,
\begin{align}\label{eq:W}
    \mathcal{W}f(\psi) =& \mathcal{B}(\psi)\cdot\nabla f(\psi)\nonumber\\&+\sum_k\int d\psi' w_k(\psi,\psi')\left[f(\psi')-f(\psi)\right].
\end{align}
Expectation values evolve in time as $\frac{d}{dt} \mathbb{E}[ f(\psi_t) ] = \mathbb{E}\left[\mathcal{W}f(\psi_t)\right]$.

Eigenfunctions of the operator ${\cal W}$ (and the associated eigenvalues) provide insight into the dynamics of unravelled quantum trajectories.  It is also useful to note that if $F$ is an eigenmatrix of ${\cal L}^\dag$ with eigenvalue $\lambda$ then there is an associated eigenfunction of ${\cal W}$;\footnote{${\cal L}^\dag$ is the adjoint of the quantum master operator with respect to the Hilbert-Schmidt scalar product $\langle A,B\rangle_{\textsc{HS}}=\Tr(A^\dag B)$.} specifically, ${\cal W} f(\psi) = \lambda f(\psi)$, where $f(\psi) = \Tr(F \psi)$ is a linear function.  The operator ${\cal L}^\dag$ typically has  $d_{\rm s}^2$ distinct eigenmatrices so this enables construction of a finite number of linear eigenfunctions of ${\cal W}$.  However, ${\cal W}$ typically has an infinite number of eigenvalues, including nonlinear functions.

If ${\cal W}$ features a weak symmetry then writing Eq.~\eqref{eq:weakU_W} as $[\mathcal{W},\Upsilon]=0$, shows that eigenfunctions of ${\cal W}$ must be eigenfunctions of $\Upsilon$.  Moreover, eigenfunctions of $\Upsilon$ are related in turn to eigenfunctions of the unitary operator $U$.  This Section explains how this structure constrains the eigenfunctions of ${\cal W}$.

We denote the eigenvectors of $U$ as $|v^{(i)}\rangle$, that is
\begin{equation}
U|v^{(i)}\rangle = e^{i\phi_i}|v^{(i)}\rangle
\end{equation} 
where $\phi_i$ is real and the $|v^{(i)}\rangle$ form a set of $d_{\rm s}$ orthonormal vectors.
A general linear function of $\psi$ is
\begin{equation}\label{def:linear_eigfuncs}
    f(\psi)=\Tr[F\psi] % = \sum_{\mu\nu} F_{\mu\nu} \psi_{\nu\mu}
\end{equation}
where $F$ is a system operator.
This is an eigenfunction of $\Upsilon$ if 
\begin{equation}
    \mathcal{U}(F) = \lambda F \,.
    \label{eq:U-F}
\end{equation}
The eigenmatrices of ${\cal U}$ can be constructed directly from those of $U$, we have 
\begin{equation}
{\cal U}(|v^{(i)}\rangle\langle v^{(j)}|) = e^{i(\phi_i-\phi_j)}|v^{(i)}\rangle\langle v^{(j)}|.
\label{eq:UU-eigen}
\end{equation} 
These eigenmatrices form a complete orthonormal basis for operators in ${\cal H}$.  Hence we can express any $\psi$ in this basis as $\psi = \sum_{ab} \psi_{ab}|v^{(a)}\rangle\langle v^{(b)}|$ for suitable coefficients $\psi_{ab}$, which are the elements of $\psi$ in the eigenbasis of $U$ (the ``symmetry-adapted basis'').

Taking $F$ in Eq.~\eqref{def:linear_eigfuncs} to be a generic eigenmatrix of $\cal U$ we  get a generic eigenfunction of $\Upsilon$ as
\begin{align}
    f^{(ij)}(\psi) %& = \Tr[F^{(ij)}\psi] \nonumber \\ 
    &= \sum_{ab} \Tr\left[|v^{(i)}\rangle\langle v^{(j)}|\psi_{ab}|v^{(a)}\rangle\langle v^{(b)}|\right] 
    = \psi_{ji}
\end{align}
which has eigenvalue $e^{i(\phi_i-\phi_j)}$. This construction gives  all the eigenfunctions of $\Upsilon$ that are linear as in Eq.~\eqref{def:linear_eigfuncs}.  
A generic linear eigenfunction of ${\cal W}$ is then given by a linear combination of eigenfunctions of $\Upsilon$, all from the same eigenspace. 

We now consider `higher-order' functions
\begin{align}\label{def:higher_eigfuncs}
    g_n(\psi) & = \Tr[G_n\psi^{\otimes n}]
    \nonumber \\ & = (G_n)_{i_1 i_2\ldots i_{2n}} \psi_{i_1i_2} \psi_{i_3i_4}\ldots \psi_{i_{2n-1}i_{2n}} \; ,
\end{align}
where $G_n$ is a tensor with $2n$ legs.
%where operator $G_n\in B(\mathcal{H}^{\otimes n})$. 
By analogy with Eq.~\eqref{eq:U-F}, this is an eigenfunction of $\Upsilon$ if
\begin{equation}
    {\mathcal{U}}^{\otimes n}(G_{n}) = \lambda_n G_{n}
\end{equation}
for which a complete set of eigen-tensors is
$G_n^{(i_1i_2i_3...i_{2n})} = |v^{(i_1)}\rangle\langle v^{(i_2)}|\otimes ... \otimes |v^{(i_{2n-1})}\rangle\langle v^{(i_{2n})}|$ with eigenvalue
$\lambda_n = \exp\left(-i\sum_{x = 1}^{2n} (-1)^x \phi_{i_x}\right)$.
It can be that more than one eigen-tensor $G$ gives rise to the same function $g$, we address this issue below.
A generic $n$th order eigenfunction of $\Upsilon$ is then
\begin{align}\label{eq:higher_eigfuncs}
    g_n(\psi) & = \Tr[G_n^{(abcd\dots yz)}\psi^{\otimes n}] 
    \nonumber\\ & = \psi_{ab}\psi_{cd} \dots \psi_{yz}
\end{align}
where the last line is a product of $n$ matrix elements, in the symmetry-adapted basis.  This is a monomial of order $n$. 

To see how more than one tensor $G$ gives the same eigenfunction, note that $G_4^{(1221)}$ generates the function $g(\psi)=\psi_{12}\psi_{21}$ and $G_4^{(2112)}$ generates the same $g(\psi)=\psi_{21}\psi_{12}$.  To construct a complete basis of (distinct) eigenfunctions at order $n$ we take
$$
g_n(\psi) = \psi_{ab} \psi_{cd} \dots \psi_{yz}
$$
where $a\leq c \leq e \leq \dots$ and if $a=c$ then $b\leq d$ etc\footnote{Considering $ab,cd$ etc as two digit numbers in base $d_{\rm s}$ then this is $ab\leq cd\leq ef \leq \dots$}.  It is explicit to construct these functions, they are all distinct, and their eigenvalues are known.

For a generic eigenfunction of $\Upsilon$ we can combine
monomials of different orders as long as the eigenvalue is the same. 
Therefore, a general eigenfunction of $\mathcal{W}$ is also an eigenfunction of $\Upsilon$ with eigenvalue $\lambda$, and this can be uniquely expressed as
\begin{align}\label{eq:eigfuncs_general}
    f(\psi) = \sum_{n=1}^\infty \sum_{\substack{i_1...i_{2n}=1: \\ \exp\left(-i\sum_{x = 1}^{2n} (-1)^x \phi_{i_x}\right) = \lambda}}^{r_{\rm s}}
    \!\!\!\!\!\! 
    c_{(i_1...i_{2n})}\prod_{x=1}^n\psi_{i_{2x-1}i_{2x}}
\end{align}
where the sum over $i_1\dots i_{2n}$ is also restricted such that $i_1i_2\leq i_3i_4 \leq \dots$ (interpreted again as two digit numbers, see above).  
%This is a fully generic form for an eigenfunction of ${\cal W}$.  
For the special case of linear eigenfunctions then $c_{(i_1...i_{2n})}=0$ for $n>1$, but generic eigenfunctions mix all values of $n$.

As Eq.~\eqref{eq:weakU_W} also implies $\left[\mathcal{W}^\dag,\Upsilon\right]=0$, the eigenspaces of $\mathcal{W}^\dag$ are constructed identically to those of $\mathcal{W}$. However, the choices of $c_{(i_1...i_{2n})}$ in Eq.~\eqref{eq:eigfuncs_general} which give eigenfunctions of $\mathcal{W}$ will not in general produce eigenfunctions of $\mathcal{W}^\dag$. {In particular, unlike in the case of $\mathcal{W}$, there is no guarantee that $\mathcal{W}^\dag$ has linear eigenfunctions and there is no simple relationship between eigenfunctions of $\mathcal{W}^\dag$ and those of $\mathcal{L}$. For example, an eigenfunction of $\mathcal{W}^\dag$ with zero eigenvalue is a stationary probability distribution, which may include Dirac delta functions, in which case the construction of this Section (with a basis of monomial functions) is not appropriate}.

\subsection{Matrix representations of ${\cal L}$ and ${\cal A}$}\label{sec:liouville}

Symmetry conditions like Eq.~\eqref{eq:weakU_QME} and \eqref{eq:weakR} constrain the representations of operators and superoperators like ${\cal L},J_k,H$ \cite{Albert2014,Mozrzymas2017,Macieszczak2021}. For example  Eq.~\eqref{eq:weakR} means that $H$ commutes with $U$, which implies a block diagonal structure for $H$ in the symmetry-adapted basis \cite{Macieszczak2021}.  Similarly, Eq.~\eqref{eq:weakU_QME} implies a particular block diagonal structure for the Liouville (matrix) representation of the superoperator ${\cal L}$ \cite{Albert2014}, see also below.

This section explores the implications of Symmetry Condition~II in Eq.~\eqref{eq:symm_cond_II} for matrix representations of superoperators such as $\mathcal{L}$ and $\cgo_\alpha$.
% the SJED jump superoperators $\cgo_\alpha$.
In Appendix \ref{app:wave+trans} we apply these results to `wave' operators, constructed from Fourier-transformations along the permutation of SJED composite action operators.

\subsubsection{Liouville representation}

 In the Liouville representation [with basis $\{|\psi_i\rangle\}_{i=1}^{d_{\rm s}}$], operators and density matrices are vectorised by stacking their rows.
 %such that $|\rho\rangle\!\rangle \equiv \sum_{ij}\langle \psi_i|\rho|\psi_j\rangle (|\psi_i\rangle\otimes|\psi_j\rangle)$ is a vector of size  $d_{\rm s}^2$.   
 Then a generic superoperator $\Phi$ becomes a matrix $\Lambda[\Phi]$ of size $d_{\rm s}^2 \times d_{\rm s}^2$ with elements
\begin{equation}
\label{eq:Lambda}
    (\Lambda[\Phi])_{mn,kl} = \big\langle\psi_m\big|\Phi\big(|\psi_k\rangle\langle\psi_l|\big)\big|\psi_n\big\rangle.
\end{equation}
The Liouville representation of ${\cal U}$ is 
\begin{equation}
\Lambda[{\cal U}]=U \otimes U^* ,
\label{eq:liouv-UU*}
\end{equation}
so for any symmetric QME one has by Eq.~\eqref{eq:weakU_QME} 
\begin{equation}
(U \otimes U^*) \, \Lambda[{\cal L}] \, (U \otimes U^*)^\dag = \Lambda[{\cal L}]
\label{eq:symm-liouv}
\end{equation}
which is a symmetry of $\Lambda$.
In the symmetry-adapted basis, the eigenspaces of ${\cal U}$ [recall Eq.~\eqref{eq:UU-eigen}] mean that Liouville matrices are naturally organised into blocks.
Then Eq.~\eqref{eq:symm-liouv} means that $\Lambda[{\cal L}]$ only has support on diagonal blocks.  That is, $(\Lambda[{\cal L}])_{mn,kl}\neq 0$ only if $|\psi_m\rangle\langle\psi_n|$ and $|\psi_k\rangle\langle\psi_l|$ share the same eigenvalue of ${\cal U}$.

\subsubsection{Choi matrices}\label{sec:choi}

In addition to their Liouville (matrix) representations, the superoperators ${\cal L}$ and ${\cal A}_\alpha$ also have corresponding Choi matrices.  We explain in this section that these Choi matrices have the same structure as the Liouville matrices discussed previously.
We also comment on the ranks of these Choi matrices.

The Choi matrix for superoperator $\Phi$ can be expressed 
in terms of the matrix elements of the Liouville representation in Eq.~\eqref{eq:Lambda} as~\cite{Wolf2012,Watrous_2018}
\begin{equation}\label{eq:choi}
    C[\Phi] = %\frac{1}{\dim(\mathcal{H})}
    \sum_{mnkl} (\Lambda[\Phi])_{mn,kl} |\psi_m\rangle\langle\psi_n|\otimes|\psi_k\rangle\langle\psi_l|.
\end{equation}
We show in Appendix~\ref{app:Choi} that
\begin{equation}\label{eq:choi_symm}
C[\mathcal{U}\Phi\mathcal{U}^\dag]  =  (U\otimes U^*) C[\Phi] (U\otimes U^*)^\dag \; ,
\end{equation}
which is analogous to Eq.~\eqref{eq:liouv-UU*} in  the Liouville representation.
It follows that Eq.~\eqref{eq:symm-liouv} holds with $\Lambda[{\cal L}]$ replaced by $C[{\cal L}]$ (under the same conditions).

Given some $\cgo_\alpha$, the rank of $C[\cgo_\alpha]$ determines how it can be purified into a sum of jump operators (see Sec.~\ref{sec:SJED_reps}, below).
From their definition, there are two types of SJED that can occur \cite{Generators}.  First,
the SJED may include jumps which are all proportional to each other.  In this case the rank of $C[\cgo_\alpha]$ is $1$.  Alternatively, the SJED may be of ``reset-type'' in which case
each jump in the SJED must reset the system to the same state, that is
\begin{equation} 
J_k = |\rs_\alpha\rangle \langle \xi_k| \sqrt{\gamma_k} \qquad \forall \, k\in S_\alpha
\end{equation} 
The SJED superoperator is then given by
\begin{equation}\label{eq:reset_SJED}
    \mathcal{A}_{\alpha}(\psi) = |\rs_\alpha\rangle\langle\rs_\alpha|\, \text{Tr}\left[\Gamma_\alpha\psi\right] \;\; \text{with} \;\; \Gamma_\alpha = \sum_{k\in S_\alpha} \gamma_k|\xi_k\rangle\langle\xi_k|.
\end{equation}
The corresponding Choi matrix is
\begin{equation}
    C[\cgo_\alpha] = |\rs_\alpha\rangle\langle\rs_\alpha| \otimes \Gamma_\alpha^*
\end{equation}
The rank of this matrix is the rank of $\Gamma_\alpha$, which is the minimal number of jumps required to represent $\cgo_\alpha$.
(This may be smaller than $|\SJED|$ if the $|\xi_i\rangle$ are linearly dependent.)  Hence, for reset-type SJED with rank$(\Gamma_\alpha)>1$, the action of $\cgo_\alpha$ is not that of any single jump operator.

\subsection{Representations of SJED under Symmetry Condition II}
\label{sec:SJED_reps}

Symmetry Condition~II constrains the SJEDs (and Hamiltonian), but not the individual jumps themselves: this is the freedom of representation of positive superoperators that has been discussed extensively above.
Each choice of jumps for a particular SJED %is called a 
gives a different representation of the QME, all related by isometric mixing.
We suppose in this section that Symmetry Condition~II holds and we discuss the choice of representations.

As noted in Sec.~\ref{sec:choi},  there are two types of SJED that can occur \cite{Generators}.
For SJED of non-reset-type, there is an operator $J^{(\alpha)}$, of rank strictly greater than one, such that $J_k=\lambda_k J^{(\alpha)}$ for all $k\in S_{\alpha}$ (with $\lambda_k=\mathbb{C}$).
% \footnote{\cb{The normalisation can be chosen such that $\Tr[{J^{(\alpha)}}^\dag J^{(\alpha)}]=1$.} \emph{\cp{Not necessarily needed here so could be removed.}}}
Different representations of the SJED can be obtained by adjusting the constants $\lambda_k$.
% The rank of $J^{(\alpha)}$ must be strictly greater than one, otherwise the SJED is of reset type.
Symmetry Condition~II implies that 
\begin{subequations}
\begin{equation}\label{eq:canon_prop_perm}
    \mathcal{U}(J^{(\alpha)}) = e^{i\theta_\alpha}J^{(\pic(\alpha))} ,
\end{equation}
% \footnote{RLJ: note I am intentionally keeping nested round brackets here}
while for reset jumps the reset states are permuted
\begin{equation}\label{eq:cond_II_reset}
    \mathcal{U}(|\rs_\alpha\rangle\langle\rs_\alpha|) = |\rs_{\pic(\alpha)}\rangle\langle\rs_{\pic(\alpha)}| , \quad \mathcal{U}(\Gamma_\alpha) = \Gamma_{\pic(\alpha)}.
\end{equation}
\end{subequations}
Hence, SJEDs of a particular type must permute to an SJED of the same type, as expected.

A \emph{canonical representation} is a minimal representation where furthermore the jump operators satisfy $\Tr[J_j^\dag J_k]=\zeta_j \delta_{jk}$ {for constants $\zeta_1,\dots,\zeta_d$}~\cite{Hall2014}.
We can also consider canonical representations of individual SJED, i.e. representations such that jumps within the same SJED are orthogonal with $\Tr[J_j^\dag J_k]=\zeta_j \delta_{jk}$ for $j,k\in \SJED$. We now discuss how these are constructed and the resulting action of the symmetry on the jumps.

As outlined in Ref.~\cite{Generators}, a canonical representation for an SJED made of proportional jumps is given by the single jump $(\sum_{k\in S_\alpha}|\lambda_k|^2)^{\frac{1}{2}}J^{(\alpha)}$. If canonical representations are taken for all SJEDs of this type, these jumps are permuted by the symmetry as in Eq.~\eqref{eq:canon_prop_perm}.
For reset SJEDs, diagonalising $\Gamma_\alpha$ gives a canonical representation of $\cgo_\alpha$ in terms of new jump operators $\hat{J}_j$.
Note $|\rs_{\pic(\alpha)}\rangle=U|\rs_\alpha\rangle$ (up to a phase which is irrelevant in the following) and $\Gamma_{\pic(\alpha)}=U\Gamma_\alpha U^\dag$ \cite{Generators}. Hence, $\Gamma_{\pic(\alpha)}$ and $\Gamma_\alpha$ have the same spectrum and their eigenvectors can be related as $|\hat{\xi}^{(\pic(\alpha))}_{i}\rangle = U|\hat{\xi}^{(\alpha)}_{i}\rangle$.  In other words we have for the representation where all SJED are taken to be the canonical choice, that the $i$th jump operator in SJED $\alpha$, denoted $\hat{J}^{(\alpha)}_{i}$, transforms as
\begin{align}\label{eq:reset_SJED_transform}
{\cal U}(\hat{J}^{(\alpha)}_{i}) & = U|\rs_\alpha\rangle\langle\hat{\xi}_{i}^{(\alpha)}|U^\dag\sqrt{\gamma_i} 
\\ \nonumber &= 
\big|\rs_{\pic(\alpha)}\big\rangle\big\langle\hat{\xi}^{(\pic(\alpha))}_{i}\big|\sqrt{\gamma_i}
\\ \nonumber &= 
\hat{J}^{(\pic(\alpha))}_{i}
\end{align}
up to an arbitrary phase. That is, individual jump operators in this representation are permuted by the symmetry operation.\footnote{As usual, eigenvectors of $\Gamma_\alpha$ are only defined up to phase factors, and degenerate eigenvalues also lead to some flexibility in the  eigenvectors.  These choices must be made consistently between different SJEDs in order to have $|\rs_{\pic(\alpha)}\rangle=U|\rs_\alpha\rangle$ and $|\hat{\xi}_{i}^{(\pic(\alpha))}\rangle = U|\hat{\xi}_{i}^{(\alpha)}\rangle$, but this is always possible.}

The considerations above show that representations (of the QME) can be constructed from canonical representations of SJEDs, to be consistent with Symmetry Condition~III.  Hence the dynamics of the dephased cMPO for such representations is symmetric (see Sec.~\ref{sec:dephased_symm}).  Moreover, we recall from Sec.~\ref{sec:measure-basis} that isometric mixing of jump operators can be achieved by changes of measurement basis in cMPOs.  The same reasoning shows that different representations \emph{of SJEDs} can be accessed by isometric mixing of their constituent jump operators. Such transformations, discussed in Sec.~\ref{sec:cond-partial}, mix quanta within SJEDs but do not change the relevant composite action superoperators $\cgo_\alpha$.  All together, this means that the QME representation $H,\hat{J}_1,\dots,\hat{J}_{d_0}$ constructed from canonical SJED representations [with $d_0=\sum_\alpha {\rm rank}(\Gamma_\alpha)$] can be related to any representation with the same $\cgo_\alpha$ superoperators via isometric mixing of quanta within SJEDs.  

In other words, any representation satisfying Symmetry Condition II can always be related to a representation satisfying Symmetry Condition III by isometric mixing of quanta within SJEDs.  Sec.~\ref{sec:cond-partial} uses an additional result, which is that a representation with $d$ jump operators can be related in this way to one with $\tilde d \geq d$ jump operators that satisfies Symmetry Condition III.  This is achieved using canonical SJED representations as an intermediate step.

In Appendix \ref{app:relation_weakly_symm}, we show how to construct a weakly symmetric representation, given any representation which satisfies Symmetry Condition III. However, the SJED composite action operators $\cgo_\alpha$ for this new representation will differ in general from those of the original one.

\section{Outlook}\label{sec:outlook}

The presence of symmetries in open quantum systems is crucial for a foundational understanding of these systems, as well as for practical implementation of quantum technologies.
This work provides a comprehensive taxonomy of weak unitary symmetries for descriptions beyond the average dynamics generated by the quantum master equation. This gives a theoretical framework which can be used for both practical application and gives insight into the interplay between symmetry, quantum measurement and environmental interactions.

This topic is also of great importance in the resource theory of asymmetry~\cite{Vaccaro2008,Marvian2013,Marvian2014a} also known as the resource theory for quantum reference frames~\cite{Gour2008,Gour2009}.
The existence of a symmetric joint discrete dynamics of system and its environment is known as  Stinespring's dilation theorem for covariant channels~\cite{Keyl1999,Marvian2012}. Our work extends Stinespring dilation theorem to continuous Markovian dynamics with weak symmetries and the context of continuous {counting} measurements. {This} construction in generalizes to weak symmetries {groups},  so that the joint dynamics is symmetric with respect to the corresponding group of separable symmetries. {This is turn provides the characterisation of local-in-time unitary symmetries of cMPSs.} %[in the rotating frame as in Eq.~\eqref{eq:UUE-dH}].

A summary of how these symmetries manifest in the different descriptions was given in Sec.~\ref{sec:discussion1}.
Several examples were then explored in which the symmetries of quantum trajectories at various levels was discussed.
We {also} gave a series of implications resulting from {weak symmetries in quantum master dynamics and unravelled quantum dynamics}, including finding eigenfunctions of the unravelled generator and how particular operators are supported in the symmetry-adapted basis. 

With the weak unitary symmetries considered so far, the techniques of this work should be extended to consider non-unitary symmetries \cite{roberts21,lieu_quad20,Sa2023} ({enabling} symmetry protected topological order \cite{Groot_2022,Paszko_2023}).
Future work should also explore also the notation of strong symmetries in open quantum dynamics~\cite{Munoz19,Tindall_2020,Zhang_2020}, which leads to multiple stationary states and thus breaking of ergodicity in the quantum master dynamics that will be inherited by the unravelled quantum dynamics. Among others, this could shed further light on spontaneous symmetry breaking in open quantum systems, such as strong-to-weak symmetry breaking \cite{Sala2024,Gu2024,Guo2025,Lessa2025}, at the level of quantum trajectories, beyond the average dynamics. This has applications in quantum computing where errors which preserve strong symmetries are dynamically restored by the system dynamics \cite{Lieu_breaking20}.

\begin{acknowledgments}
We thank Juan Garrahan, M\u{a}d\u{a}lin Gu\c{t}\u{a} and Berislav Bu\v{c}a for helpful discussions. This work was supported by the Engineering and Physical Sciences Research Council [grant number EP/T517847/1].
\end{acknowledgments}

\begin{appendix}

\section{Non-uniqueness of the matrix $\mathbf{U}$}
\label{app:boldU}

\subsection{General case}
\label{app:boldU-gen}

If the (traceless) jump operators $\{J_k'\}$ in Eq.~\eqref{eq:traceless_rep} are linearly dependent then the matrix $\mathbf{U}$ in Eq.~\eqref{eq:symm_cond_I} is not unique.  To see this, suppose that
\begin{equation}
\sum_j b_j^* J_j' = 0
\label{equ:J-LI}
\end{equation}
and ${\cal U}(J_k') = \sum_j \mathbf{U}_{kj} J_j'$ as required by Eq.~\eqref{eq:symm_cond_I}.   Then  Eq.~\eqref{eq:symm_cond_I} still holds on replacing $\mathbf{U}$ by $\mathbf{X}$ with elements
\begin{equation}
{\mathbf{X}}_{kj} = \mathbf{U}_{kj} + c_k b_j^* 
\label{eq:Uca}
\end{equation}
for arbitrary complex constants $\{ c_k \}$.  We additionally require $\mathbf{X}$ to be unitary: this constrains $c_k = z \sum_j \textbf{U}_{kj} b_j$ where the complex constant $z$ obeys $z + z^* + |z|^2 \sum_j |b_j|^2=0$.\footnote{To see this, note that unitarity implies  $0=\textbf{v}^\dag ( \textbf{X} \textbf{X}^\dag - \textbf{U} \textbf{U}^\dag ) \textbf{v}$ for all $\textbf{v}$,
which means $0=\textbf{v}^\dag ( \mathbf{U}\mathbf{bc}^\dag + \mathbf{cb}^\dag \mathbf{U}^\dag \ +  \mathbf{cb}^\dag\mathbf{bc}^\dag ) \textbf{v}$.  Resolving $\textbf{v}$ parallel and perpendicular to $\mathbf{U}\mathbf{b}$ shows that $\mathbf{c}$ is parallel to $ \mathbf{U}\mathbf{b}$ with proportionality coefficient as above.}
Solutions for $z$ lie on a circle in the complex plane.
An example of this non-uniqueness is shown at the end of the single qubit example in Sec.~\ref{sec:example_qubit}.

Writing Eq.~\eqref{equ:J-LI} as $\vec{b}^\dag \vec{J}'=0$ with $\vec{b}=(b_1,\dots,b_d)$ and $\vec{J}'=(J_1,\dots,J_d)$, we can generalise to situations where several linearly independent vectors $\vec{b}$ satisfy this relationship.  Specifically, suppose that
$
\mathbf{B}^\dag \vec{J}' = 0
$
where $\mathbf{B} \in \mathbb{C}^{d\times n}$ is a matrix whose columns are an orthonormal basis for all possible vectors $\vec{b}$.   Then Eq.~\eqref{eq:Uca} generalises to
$
\mathbf{X} = \mathbf{U} +  \mathbf{C} \mathbf{B}^\dag
$
with arbitrary $\mathbf{C} \in \mathbb{C}^{d\times n}$.

\subsection{Unitary mixing under Symmetry Condition~\IIb}

We discuss the structure of the unitary matrix $\mathbf{U}$ that appears in Eq.~\eqref{eq:not_cond_IIb}.  Under Symmetry Condition~\IIb, we show that there exist
matrices $\mathbf{C} , \mathbf{B} \in \mathbb{C}^{d\times d'}$ such that 
\begin{subequations}
\label{eq:UU-VV-II}
\beq
\mathbf{U} = \mathbf{X} + \mathbf{C} \mathbf{B}^\dag
\label{eq:UCB-a}
\eeq
is unitary; also,
\beq
\sum_{k=1}^d \mathbf{B}_{kj}^* J_k = 0
\label{eq:B-works}
\eeq
and for every column $j$ of $\mathbf{B}$ there exists some SJED $\alpha$ such that 
\beq
\mathbf{B}_{kj} = 0 \qquad k \notin S_\alpha \; ,
\label{eq:B-block}
\eeq
\end{subequations}
If Eq.~\eqref{eq:B-block} holds then we say that ``the $j$th column of $\mathbf{B}$ is supported on SJED $\alpha$''.
The matrix $\mathbf{B}$ has $d'$ columns; the value of $d'$ will be discussed below.  (It can be that $d'=0$ in which case $\mathbf{U} = \mathbf{X}$ is unitary.)
The unitary $\mathbf{U}$ satisfying Eq.~\eqref{eq:UU-VV-II} is useful because it implies Eq.~\eqref{eq:UU-partial-lemma} of the main text, see also Eq.~\eqref{eq:UU-not-lemma} below.

To show Eq.~\eqref{eq:UU-VV-II}, recall from (\ref{eq:UIIR},\ref{eq:not_cond_IIb})
that we are guaranteed that some unitary $\mathbf{U}$ exists, with $\sum_k (\mathbf{U}-\mathbf{X})_{jk} J_k=0$.  Repeating the argument of Appendix~\ref{app:boldU-gen} shows that $\mathbf{U}$ must be of the form
\beq
\mathbf{U} = \mathbf{X} + \mathbf{C} \mathbf{B}_{\rm all}^\dag
\label{eq:UCB}
\eeq
where the columns of $ \mathbf{B}_{\rm all}$ are an orthonormal basis for \emph{all} vectors $\vec{b}=(b_1,\dots,b_d)$ such that 
\beq
\sum_{j=1}^d b_j^* J_k =0
\eeq
These columns describe the linear dependencies of the jump operators $\{J_k\}_{k=1}^d$.  

The content of Eq.~\eqref{eq:UU-VV-II} is that the matrix $\mathbf{B}_{\rm all}$ in Eq.~\eqref{eq:UCB} may be replaced by a matrix $\mathbf{B}$ with the property Eq.~\eqref{eq:B-block}.
To build this matrix, define $\mathbf{B}^{(\alpha)}$ as a matrix whose columns are an orthonormal basis for vectors $\vec{b}$ satisfying
\beq
\sum_{k\in S_\alpha} b_k^* J_k =0, \quad \text{and} \quad b_k = 0 \;\; \text{for} \; k \notin S_\alpha
\label{eq:b-alpha}
\eeq
These columns are supported on SJED $\alpha$ and describe the linear dependencies that relate jump operators in that set.
%
%Now construct the matrices $\mathbf{B}^{(\alpha)}$ for all SJEDs $\alpha$.  
The matrix $\mathbf{B}$ is defined by concatenating the columns of the matrices $\mathbf{B}^{(\alpha)}$ for $\alpha=1,\dots,\dC$.  This $\mathbf{B}$ is consistent with (\ref{eq:B-works},\ref{eq:B-block}).  The total number of such columns is the value of $d'$. 

It remains to show that one can choose $\mathbf{C}$ such that $\mathbf{U}$ in Eq.~\eqref{eq:UCB-a} is unitary.  This is demonstrated in Appendix~\ref{app:unit-ucb}.  The remainder of this Section uses the above construction to derive Eq.~\eqref{eq:UU-partial-lemma} of the main text, which is a useful property of $\mathbf{U}$.
We first note that
\beq
 \sum_{m\in S_\alpha} B_{mn}^* J_m = 0  \quad \forall \, n
 \label{eq:B-lemma}
\eeq
To show this: Either the $n$th column of $\mathbf{B}$ is supported on SJED $\alpha$ in which case it is a column of $\mathbf{B}^{(\alpha)}$ and the sum is zero by Eq.~\eqref{eq:b-alpha}; or the $n$th column of $\mathbf{B}$ is supported on some SJED $\beta\neq\alpha$ in which case $B_{mn}=0$ for $m\in S_\alpha$.
Now use (\ref{eq:UCB-a}) to write
\beq
        \sum_{m\in S_\alpha} \mathbf{U}_{jm} J_m = \sum_{m\in S_\alpha} {\bf X}_{jm} J_m 
    + \sum_{p=1}^d \mathbf{C}_{jp} \sum_{m\in S_\alpha}  \mathbf{B}^*_{mp} J_m
\eeq
The second term on right hand side is zero by Eq.~\eqref{eq:B-lemma}.  Then using 
%(\ref{eq:Va0},\ref{eq:Va1},\ref{eq:UIIR}) 
(\ref{eq:X},\ref{eq:Va1}) 
one obtains
\beq
    \sum_{m \in S_{\alpha}} \mathbf{U}_{jm} J_m = 
    \begin{cases} {\cal U}(J_j) , \quad &  j \in S_{\pi^{-1}(\alpha)}
    \\
    0 ,  &  \text{otherwise} \; .
    \end{cases}
    \label{eq:UU-not-lemma}
\eeq
Multiplying both sides by ${\bf U}^*_{jk}$, summing over $k=1,\dots,d$ and using unitarity of $\mathbf{U}$, one obtains Eq.~\eqref{eq:UU-partial-lemma} of the main text, which is the most important implication of this Section.

\subsection{Unitarity in \eqref{eq:UCB-a}}
\label{app:unit-ucb}

We show that  $\mathbf{C}$ can be chosen in Eq.~\eqref{eq:UCB-a}  such that $\mathbf{U}$ is unitary.  The idea is to introduce a representation of an auxiliary QME in which the only linear dependencies of the jump operators are those encoded by $\mathbf{B}$.  
The auxiliary QME has a symmetry under which its  jump operators transform according to a unitary matrix of the form Eq.~\eqref{eq:UCB-a}, so there must exist a matrix of this form,.

We first introduce a new representation $H,\tilde{J}_1,\dots,\tilde{J}_D$ of the QME.  The SJEDs of this representation are denoted $\tilde S_\alpha$.  It is chosen such that all SJED action operators ${\cal A}_\alpha$ have canonical representations, that is:
\beq
{\cal A}_\alpha(\cdot) = \sum_{k \in \tilde S_\alpha} \tilde{J}_k (\cdot) \tilde{J}_k^\dag
\label{eq:A-tildeJ}
\eeq
with $\Tr[\tilde J_k^\dag \tilde J_j]=0$ if the two indices $j\neq k$ are in the same SJED.   The representation has the same ${\cal A}_\alpha$ so it obeys Symmetry Condition~II.  Hence by
Symmetry Condition~\IIb, the symmetry acts in this representation as 
\beq \mathcal U(\tilde J_k) = \sum_j \tilde{\mathbf{X}}_{kj} \tilde J_{j}
\label{eq:V-min}
\eeq 
Moreover, $\tilde{\mathbf{X}}$ is unitary, because $|\tilde S_\alpha| = |\tilde S_{\pic(\alpha)}|$ so both conditions (\ref{eq:Va2},\ref{eq:Va3}) hold in Symmetry Condition \IIb.
Since Eq.~\eqref{eq:A-tildeJ} is a representation of ${\cal A}_\alpha$ we have %[similar to %\eqref{eq:Va0}
%\eqref{eq:X}]
that
\beq
J_k = \sum_{j\in \tilde{S}_\alpha}  \mathbf{V}_{kj}^{(\alpha)} \tilde{J_j}
\label{eq:X-isom}
\eeq
where $\mathbf{V}^{(\alpha)} $ is isometric in the sense that $\sum_{k\in S_\alpha} (\mathbf{V}_{kj}^{(\alpha)})^* \mathbf{V}^{(\alpha)}_{kj'}=\delta_{j,j'}$ for $j\in \tilde S_\alpha$. Note that all linear dependencies of jump operators in set $\alpha$ are encoded in $\mathbf{V}^{(\alpha)}$.

We now introduce the auxiliary QME, which is defined on a separate Hilbert space, as follows.  Let $\tilde K_1,\dots,\tilde K_D$ be a set of orthogonal jump operators so $\Tr[\tilde K_k^\dag \tilde K_j]=0$ for $j\neq k$.  
(Note, this is different from the original QME, where $\tilde J_k$ operators from different SJEDs may be linearly dependent.)  
Define a unitary operation ${\cal U}'$ on this Hilbert space such that ${\cal U}'(\tilde K_j) = \sum_k \tilde{\mathbf{X}}_{jk} K_k$ with the same (unitary) matrix $\tilde{\mathbf{X}}$ that appears in Eq.~\eqref{eq:V-min}.
Also define 
\beq
K_k = \sum_{j\in \tilde{S}_\alpha}  \mathbf{V}_{kj}^{(\alpha)} \tilde{K_j}
\eeq
where $\tilde S_\alpha$ and $\mathbf{V}_{kj}^{(\alpha)}$ are the same objects that appear in Eq.~\eqref{eq:X-isom}.  
This implies that the $K_k$ operators share the linear dependencies encoded by the $\mathbf{V}^{(\alpha)}$, which are exactly the linear dependencies encoded by the columns of $\mathbf{B}$.  Moreover, orthogonality of the underlying $\tilde K_k$ operators means that these are the only linear dependencies of the $K_k$ operators.

With all these definitions we have that
\beq
{\cal U}'(K_k) = \sum_{j=1}^d \mathbf{X}_{kj} K_j
\eeq
where $\mathbf{X}$ is the matrix that appears in Eq.~\eqref{eq:UCB-a}.  The auxiliary QME has the weak symmetry $\cal U'$ so there must be a unitary $\mathbf{U}'$ such that ${\cal U}'(K_k) = \sum_{j=1}^d \mathbf{U}'_{kj} K_j$, and the linear dependences of the $K_j$ operators means that
\beq
\mathbf{U}' = \mathbf{X} + \mathbf{CB}^\dag
\eeq
for some $\mathbf{C}$.  Since the same $\mathbf{X},\mathbf{B}$ also appear in Eq.~\eqref{eq:UCB-a}, one may take $\mathbf{U}=\mathbf{U}'$ which is unitary, as required.

\section{Action of symmetry on unravelled generators}\label{app:UWU}

We derive the action of the symmetry operator $\Upsilon$ on the generator $\mathcal{W}^\dag$, showing that it yields a new generator $\Upsilon\mathcal{W}^\dag\Upsilon^\dag$ in which the Hamiltonian and jumps are transformed by Eq.~\eqref{eq:UJH}. 
Applying $\Upsilon$ to Eq.~\eqref{eq:generator_trajectories} and using the definition of $\nabla$, we obtain
\begin{multline}\label{eq:Up_applied_W}
    [\Upsilon\mathcal{W}^\dag\Upsilon^\dag (P)](\psi) = -\Upsilon \sum_{jk}\frac{\partial}{\partial\psi_{jk}} \left[P(\mathcal{U}\psi)\mathcal{B}(\psi)_{jk} \right] 
    \\
    + \int d\psi' P(\mathcal{U}\psi')W(\psi',\mathcal{U}^\dag\psi)-P(\psi)W(\mathcal{U}^\dag\psi,\psi')
\end{multline}
where  $\psi_{jk}$ is an element of the density matrix in some (arbitrary) basis and similarly $\mathcal{B}(\psi)_{jk}$ is an element of the matrix $\mathcal{B}(\psi)$.
Writing out the second term on the right hand side and changing the integration variable from $\psi'$ to ${\cal U}(\psi')$ yields
\begin{multline}
\label{eq:B-jumps-ugly}
    \sum_j \int d\psi' P(\psi') \delta\!\left(\psi - \frac{\mathcal{U}{\cal J}_j\mathcal{U}^\dag(\psi')}{\Tr[\mathcal{U}{\cal J}_j\mathcal{U}^\dag(\psi')]}\right)\Tr[\mathcal{U}{\cal J}_j\mathcal{U}^\dag(\psi')]
    \\ - P(\psi)\delta\left(\psi'-\frac{\mathcal{U}{\cal J}_j\mathcal{U}^\dag(\psi)}{\Tr[\mathcal{U}{\cal J}_j\mathcal{U}^\dag(\psi)]}\right)\Tr[\mathcal{U}{\cal J}_j\mathcal{U}^\dag(\psi)]
\end{multline}
where the operator $\mathcal{U}{\cal J}_j\mathcal{U}^\dag(\cdot)$ is given by Eq.~\eqref{eq:superJ} with $J_j,J_j^\dag$ replaced by ${\cal U}(J_j),{\cal U}(J_j^\dag)$.
Hence Eq.~\eqref{eq:B-jumps-ugly}
coincides with the second term on right hand side of  Eq.~\eqref{eq:generator_trajectories}, except that 
$J_k$ has been replaced by $\mathcal{U}(J_k)$, as required.

To deal with the other term in Eq.~\eqref{eq:Up_applied_W}, we 
note the operator relation
\beq
\Upsilon \frac{\partial}{\partial \psi_{jk}} = \sum_{lm} (\mathcal{U})_{lmjk}\frac{\partial}{\partial \psi_{lm}} \Upsilon 
\eeq
with $(\mathcal{U})_{lmjk} = \langle l|U|j\rangle\langle k|U^\dag|m\rangle $
Using this formula, the first (``advection'') term on right hand side of Eq.~\eqref{eq:Up_applied_W} becomes
\begin{align}
    -\sum_{jklm}(\mathcal{U})_{lmjk}\frac{\partial}{\partial\psi_{lm}}P(\psi)\mathcal{B}(\mathcal{U}^\dag\psi)_{jk}
    \label{eq:appB-intermediate}
\end{align}
Now observe that $\sum_{jk}\mathcal{B}(\mathcal{U}^\dag\psi)_{jk} (\mathcal{U})_{lmjk}=(\mathcal{U}\mathcal{B}[\mathcal{U}^\dag\psi])_{lm}$ so Eq.~\eqref{eq:appB-intermediate} reduces to
\begin{equation}
    -\sum_{jk} \frac{\partial}{\psi_{jk}} P(\psi)(\mathcal{U}\mathcal{B}[\mathcal{U}^\dag\psi])_{jk};
\end{equation}
Comparing with Eq.~\eqref{eq:generator_trajectories}, one sees that the effect of $\Upsilon$ on the advection term in  is to replace ${\cal B}[\psi]$ by $\mathcal{U}\mathcal{B}[\mathcal{U}^\dag\psi]$.
Also note from Eq.~\eqref{eq:B} that  %=\mathcal{B}[\psi]$.
\begin{multline}
    \mathcal{U}\mathcal{B}(\mathcal{U}^\dag\psi) = -i\mathcal{U}(\He)\psi+i\psi\mathcal{U}(\He)
    \\ -\psi\Tr[-i\mathcal{U}(\He)\psi+i\psi\mathcal{U}(\He)]
\end{multline}
Comparing this with Eq.~\eqref{eq:B} one sees that the effect of the transformation is to replace $\He$ with ${\cal U}(\He)$.

Finally, noting from Eq.~\eqref{eq:H_eff} that ${\cal U}(H) = {\cal U}(\He) - \frac{i}{2}\sum_j {\cal U}(J_k^\dag) {\cal U}(J_k)$ and combining the results above, one sees that indeed $\Upsilon\mathcal{W}^\dag\Upsilon^\dag$ is the generator for the unravelled {quantum} dynamics of the representation obtained by Eq.~\eqref{eq:UJH}.

\bigskip

\section{Joint dynamics (cMPS)}

\subsection{Transformation to rotating frame}
\label{app:rotate}

We derive Eq.~\eqref{eq:dH'} of the main text, working in the
notation of Sec.~\ref{sec:stoch-ham}.  
Using that environmental operators at different times commute with each other, Eq.~\eqref{eq:dH-dH'} implies
\begin{equation}
    \mathcal{T}e^{-i\int_{t'}^t dH'_\tau} =  \mathcal{T}  \left( \mathbb{1}\otimes  e^{-i\int_{t'}^tdQ_\tau} \right)e^{-i\int_{t'}^t dH_\tau}
\end{equation}
Using the Baker-Cambell-Hausdorff (BCH) formula, we obtain
\begin{multline}
    \mathcal{T}e^{-i\int_{t'}^t dH'_\tau} = \mathcal{T}\exp\Big\{-i\int_{t'}^t (\mathbb{1}\otimes dQ_\tau)  
     + dH_\tau \\ - \frac12[ \mathbb{1}\otimes dQ_\tau, dH_\tau ] \Big\}
\end{multline}
Also observe [by the Ito table, Eq.~\eqref{eq:Ito}] that
\begin{equation}
    [\mathbb{1}\otimes dQ_\tau,dH_\tau] = \frac{1}{\dS}\sum_j\left[\Tr(J_j)J_j^\dag-\Tr(J_j^\dag)J_j\right]\otimes\IE d\tau \,.
\end{equation}
Combining these results we obtain  Eq.~\eqref{eq:dH'}.

\subsection{Symmetries of joint dynamics with field in coherent state}\label{app:non-vac}

It was assumed in Sec.~\ref{sec:cMPS} that the initial state of the environment was the vacuum.  Here we consider the case where the initial state is a coherent state, but the system evolution still obeys the QME Eq.~\eqref{eq:QME}.  We show that the resulting cMPS still features a weak (time-dependent) symmetry.

Suppose that the field is initially in a coherent state:
\begin{equation}
    |\coh\rangle = \disp_{\infty,0}(\bm a) |\vac\rangle , %(\{a_j\}_j) |\vac\rangle ,
\end{equation} 
where (assuming $t\geq t')$
\beq
\disp_{t,t'}(\bm a) = %\mathcal{T}
\exp\left(\int_{t'}^t\sum_j(a_j dB_{j,\tau}^\dag - a_j^* dB_{j,\tau})\right)
\eeq
is a coherent displacement of the field, and $a_j = ({\bm a})_j$ is the displacement of the $j^{th}$ emission mode in the phase space.
(This $\disp_{t,t'}(\bm{a})$ is used only in this appendix: one recovers $D_t$ of the main text as $\disp_{t,0}(-\bm T / \dS)$ with $({\bm T})_j = \Tr[J_j]$.)

To ensure that the system evolves according to the QME Eq.~\eqref{eq:QME}, the cMPS evolves as
\begin{equation}\label{eq:cMPS_coh}
    |\Psi_t^\text{coh}\rangle=\mathcal{T} e^{-i\int_0^t d H_{\tau}^\text{coh}} \left( |\psi_0\rangle\otimes |\text{coh}\rangle \right) 
\end{equation}
with stochastic Hamiltonian
\begin{multline}\label{eq:dH_tau_coh}
    dH_{\tau}^\text{coh} =  \Big(H - \frac{i}{2}\sum_j(a_j^* J_j - a_j J_j^\dag)\Big)\otimes\IE d\tau 
    \\+ i \sum_j\left[(J_j-a_j \mathbb{1})\otimes dB_{j,\tau}^\dag - (J_j^\dag-a_j^*\mathbb{1})\otimes dB_{j,\tau}\right]\,.
\end{multline}

This stochastic Hamiltonian can be related to $dH_\tau$ of Eq.~\eqref{eq:dH_tau} by an argument similar to Appendix~\ref{app:rotate}:
write $D_{t,t'}(\bm a)=e^{-i\int_{t'}^t dQ_\tau(\bm a)}$ where
\begin{equation}
    dQ_\tau(\bm{a}) = \mathbb{1} \otimes i \sum_j \left(a_j dB_{j,\tau}^\dag - a_j^* dB_{j,\tau}\right)
\end{equation}
Then one finds that
\beq \label{eq:cMPS_coh3}
\mathcal{T} e^{-i\int_{t'}^t d H_{\tau}^\text{coh}} = \big(\mathbb{1}\otimes \disp_{\infty,t}(\bm a) \big)  \mathcal{T} e^{-i\int_{t'}^t d H_{\tau} } 
 \big(\mathbb{1}\otimes\disp_{\infty,t'}(\bm a)\big)^\dag ,
\eeq
which may be verified by writing the RHS as
$
{\cal T}  e^{-i\int_{t'}^t dH_\tau} e^{i\int_{t'}^t dQ_\tau(\bm a)} 
$
and using the BCH formula along with 
\beq
[ dQ_\tau(\bm{a}),dH_\tau ] = -\sum_j(a_j J_j^\dag- a_j^* J_j)\otimes\IE d\tau.
\eeq
Finally, combining Eq.~\eqref{eq:cMPS_coh3} with 
the weak symmetry Eq.~\eqref{eq:symm_dH_Ut_expanded-v2} of $dH_\tau$ implies the promised symmetry of $dH^{\rm coh}$:
 \beq
 \mathcal{T} e^{-i\int_{t'}^t dH^{\rm coh}_\tau} 
 =
 (U\otimes \UEt^{\rm coh}) \big( \mathcal{T} e^{-i\int_{t'}^t dH_\tau^{\rm coh}} \big) (U\otimes \UEtp^{\rm coh})^\dag 
 \eeq
 with 
$
 \UEt^{\rm coh} = \disp_{\infty,t}(\bm a) \, \UEt  \, \disp_{\infty,t}(\bm a)^\dag
$.
 Symmetry properties of dephased cMPOs etc can be analysed similarly.

\onecolumngrid

\vspace{18pt}\hrule \vspace{-6pt}

\section{Technical material for cMPS and cMPOs}\label{app:dephased_cmps}

This Appendix collects results for dynamics of cMPS and cMPO joint descriptions of  system and environment.

\subsection{$\UEt=\UE$ under Symmetry Condition~II}\label{app:UEt=UE}

The following result is useful throughout this Appendix.  Note from Eq.~\eqref{eq:UEt} that $\UEt=\UE$ if and only if $[D_t,\UE]=0$,
which requires in turn that
\begin{equation}
    \sum_k \left[ \left(dB^\dag_{k,\tau}\Tr(J_k)-dB_{k,\tau}\Tr(J_k^\dag)\right) , \UE \right]  = 0 \,.
    \label{eq:UE-comm}
\end{equation}
This reduces to
\begin{equation}
    \sum_k \superUE(dB^\dag_{k,\tau})  \Tr(J_k) = \sum_k dB^\dag_{k,\tau} \Tr(J_k)
\end{equation}
where $\superUE(\cdot) = \UE(\cdot)\UE^\dag$.  Using Eq.~\eqref{eq:UE} and comparing coefficients of $dB^\dag$, this condition is equivalent to Eq.~\eqref{eq:U-Tr}.  Symmetry Condition~II is sufficient for  Eq.~\eqref{eq:U-Tr} so it therefore implies $\UEt=\UE$.

\subsection{Isometric change of measurement basis}
\label{app:mpo-UCB}

We discuss Eq.~\eqref{eq:Yt} and its relationship to Eq.~\eqref{eq:general_rep_change}.  We consider two representations of a symmetric QME: $H,J_1,\dots,J_d$ and $\tilde H, \tilde J_1, \dots, \tilde J_{\tilde d}$.   The corresponding traceless representations are obtained via Eq.~\eqref{eq:traceless_rep} and denoted 
$H',J_1',\dots,J_d'$ and $\tilde H', \tilde J_1', \dots, \tilde J_{\tilde d}'$.
Symmetry of the QME means that the stochastic Hamiltonians must be also symmetric.  In the traceless representation, this means that
\beq
\label{eq:tilde-symm}
( U \otimes \tilde{U}_E ) d\tilde{H}'_\tau ( U \otimes\tilde{U}_E )^\dag = d\tilde{H}'_\tau
\eeq
where $\tilde{U}_E$ is the (unitary) environmental symmetry operator (not unique in general), which acts in the same way as Eq.~\eqref{eq:UE} with an associated $\tilde d \times \tilde d$ unitary matrix $\tilde{\mathbf{U}}$.
Eq.~\eqref{eq:Yt} has a similar form
\beq
\label{eq:VUV-traceless}
( U \otimes V_{E} U_{E} V_{E}^\dag ) \, d\tilde{H}'_\tau \, ( U \otimes V_{E} U_{E}^\dag V_{E}^\dag ) = d\tilde{H}'_\tau 
\eeq
The question is how to relate $ \tilde{U}_E$ to $V_{E} U_{E} V_{E}^\dag$.   In particular, we seek to construct $ \tilde{U}_E$ from $U_E$, using the change of measurement basis encoded by $V_E$.

For $\tilde d=d$ then $V_{E} U_{E} V_{E}^\dag$ is unitary so one simply takes $\tilde{U}_E=V_{E} U_{E} V_{E}^\dag$ (see main text).  Here
we consider the case $\tilde d > d$ so the transformation $V_E$ in Eq.~\eqref{eq:general_rep_change} is isometric (not unitary).  Note that the 
 conditions for Eqs.~(\ref{eq:tilde-symm},\ref{eq:VUV-traceless}) to hold simultaneously are that ${\cal U}(\tilde H)=\tilde H$ and 
\begin{align}
{\cal U}(\tilde J_j') & = \sum_{j=1}^d \tilde{\mathbf{U}}_{jk} \tilde{J}_k' 
\nonumber\\& = \sum_{j=1}^d (\mathbf{V} \mathbf{U} \mathbf{V}^\dag)_{jk} \tilde{J}_k' 
\end{align}
[recall Eqs.~(\ref{eq:VE},\ref{eq:UE}), see also Eq.~\eqref{eq:UUE-dH} and the associated discussion].  Hence we recognise the situation discussed in Appendix~\ref{app:boldU}, that ${\cal U}(\tilde J_j')$ can be expressed in terms of the jumps $\tilde J_k'$ using a non-unitary matrix (here $\mathbf{V} \mathbf{U} \mathbf{V}^\dag$) or a unitary one (here $\tilde{\mathbf{U}}$).  As explained there, these matrices can always be related as
\beq
\tilde{\mathbf{U}} = \mathbf{V} \mathbf{U} \mathbf{V}^\dag + \mathbf{C} \mathbf{B}^\dag
\label{eq:ucb-tilde}
\eeq
where the matrix $\mathbf{B}$ is constructed by considering linear dependence of the $\tilde J_j'$, and $\mathbf{C}$ is a suitably chosen matrix of coefficients.  (There should be no confusion between $\mathbf{B}$ and the quanta $dB_{k,t}$.)  
This allows construction of $\tilde{\mathbf{U}}$ which then prescribes the symmetry of $d\tilde{H}_\tau$, in terms of the symmetry operator of $dH_\tau$ and the change of measurement basis.

To express the operator $\tilde{U}_E$ directly in terms of $U_E,V_E$, use that the matrix elements of $\mathbf{B}\in \mathbb{C}^{\tilde d \times n}$ obey $\sum_{k=1}^{\tilde d} (\mathbf{B})^*_{kj} \tilde J'_k = 0$ for all $j$ [cf. Eq.~\eqref{eq:B-works}] and
consider a (linear but not isometric or unitary) environmental operator $Z_E$ which acts as 
\beq
\label{eq:ZE}
{ Z}_E( d\tilde B^\dag_{j,t} ) Z_E^{-1} = \sum_{i=1}^{\tilde d} \mathbf{Z}_{jk}^* d\tilde B^\dag_{k,t} 
%+ \sum_{k=1}^n \sum_{i=1}^{\tilde d} b^*_{jk} c_{ik}  d\tilde B^\dag_{i,t}, 
\qquad Z_E | \tilde{\rm vac}\rangle =  | \tilde{\rm vac}\rangle
\eeq
where $\mathbf{Z} = \mathbf{I} - \mathbf{B} \tilde{\mathbf{C}}^\dag$ 
and $\tilde{\mathbf{C}} \in \mathbb{C}^{\tilde d \times n}$;
%= \delta_{kj} + \sum_{m=1}^n b^*_{jm} c_{km}$ and
the matrix $\mathbf{Z}$ is assumed to be non-singular.
Then a direct calculation shows that 
\beq
\label{eq:ZE-symm}
(\mathbb{1}\otimes Z_E) \, d\tilde H_t' \, (\mathbb{1}\otimes Z_E^{-1}) = d \tilde H_t'.
\eeq  
independent of $\tilde{\mathbf{C}}$.
This invariance of the stochastic Hamiltonian is analogous to the non-uniqueness of the matrix $\mathbf{U}$ discussed in Appendix~\ref{app:boldU-gen}.
Using this with Eq.~\eqref{eq:VUV-traceless} one sees that 
\beq
( U \otimes  V_{E} U_{E} V_{E}^\dag Z_E  ) \, d\tilde{H}'_\tau \, ( U \otimes Z_E^{-1} V_{E} U_{E}^\dag V_{E}^\dag  ) = d\tilde{H}'_\tau 
\eeq
and it remains to choose $Z_E$ such that $Z_E^{-1} V_{E} U_{E}^\dag V_{E}^\dag  $ is unitary (and forms a suitable choice for $\tilde{U}_E^\dag$).  That means $V_{E} U_{E}^\dag V_{E}^\dag = Z_E\tilde{U}_E^\dag  $ which recovers  Eq.~\eqref{eq:ucb-tilde} if one identifies $\mathbf{\tilde U}\tilde{\mathbf{C}}=\mathbf{C}$.

\subsection{{Full dephasing and Symmetry Condition~III}}\label{app:full_deph}

We show that the dephased cMPO has a symmetry Eq.~\eqref{eq:symm-deph-rot} if and only if Symmetry Condition~III holds for some $\pi$ (see Sec.~\ref{sec:cond-deph}).

The proof occurs in several stages:
Sec \ref{app:full_deph_sufficient} shows that Symmetry Condition~III is sufficient for a symmetry of the form of Eq.~\eqref{eq:symm-deph-rot}.
Then Sec \ref{app:UEt-UE} establishes that $\UEt=\UE$ is necessary for such a symmetry.  Finally Sec \ref{app:full_cond_necessary} uses this result to show that Symmetry Condition~III is necessary for the symmetry Eq.~\eqref{eq:symm-deph-rot}.

\subsubsection{Symmetry condition III is sufficient for Eq.~\eqref{eq:symm-deph-rot}.} 
\label{app:full_deph_sufficient}

Assume that Symmetry Condition~III holds for some permutation $\pi$.
% with $\bm{U}$ chosen as in \eqref{eq:UIIIb}. 
Observe that
\beqOpt
\label{eq:UUE-dL-app}
    [\UUE \, \dL_t \, \UUE^\dag](\deph) = -i[({\cal U}(\He)\otimes\IE)  \deph - \deph({\cal U}(\He^\dag)\otimes\IE) ]
%    \\
    + \sum_k\left(\mathcal{U}(J_k)\otimes\superUE(dB^\dag_{k,t})\right)\deph\left(\mathcal{U}(J_k^\dag)\otimes\superUE(dB_{k,t})\right) .
\eeqOpt
Under the stated conditions, ${\cal U}(J_k) = e^{i\delta_k} J_{\pi(k)}$
and ${\cal U}(\He)=\He$; we also choose $\mathbf{U}$ in Eq.~\eqref{eq:UE} such that $\superUE(dB_{k,t}) = e^{i\delta_k} dB_{\pi(k),t}$.
Hence 
\beqOpt
\label{eq:UUE-dL-int}
    [\UUE \, \dL_t \, \UUE^\dag](\deph) = -i[(\He\otimes\IE) \deph - \deph(\He^\dag\otimes\IE) ]
%    \\
    + \sum_k\left(J_{\pi(k)}\otimes dB^\dag_{\pi(k),t}\right)\deph\left(J_{\pi(k)}^\dag\otimes dB_{\pi(k),t}\right)
\eeqOpt
Relabelling the sum, this coincides with $d\mathbb{L}_t$ in Eq.~\eqref{eq:dLf}, so Eq.~\eqref{eq:dLsymm} holds.  
From Appendix \ref{app:UEt=UE} (and using that Symmetry Condition III implies Symmetry Condition II) we have $\UEt=\UE$. % [recall Appendix \ref{app:UEt=UE}]. %\eqref{eq:UE-comm} and the associated discussion].
Hence $\mathbb{U}_t=\mathbb{U}$ so both Eqs.~\eqref{eq:dLsymm} and \eqref{eq:symm-deph-rot} hold.

\subsubsection{$\UEt=\UE$ is necessary for Eq.~\eqref{eq:symm-deph-rot}.}
\label{app:UEt-UE}

We show that Eq.~\eqref{eq:symm-deph-rot} implies $\UEt=\UE$ (and hence $\mathbb{U}_t=\mathbb{U}$).  
Taking $t'=0$ in Eq.~\eqref{eq:symm-deph-rot}, one has that
\beqOpt
\mathbb{U} \left\{ (\mathbb{1}\otimes D_t) \big[ \mathcal{T} e^{-i\int_{0}^t d\mathbb{L}_\tau}(\mathbb{U}^\dag R_{0}) \big] (\mathbb{1}\otimes D_t)^\dag \right\}  
%\\ 
= 
 (\mathbb{1}\otimes D_t) \left[ \mathcal{T} e^{-i\int_{0}^t d\mathbb{L}_\tau}(R_{0}) \right] (\mathbb{1}\otimes D_t)^\dag  
 \label{eq:symm-ugly}
\eeqOpt
must hold for all admissible $R_0$ (states of the form $\psi_0 \otimes |\rm{vac}\rangle$).
Adapting the idea of Sec.~\ref{sec:cmps_symm}, we define a Liouvillian in a rotating frame, denoted by $d\mathbb{L}_\tau'$.  That is,
\begin{equation}
\label{eq:dL'-dL}
 {\cal T} e^{-i\int_{0}^t d\mathbb{L}_\tau'}(R_{0}) 
= (\mathbb{1}\otimes D_t) \left[ \mathcal{T} e^{-i\int_{0}^t d\mathbb{L}_\tau}(R_{0}) \right] (\mathbb{1}\otimes D_t)^\dag \, .
\end{equation}
Hence we find
$
d\mathbb{L}_\tau' = d\mathbb{L}_\tau + d\mathbb{Q}_\tau
$
with 
\begin{equation}
d\mathbb{Q}_\tau(R) = \sum_k \big[ \mathbb{1}\otimes({-dB^\dag_{k,\tau} \Tr(J_k) + dB_{k,\tau} \Tr(J^\dag_k)}) , R \big] 
\end{equation}
such that $e^{\int_0^t d\mathbb{Q}_\tau}(R) = (\mathbb{1}\otimes D_t) R (\mathbb{1}\otimes D_t)^\dag$.  (Obtaining $d\mathbb{L}_\tau'$ from Eq.~\eqref{eq:dL'-dL} requires a BCH formula but $[d\mathbb{L},d\mathbb{Q}]$ is negligible so this is simple.)
Then Eq.~\eqref{eq:symm-ugly} implies
$
\mathbb{U} ( d\mathbb{L}_\tau + d\mathbb{Q}_\tau)  \mathbb{U}^\dag = ( d\mathbb{L}_\tau + d\mathbb{Q}_\tau).
$
Inspecting the definitions of $d\mathbb{L}_\tau,d\mathbb{Q}_\tau$ one sees that all terms in $d\mathbb{L}_\tau$ involve even numbers of $dB,dB^\dag$ operators while $d\mathbb{Q}_\tau$ involves  single operators.  The symmetry $\mathbb{U}$ does not change the number of these operators so it follows that
\begin{equation}
\mathbb{U} d\mathbb{L}_\tau \mathbb{U}^\dag  =  d\mathbb{L}_\tau, \qquad  \mathbb{U} d\mathbb{Q}_\tau  \mathbb{U}^\dag = d\mathbb{Q}_\tau
\label{eq:double-symm}
\end{equation}
separately.  The second of these equalities is exactly Eq.~\eqref{eq:UE-comm} so it follows that $\mathbb{U}_t=\mathbb{U}$ as required.

\subsubsection{Symmetry Condition~III is necessary  for Eq.~\eqref{eq:symm-deph-rot}}
\label{app:full_cond_necessary}

We now finish the proof that Eq.~\eqref{eq:symm-deph-rot} implies that Symmetry Condition~III holds (for some permutation $\pi$).  We have shown that Eq.~\eqref{eq:symm-deph-rot} implies Eq.~\eqref{eq:double-symm}.  Using (\ref{eq:UUE-dL-app},\ref{eq:UE}) we have
\beqOpt
 [\UUE \, \dL_t \, \UUE^\dag](\deph)    = -i[(\mathcal{U}(\He)\otimes\IE)  \deph - \deph (\mathcal{U}(\He^\dag)\otimes\IE)]
 % \\
    + \sum_k\sum_{m,n} \mathbf{U}^*_{km}\mathbf{U}_{kn} ({\cal U}({J}_k)\otimes dB_{m,t}^\dag)\deph ({\cal U}({J}_k^\dag)\otimes dB_{n,t})
    \label{eq:L-ugly}
\eeqOpt
which is equal to $d\mathbb{L}_t(R)$ by Eq.~\eqref{eq:double-symm}.
%The dynamical symmetry Eq.~\eqref{eq:dLsymm} states that $\UUE \, \dL_t \, \UUE^\dag = \dL_t$.  
Comparing terms which act trivially on the environment, we see that 
\beq 
\mathcal{U}(\He)=\He \label{equ:HeHe-deph} 
\eeq
Comparing the terms that involve non-trivial environmental operators we obtain
\beqOpt
\label{eq:dephased_symm_condition}
    \sum_k \left({\cal U}({J}_k) \otimes dB^\dag_{m,t}\right) \deph \left( {\cal U}({J}_k^\dag) \otimes dB_{n,t}\right) \mathbf{U}^*_{km}\mathbf{U}_{kn} 
%\\ 
=    \delta_{mn} (J_m\otimes dB^\dag_{m,t}) \deph (J_m^\dag \otimes dB_{n,t})
\eeqOpt
which we rearrange as
\beqOpt
    (\mathbb{1}\otimes dB^\dag_{m,t})\Big[\sum_k({\cal U}({J}_k)\otimes\IE)R({\cal U}({J}_k^\dag)\otimes\IE)\mathbf{U}^*_{km}\mathbf{U}_{kn}
    %\\ 
    -(J_m\otimes\IE)R(J_m^\dag\otimes\IE)\delta_{mn}\Big](\mathbb{1}\otimes dB_{n,t})=0
\eeqOpt
This holds for all $R$ and so
\beqOpt
    \sum_k({\cal U}({J}_k)\otimes\IE)R({\cal U}({J}_k^\dag)\otimes\IE)\mathbf{U}^*_{km}\mathbf{U}_{kn}
    %\\ 
    -(J_m\otimes\IE)R(J_m^\dag\otimes\IE)\delta_{mn}=0% \quad \forall\;R \; .
    \label{eq:dephased-cond3}
\eeqOpt
Multiplying by $\mathbf{U}^*_{ln}$, summing over $n$, and using unitarity of $\mathbf{U}$, we obtain
\beqOpt
\label{eq:U_from_dephased}
    \Big[({\cal U}(J_l)\otimes\IE)\,R\,({\cal U}({J}_l^\dag)\otimes\IE) 
    %\\
    - (J_m\otimes\IE)R(J_m^\dag\otimes\IE)\Big]\mathbf{U}^*_{lm} = 0
\eeqOpt
This must hold for all $R$ which implies that either $\mathbf{U}_{lm}=0$ or ${\cal U}({J}_l) = e^{i\phi_{lm}}J_m$.  
For any fixed $l$, unitarity of $\mathbf{U}$ means that is not possible that $\mathbf{U}_{lm}=0$ for all $m$ so there must be some $m,\phi_m$ such that $\mathcal{U}(J_l) = J_{m}e^{i\phi_m}$.  This holds for all $l$ so there must be a permutation $\pi$ such that
$\mathcal{U}(J_j) = J_{\pi(j)}e^{i\phi_j} \forall j$,
 as in 
Symmetry Condition~III.  Hence one also has $\sum_j \mathcal{U}(J_j^\dag) \mathcal{U}(J_j) = \sum_j J_j^\dag J_j $ which combined with Eq.~\eqref{equ:HeHe-deph} gives ${\cal U}(H)=H$.  Hence we have shown that Eq.~\eqref{eq:symm-deph-rot} implies Eq.~\eqref{eq:symm_cond_III}, which is Symmetry Condition~III, the relevant permutation is determined by the matrix $\bm{U}$ which appears in the definition of $\mathbb{U}$, via Eq.~\eqref{eq:UE}.

\subsection{Partial dephasing and Symmetry Condition~II}\label{app:partial_deph}

We show that the partially-dephased cMPO has a symmetry Eq.~\eqref{eq:LP-symm_t} if and only if Symmetry Condition~II holds (see Sec.~\ref{sec:cond-partial}).  The argument is similar to Appendix~\ref{app:full_deph}.  In particular, the argument of Appendix \ref{app:UEt-UE} still applies if $d\mathbb{L}_\tau$ is replaced by $d\mathbb{L}_\tau^\textsc{p}$.  That means that 
$\UUE=\UUE_t$ is necessary for Eq.~\eqref{eq:LP-symm_t} so we only need to consider symmetries of the form Eq.~\eqref{eq:LP-symm}.
Then expand $\UUE \, \dL_t^\textsc{p} \, \UUE^\dag$ as
\beqOpt
    \UUE \, d\mathbb{L}_t^\textsc{p} \, \UUE^\dag(R_t^\textsc{p}) = -i \Big[(\mathcal{U}(H_\text{eff})\otimes\IE)\deph^\textsc{p}_t
    %\\
    -\deph^\textsc{p}_t(\mathcal{U}(H_\text{eff}^\dag)\otimes\IE)\Big] dt 
    %\\ 
    + \sum_\alpha\sum_{m,n}\sum_{j,k\in S_{\pic(\alpha)}} \left[ \hat{\mathbb{J}}_{j,m,t} \, R_t^\textsc{p}
    \, \hat{\mathbb{J}}_{k,n,t}^\dag \right] % \left(\mathcal{U}(J_{k}^\dag)\otimes dB_{n,t}\right)
    \mathbf{U}^*_{jm}\mathbf{U}_{kn}
    \label{equ:LP-ugly}
\eeqOpt
where $\mathbf{U}$ is the unitary matrix appearing in the environmental symmetry operator Eq.~\eqref{eq:UE} and we have introduced (only in this Appendix) the shorthand $\hat{\mathbb{J}}_{j,m,t} = \mathcal{U}(J_j)\otimes dB_{m,t}^\dag$, for compactness of notation.

\subsubsection{Symmetry Condition~II is sufficient}

Symmetry Condition~II implies Symmetry Condition~\IIb, which further implies the existence of a matrix $\mathbf{U}$ with the properties (\ref{eq:not_cond_IIb},\ref{eq:UU-partial-lemma}).  We use this matrix in Eq.~\eqref{eq:UE}.  Note that
Eq.~\eqref{eq:symm_cond_II_to_Ia} also follows from Symmetry Condition~II: using this with Eq.~\eqref{eq:UU-partial-lemma}, we simplify Eq.~\eqref{equ:LP-ugly} as
\beqOpt
    \UUE d\mathbb{L}_t^\textsc{p} \UUE^\dag(R_t^\textsc{p}) = -i \left[ (H_\text{eff}\otimes\IE) \deph^\textsc{p}_t-\deph^\textsc{p}_t(H_\text{eff}^\dag\otimes\IE)\right] dt 
 %   \\
     + \sum_\alpha\!\sum_{m,n \in S_\alpha }\! %\left[
    \left( J_m \otimes dB_{m,t}^\dag\right)R_t^\textsc{p}\Big(J_n^\dag\otimes dB_{n,t}\Big) 
       .
\eeqOpt
The right hand side coincides with $d\mathbb{L}_t^\textsc{p}$, which verifies Eq.~\eqref{eq:LP-symm}, as required.

\subsubsection{Symmetry Condition~II is necessary}

We now show that Eq.~\eqref{eq:LP-symm} implies Symmetry Condition~II.
%where again $\mathbf{U}$ is the unitary matrix appearing in \eqref{eq:UE}.
Eq.~\eqref{eq:LP-symm} states that Eq.~\eqref{equ:LP-ugly} is equal to $d\mathbb{L}_t^\textsc{p}(R_t^\textsc{p})$. Comparing the terms which act trivially on the vacuum we obtain 
\beq
\mathcal{U}(\He)=\He
\label{equ:HeHe-partial}
\eeq
Comparing the remaining terms we obtain
\beqOpt
     \sum_\alpha\sum_{m,n}\sum_{j,k\in S_\alpha}\left[ \hat{\mathbb{J}}_{j,m,t} \, R_t^\textsc{p}
    \, \hat{\mathbb{J}}_{k,n,t}^\dag \right]
 \mathbf{U}^*_{jm}\mathbf{U}_{kn} 
 % \\
    = \sum_\alpha\sum_{m,n\in S_\alpha}\left[\left(J_m\otimes dB_m^\dag\right)R_t^\textsc{p}\left(J_n^\dag\otimes dB_n\right)\right] 
\eeqOpt
which can be expressed as
\begin{multline}
    0 = \sum_{\beta,\gamma}\sum_{m\in S_\beta}\sum_{n\in S_\gamma}(\mathbb{1}\otimes dB_{m,t}^\dag)
    %\\ \times 
    \Bigg[\sum_\alpha\sum_{j,k\in S_\alpha}\left(\mathcal{U}(J_j)\otimes\IE\right)R_t^\textsc{p}(\mathcal{U}(J_k^\dag)\otimes\IE)\mathbf{U}_{jm}^*\mathbf{U}_{kn} 
    \\
    -\delta_{\beta\gamma}\left(J_m\otimes\IE\right)R_t^\textsc{p}\left(J_n^\dag\otimes\IE\right)\Bigg]\left(\mathbb{1}\otimes dB_{n,t}\right)
\end{multline}
As this holds for all $R_t^\textsc{p}$, the object in square brackets must vanish for all $m,n$. 
 Then taking %\emph{and the $dB$'s are orthogonal (?)},
$R_t^\textsc{p}=\psi\otimes\IE$ %/\text{dim}(\mathcal{H}_E)$
 where $\psi$ is a pure density matrix, this reduces to
\begin{equation}\label{eq:simplified_partial}
    0 = \sum_\alpha\sum_{j,k\in S_\alpha} \mathcal{U}(J_j)\psi\mathcal{U}(J_k^\dag)\mathbf{U}_{jm}^*\mathbf{U}_{kn} - \delta_{\beta_m\gamma_n}J_m\psi J_n^\dag
\end{equation}
where the symbol $\delta_{\beta_m\gamma_n}$ is defined as $\delta_{\beta_m\gamma_n}=1$ if $m,n$ come from the same SJED and $\delta_{\beta_m\gamma_n}=0$ otherwise.  

Taking $m=n$, we obtain the super-operator equation
\beq
\label{equ:JKa}
\sum_\alpha K_{\alpha,m}(\cdot)K_{\alpha,m}^\dag  
=  
J_m (\cdot) J_m^\dag 
\eeq 
with
$K_{\alpha,m} = \sum_{j\in S_\alpha}  
\mathbf{U}_{jm}^* {\cal U}(J_j)
%\mathbf{U}_{kn}
$
These are representations of a completely positive super-operator so the operators $J_m$ and $\{K_{\alpha,m}\}_\alpha$ are related by an isometry, which means here that either $K_{\alpha,m}=0$ or $K_{\alpha_m}\propto J_m$.  By the definition of SJEDs, it cannot be that $K_{\alpha,m}\propto K_{\alpha',m}$ for $\alpha\neq\alpha'$.  Hence there is only one $\alpha$ for which $K_{\alpha,m}\neq0$.  The structure of the SJEDs means that this is $\alpha$ is the same for all $m$ within a given SJED.  Using that Eq.~\eqref{equ:JKa} holds for all $m$, this yields
\begin{equation}\label{eq:partial_perm}
    \sum_{j,k\in S_\alpha}\mathcal{U}(J_j)(\cdot) \mathcal{U}(J_k^\dag)\mathbf{U}_{jm}^*\mathbf{U}_{km} =
    \begin{cases} J_m (\cdot) J_m^\dag \; &  m\in S_{\pic(\alpha)}
    \\ 0 & \text{otherwise} \end{cases}
\end{equation}
where $\pi$ is a permutation of $\{1,2,\dots,\dC\}$.
Finally sum over $m$ (recalling that $\sum_m \mathbf{U}_{jm}^*\mathbf{U}_{km}=\delta_{jk}$), which yields
\beq\label{eq:final_partial_app}
\sum_{j\in S_\alpha}\mathcal{U}(J_j)(\cdot) \mathcal{U}(J_j^\dag) = \sum_{m\in S_{\pic(\alpha)}} J_m(\cdot) J_m^\dag \; .
\eeq
This says that ${\cal U A}_\alpha{\cal U}^\dag = {\cal A}_{\pic(\alpha)}$.  
Summing Eq.~\eqref{eq:final_partial_app} over $\alpha$ and taking the trace yields the operator equation $\sum_j \mathcal{U}(J_j^\dag) \mathcal{U}(J_j) = \sum_j J_j^\dag J_j $, which combined with Eq.~\eqref{equ:HeHe-partial} gives 
${\cal U}(H)=H$.  Hence we have shown that Eq.~\eqref{eq:LP-symm} implies Eq.~\eqref{eq:symm_cond_II}, which is Symmetry Condition~II, as required.

\subsection{Coarse-{graining and Symmetry Condition~II}}\label{app:coarse_deph}

We prove the conditions given in \ref{sec:cond-cg} for symmetry of the coarse-grained cMPO.

\subsubsection{Time independence of symmetry}

We define time-dependent unitary (super)-operators for the coarse-grained cMPS as
\begin{align}
\UEt^{\textsc{c}}  = D_t^\textsc{c} \UE^\textsc{c} (D_t^\textsc{c})^\dag, %\qquad
\qquad 
%\nonumber\\ 
\mathcal{U}_{E,t}^\textsc{c} (\cdot)  = \UEt^\textsc{c}(\cdot) (\UEt^\textsc{c})^\dag ,
\qquad 
%\nonumber\\ 
\mathbb{U}_t^\textsc{c}  = \mathcal{U}\otimes\mathcal{U}_{E,t}^\textsc{c}, 
\label{equ:UC-defs}
\end{align}
where
\begin{equation}
    D_t^\textsc{c} = \exp\left[-\int_0^t \frac{1}{\dS}\sum_\alpha\left(c_\alpha dC_{\alpha,t}^\dag - c_\alpha^* dC_{\alpha,t}\right)\right]
\end{equation}
is a coherent displacement with $\bm{c}\in \mathbb{C}^{d_c}$.  (The dependence on the parameters $c_\alpha=(\bm{c})_\alpha$ is implicit.)
A time-dependent symmetry of the coarse-grained cMPO is 
\begin{equation}
    \mathbb{U}_{t}^\textsc{c}\left(\mathcal{T}e^{-i\int_{t'}^t d\mathbb{L}_\tau^\textsc{c}}\right){\mathbb{U}_{t}^\textsc{c}}^\dag = \mathcal{T}e^{-i\int_{t'}^t d\mathbb{L}_\tau^\textsc{c}}
\end{equation}
Repeating the arguments of Appendix~\ref{app:UEt-UE} shows that this symmetry can only occur if $\UEt^\textsc{c} = \UE^\textsc{c}$, that is, all symmetries are stationary, and described by Eq.~\eqref{eq:symm-LC}.

\subsubsection{Necessary and sufficient conditions for symmetry of coarse-grained cMPO}

We show that a symmetry Eq.~\eqref{eq:symm-LC} holds 
if and only if Symmetry Condition~II is satisfied.  
First, use (\ref{eq:dLc},\ref{eq:UEC_V},\ref{equ:UC-defs}) to obtain [cf.~\eqref{eq:L-ugly}]
\begin{multline}\label{eq:UdLcU}
    [\UUE^\textsc{c} d\mathbb{L}_t^\textsc{c} {\UUE^\textsc{c}}^\dag](R_t^\textsc{c}) =  -i 
    \Big[(\mathcal{U}(H_\text{eff})\otimes\IE)\deph^\textsc{p}_t
   % \\ 
    -\deph^\textsc{p}_t(\mathcal{U}(H_\text{eff}^\dag)\otimes\IE)\Big] dt 
    \\ 
    + \sum_\alpha\sum_{k\in S_\alpha}\sum_{\beta,\gamma} (\mathbf{U}^{\textsc c}_{\alpha\beta})^*\mathbf{U}^{\textsc c}_{\alpha\gamma}
    (\mathcal{U}(J_k)\otimes dC_{\beta,t}^\dag)
    R_t^\textsc{c}
    (\mathcal{U}(J_k)\otimes dC_{\gamma,t})
\end{multline}
where $\mathbf{U}^{\textsc c}$ is the matrix appearing in the environmental symmetry operator Eq.~\eqref{eq:UEC_V}.
To show
that Symmetry Condition~II is sufficient for a symmetry of the form Eq.~\eqref{eq:symm-LC}, we choose the matrix $\mathbf{U}^{\textsc c}$ in terms of the permutation appearing in Eq.~\eqref{eq:symm_cond_II}, as 
\beq
\mathbf{U}^{\textsc c}_{\alpha\beta} = \delta_{\beta,\pic(\alpha)} 
\label{eq:Vc-simple}
\eeq
so that $\superUE(dC_{\alpha,t}) = dC_{\pic(\alpha),t} $ by Eq.~\eqref{eq:UEC_V}.
Symmetry Condition~II also implies Eq.~\eqref{eq:symm_cond_II_to_Ia} so Eq.~\eqref{eq:UdLcU} becomes
\beqOpt
    [\UUE^\textsc{c} d\mathbb{L}_t^\textsc{c} {\UUE^\textsc{c}}^\dag](R_t^\textsc{c}) =  
    -i \left[(H_\text{eff}\otimes\IE)\deph^\textsc{p}_t-\deph^\textsc{p}_t(H_\text{eff}^\dag\otimes\IE)\right] dt 
  %  \\ 
    + \sum_\alpha\sum_{k\in S_\alpha}
    (\mathcal{U}(J_k)\otimes dC_{\pic(\alpha),t}^\dag)
    R_t^\textsc{c}
    (\mathcal{U}(J_k)\otimes dC_{\pic(\alpha),t})
\eeqOpt
Relabelling the sum over $\alpha$ recovers $d\mathbb{L}_t^\textsc{c}$ as in Eq.~\eqref{eq:dLc} so indeed Symmetry Condition~II is sufficient for the symmetry.

The remainder 
of this section shows that Symmetry Condition~II is necessary for  Eq.~\eqref{eq:symm-LC}.  That is,
we suppose that the RHS of Eq.~\eqref{eq:UdLcU} coincides with $d\mathbb{L}_t^\textsc{c}$ from Eq.~\eqref{eq:dLc} and we derive Eq.~\eqref{eq:symm_cond_II}.
Note that Eq.~\eqref{eq:UdLcU} strongly resembles Eq.~\eqref{eq:L-ugly} and the following derivation follows that of Appendix~\ref{app:full_cond_necessary}, where the role of $\mathbf{U}$ is here played by $\mathbf{U}^{\textsc c}$.

Equating $d\mathbb{L}_t^\textsc{c}$ with $\UUE^\textsc{c} d\mathbb{L}_t^\textsc{c} {\UUE^\textsc{c}}^\dag$ and comparing
the terms which act trivially on the environment we obtain that 
\beq
\mathcal{U}(\He) = \He.
\label{equ:HeHe-coarse}
\eeq
Comparing the remaining terms, we obtain
\beqOpt
    0 = \sum_{\alpha,\beta,\gamma} \sum_{k\in S_\alpha}
    %\sum_{\beta,\gamma}
     [\mathcal{U}(J_k)\otimes dC_{\beta,t}^\dag]R_t^\textsc{c}[\mathcal{U}(J_k)\otimes dC_{\gamma,t}](\mathbf{U}^{\textsc c}_{\alpha\beta})^*\mathbf{U}^{\textsc c}_{\alpha,\gamma} 
    %\\ 
    - \sum_\sigma\sum_{j\in S_\sigma} (J_j\otimes dC_{\sigma,t}^\dag)R_t^\textsc{c}(J_j^\dag\otimes dC_{\sigma,t}) \; .
\eeqOpt
analogous to Eq.~\eqref{eq:dephased_symm_condition}.
Repeating the steps between Eq.~\eqref{eq:dephased_symm_condition} to Eq.~\eqref{eq:U_from_dephased} we obtain
\beqOpt
    0 = \Bigg[
    \sum_{k\in S_\sigma}(\mathcal{U}(J_k)\otimes\IE) R_t^\textsc{c}(\mathcal{U}(J_k)\otimes\IE)
   %\\ 
   - \sum_{j\in S_\beta}(J_j\otimes\IE)R_t^\textsc{c}(J_j^\dag\otimes\IE)
    \Bigg]
    (\mathbf{U}^{\textsc c}_{\sigma\beta})^* \; ,
\eeqOpt
which is analogous to Eq.~\eqref{eq:U_from_dephased}.
Again, similar to Appendix \ref{app:full_cond_necessary}, it must be that either $\mathbf{U}^{\textsc c}_{\sigma\beta}=0$, or the term in square brackets vanishes.  Using Eq.~\eqref{eq:superA} and that all terms in this bracket act trivially on the environment, we obtain 
\begin{equation}\label{eq:CG_J_relation}
   \mathbf{U}^{\textsc c}_{\sigma\beta}=0 \quad \text{or} \quad   \mathcal{U}{\cal A}_\sigma \mathcal{U}^\dag = {\cal A}_\beta  \,.
\end{equation}
For any fixed $\sigma$, unitarity of $\mathbf{U}^{\textsc c}$ means that
it is not possible that $\mathbf{U}^{\textsc c}_{\sigma\beta}=0$ for all $\beta$, so there must exist $\beta$ such that $ \mathcal{U}{\cal A}_\sigma  \mathcal{U}^\dag = {\cal A}_\beta$.  The definition of the SJEDs means that each ${\cal A}_\alpha$ is distinct, so this $\beta$ is unique.
This argument applies for all $\sigma$ so there must exist a permutation $\pic$ such that
\begin{equation}\label{eq:CG_J_relation2}
    \mathcal{U}{\cal A}_\alpha \mathcal{U}^\dag = {\cal A}_{\pic(\alpha)}
\end{equation}
Together with Eq.~\eqref{equ:HeHe-coarse}, 
this is Symmetry Condition~II Eq.~\eqref{eq:symm_cond_II}: we have shown that this is necessary for the symmetry Eq.~\eqref{eq:symm-LC}, as required.

As a final remark, note that combining Eq.~\eqref{eq:CG_J_relation} with Eq.~\eqref{eq:CG_J_relation2} shows that $\mathbf{U}^{\textsc c}_{\alpha\beta}=0$ unless $\beta=\pic(\alpha)$.  Then unitarity of $\mathbf{U}^{\textsc c}$ means that $\mathbf{U}^{\textsc c}_{\alpha,\beta} =  \delta_{\beta,\pic(\alpha)} e^{i\phi_\alpha}$, similar to Eq.~\eqref{eq:Vc-simple} but now with arbitrary phases $\phi_1,\dots,\phi_{\dC}\in \mathbb{R}$.  Hence by Eq.~\eqref{eq:UEC_V}
\begin{equation}\label{eq:dC_perm}
    \superUE^\textsc{c}(dC_{\alpha,t})=dC_{\pic(\alpha),t} e^{i\phi_\alpha}
\end{equation}
which is Eq.~\eqref{eq:UEC} of the main text.
This means that in addition to Symmetry Condition~II, it is also necessary for Eq.~\eqref{eq:symm-LC} that the environmental operator $\superUE$ acts to permute the emitted quanta and multiply them by phases; the relevant permutation is the same one that appears in Eq.~\eqref{eq:symm_cond_II}.

% one or the other of these
\bigskip\hrule\bigskip
%\newpage

\twocolumngrid

\section{Technical details for implications of symmetric unravelled dynamics}\label{app:implications}

This Appendix provides additional technical material to supplement Sec.~\ref{sec:implications}.

\subsection{SJED wave operators and translational symmetry}\label{app:wave+trans}

We discuss how the SJED composite action operators $\cgo_\alpha$ can used to construct eigenoperators of the symmetry (``wave operators''). We then describe how these operators are supported on symmetry eigenspaces, which gives insight into the structure of the $\cgo_\alpha$.

\subsubsection{SJED wave operators}\label{app:wave}

For simplicity, we consider the (simple) case where the permutation $\pic$ in Eq.~\eqref{eq:symm_cond_II} is a single cycle of length $\dC$.
Then the jump superoperators are 
\begin{equation}
{\cal A}_{\alpha+1} = {\cal A}_{\pic^{\alpha}(1)}
\label{eq:perm-cycle}
\end{equation} 
for $\alpha=0,\dots,\dC-1$, where $\pic^{\alpha}$ denotes $\alpha$ operations of the permutation.

We define a set of superoperators by ``Fourier-transforming along the permutation'':
\begin{equation}\label{eq:X_wave}
    \cgo^{(k)}_\text{wave}
    = 
    \sum_{j=0}^{\dC-1}
    \cgo_{\pic^j(1)}
    e^{-2 i \pi k  j/\dC}
\end{equation}
for $k=0,\, ..., \, \dC-1$.
Note that the inverse transform of Eq.~\eqref{eq:X_wave} is
\begin{equation}
	\cgo_{\pic^j(\alpha)} 
 = \frac{1}{\dC}\sum_{k=0}^{\dC-1}e^{2 i \pi k/\dC}\cgo^{(k)}_\text{wave} \; .
 \label{eq:X-phases}
\end{equation}

The construction of $\cgo^{(k)}_\text{wave}$ ensures that
\begin{equation}
{\cal U} \cgo^{(k)}_\text{wave} {\cal U}^\dag= {\rm e}^{2 i \pi k/\dC} \cgo^{(k)}_\text{wave} 
\label{eq:eigen-Jwave}
\end{equation} 
We recognise ${\rm e}^{2 i \pi k/\dC} $ as an eigenvalue of the (superoperator-valued) map ${\cal U}(\cdot){\cal U}^\dag$.
This has implications for matrix representations of $\cgo^{(k)}_\text{wave}$.

Writing Eq.~\eqref{eq:eigen-Jwave} as 
\begin{equation}
(U \otimes U^*)  \Lambda[\cgo^{(k)}_\text{wave}] (U \otimes U^*)^\dag = {\rm e}^{-2 i \pi k/\ell}\Lambda[\cgo^{(k)}_\text{wave}],
\label{eq:eigen-wave-liouv}
\end{equation} 
and comparing with Eq.~\eqref{eq:symm-liouv}, the additional phase factor means that the matrix representation of  $\cgo^{(k)}_\text{wave}$ has different support from that of ${\cal L}$ (which we recall was supported on diagonal blocks in the symmetry-adapted basis, see also below).
To illustrate this in a simple scenario,
we now consider the case of a translational symmetry.

\subsubsection{Translational symmetry}\label{app:trans}

In addition to Eq.~\eqref{eq:perm-cycle}, we now assume that the permutation $\pic$ in that equation is directly related to the symmetry operation $U$.  Specifically, we assume that the eigenvalues of $U$ are $e^{i\phi_j}$ with $\phi_j=2\pi j/\dC$ and all its eigenspaces have the same degeneracy.  This situation arises naturally if one considers a chain of $\dC$ qubits with periodic boundaries, where $U$ corresponds to a translation along the chain, and the system is symmetric under these translations.  
Write ${\cal H}_j$ for the eigenspace of $U$ with eigenvalue $e^{i\phi_j}$.
The corresponding eigenspaces of ${\cal U}$  are ${\cal H}_j \otimes {\cal H}_{j'}$ and the corresponding eigenvalue is $e^{i\delta_k}$ with $\delta_k=\phi_j-\phi_{j'}$.  Note that $e^{i\delta_k}$ is also a $\dC$th root of unity. 
Similarly the eigenvalues of the map ${\cal U}(\cdot){\cal U}^\dag$ are $\dC$th roots of unity, specifically $e^{i\Delta_{kk'}}$ with $\Delta_{kk'}=\delta_k - \delta_{k'}$.

Working in the Liouville representation with the symmetry adapted basis, the condition
%We have 
$[{\cal L},{\cal U}]=0$, means that $\Lambda[\mathcal{L}]$ is supported on the diagonal blocks.
This structure is shown below, where each block is labelled by its $\delta_k$:
\definecolor{deltacolour}{RGB}{19, 115, 232}
\newcommand{\dc}[1]{{\color{deltacolour}#1}}
\begin{equation}
	\frac{2\pi}{\dC}\begin{pmatrix}
	\dc{0}& & & \\
	 &\dc{1}& & \\
	 & & \dc{\ddots} &  \\
	   & & & \dc{\dC-1}
	\end{pmatrix}
 \label{eq:diag-liou}
\end{equation}

Since ${\cal U}$ is diagonal in this basis, it is easily checked that for any superoperator ${\cal O}$, the corresponding  ${\cal U}{\cal O}{\cal U}^\dag$ is obtained by multiplying each of its blocks by a suitable
eigenvalue of ${\cal U}(\cdot){\cal U}^\dag$, for example ${\rm e}^{i\Delta_{kk'}}$.  In the case $\dC=4$, all blocks of a generic Liouville matrix can be labelled as
\definecolor{Deltacolour}{RGB}{155, 66, 245}
\newcommand{\Dc}[1]{{\color{Deltacolour}#1}}
\begin{equation}
    \frac{\pi}{2} \begin{pmatrix}
        \dc{0} & \Dc{1} & \Dc{2} & \Dc{3} \\
        \Dc{3} & \dc{1} & \Dc{1} & \Dc{2} \\
        \Dc{2} & \Dc{3} & \dc{2} & \Dc{1} \\
        \Dc{1} & \Dc{2} & \Dc{3} & \dc{3} \\
    \end{pmatrix}
    \label{eq:d=4mat}
\end{equation}
where blue numbers give the value of $\delta$ on each diagonal block and purple the value of $\Delta$, for each off-diagonal block. The diagonal blocks have $\Delta=0$, this is not indicated.  We take $\Delta\in[0,2\pi)$ which is always possible since only $e^{i\Delta}$ enters the calculations.

It follows that any eigenmatrix of ${\cal U}(\cdot){\cal U}^\dag$ with eigenvalue $e^{i\Delta_k}$ is supported on blocks with that value of $\Delta_k$.  Since ${\cal L}$ and ${\cal U}$ are both eigenmatrices with eigenvalue unity, they are supported on the diagonal blocks.
From Eq.~\eqref{eq:eigen-Jwave}, $\cgo^{(k)}_\text{wave}$ has eigenvalue $e^{i\Delta_k}$ with $\Delta_k=2\pi k/\dC$ so it is supported on the $k$th and $(\dC-k)$th diagonals in this representation.  Then, Eq.~\eqref{eq:X-phases} shows that each $\cgo_\alpha$ has support on the combined support of all the $\cgo_\text{wave}$ operators (including all blocks in general), and the various $\cgo_\alpha$ differ by block-dependent phase factors.
When computing $\Lambda[{\cal L}]$, the off-diagonal contributions from different $\cgo_\alpha$ all cancel each other, leaving a (block)-diagonal result, as required.

For the example with translational symmetry given in Sec.~\ref{sec:example_translate}, the
eigenspaces of the translation operator $U_T$ are
\begin{equation}
    \mathcal{H}_j = \left\{\frac{1}{\sqrt{L}}\sum_{k=0}^{L-1}e^{-2i\pi j k/L}U_T^j|\psi\rangle : |\psi\rangle\in\mathcal{H}\right\} %, \quad j = 0,1,\dots,L-1
\end{equation}
where $\mathcal{H}$ is the total Hilbert space of the system. The corresponding eigenvalues are $e^{2i\pi j/L}$. The model exhibits the same support of $\Lambda[\mathcal{L}]$ and $\mathcal{A}^{(k)}_\text{wave}$ as described above (with $\dC = L$).

\subsection{Action of ${\cal U}$ on the Choi matrix}
\label{app:Choi}

We first derive Eq.~\eqref{eq:choi_symm} of the main text, before commenting on the structure of Choi matrices in the symmetry-adapted basis.
The matrix elements of the Choi matrix are 
\begin{equation}
    (C[\Phi])_{mn,kl} = (\langle\psi_m| \otimes \langle\psi_n|) \, C[\Phi] \, (|\psi_k\rangle \otimes |\psi_l\rangle) \,.
\end{equation}
Combined with Eq.~\eqref{eq:choi} this gives
\begin{equation}
(C[\Phi])_{mn,kl} = (\Lambda[\Phi])_{mk,nl} \,.
\label{eq:C-Lambda}
\end{equation}
By Eq.~\eqref{eq:Lambda}, Liouville matrices are a representation for the algebra of superoperators, so $\Lambda[{\cal U}\Phi{\cal U}^\dag] = \Lambda[{\cal U}] \Lambda[\Phi] \Lambda[{\cal U}^\dag]$.  Then using Eq.~\eqref{eq:liouv-UU*} 
and the convention $(A\otimes B)_{ab,pq}=A_{ap}B_{bq}$
gives
\begin{align}
(\Lambda[{\cal U}\Phi{\cal U}^\dag])_{mk,nl} 
%& = \sum_{ab,pq} (U \otimes U^*)_{mk,ab}  \Lambda[\Phi]_{ab,pq} (U \otimes U^*)^\dag_{pq,nl}
%\nonumber\\
& =  \sum_{ab,pq}  U_{ma} U^*_{kb}  \Lambda[\Phi]_{ab,pq} U^*_{np} U_{lq} \,.
\label{eq:uu1}
\end{align}
Using Eq.~\eqref{eq:C-Lambda} and changing the order of some multiplicative factors, one obtains
\begin{align}
(C[{\cal U}\Phi{\cal U}^\dag])_{mn,kl} 
& = \sum_{ab,pq}  U_{ma} U^*_{np} C[\Phi]_{ap,bq} U^*_{kb} U_{lq} 
%\nonumber\\
%& =   \sum_{ab,pq}  (U \otimes U^*)_{mn,ap} C[\Phi]_{ap,bq}  (U \otimes U^*)^\dag_{bq,kl}
\label{eq:uu2}
\end{align}
which is the same as Eq.~\eqref{eq:choi_symm}, as required.

We now comment briefly on the structure of Choi matrices for $\cgo_\text{wave}^{(k)}$ superoperators. It follows from Sec.\ref{sec:choi} that $C[{\cal L}]$ is supported on the same diagonal blocks shown in Eq.~\eqref{eq:diag-liou}.
Similarly, for Eq.~\eqref{eq:eigen-wave-liouv}, we have that
\begin{equation}
(U \otimes U^*)  C[\cgo^{(k)}_\text{wave}] (U \otimes U^*)^\dag = {\rm e}^{-2 i \pi k/\ell}C[\cgo^{(k)}_\text{wave}],
\label{eq:eigen-wave-choi}
\end{equation} 
Hence, $C[\cgo_{\rm wave}^{(k)}]$ is supported on specific off-diagonal blocks as in Eq.~\eqref{eq:d=4mat}, and the action of translation symmetry on $\cgo_\alpha$ corresponds to mutiplication by block-dependent phases.

\subsection{Relation to weakly symmetric representation}\label{app:relation_weakly_symm}

We now discuss how, from a representation which satisfies Symmetry Condition~III, we can build a weakly symmetric representation by Fourier-like transformations of the jumps. We make the following assumptions, which are sufficient to illustrate the ideas:
assume (i)~$U$ is a permutation (translation) with exactly $n$ eigenvalues given by $\exp(2 i \pi k/n)$; 
(ii)~there is a cycle of $n$ jump operators which obey this translation symmetry, specifically:
\begin{equation}\label{init_rep}
	\mathcal{U}(J_k) = J_{k+1} e^{i\delta_k}
\end{equation}
so that
the action of the symmetry on each jump produces the next jump in the cycle up to an arbitrary phase $e^{i\delta_k}$; (iii) these are the only jumps, in particular each jump is an SJED on its own.
%I.e. we are in a canonical representation.
Under these conditions, we show that construction of ``jump waves'' by Fourier corresponds to transforming to the symmetric representation.

First note that
\begin{equation}
	\mathcal{U}^n(J_k) = J_k {\rm e}^{i\sum_j \delta_j}
\end{equation}
Hence, $\exp(i\sum_j \delta_j/n)$ is an eigenvalue of $\mathcal{U}$ and therefore $\sum_j \delta_j = 2 i \pi l$ for some $l\in \mathbb{Z}$.
We build a representation that respects Symmetry Condition III in the form $\mathcal{U}(\tilde J_k) = \tilde J_{k+1}$. Let $\tilde J_k = J_k e^{i\beta_k}$: then $\mathcal{U}(J_k') = J_{k+1} e^{i(\delta_k+\beta_k - \beta_{k+1})}$. Hence, we need $\delta_k+\beta_k - \beta_{k+1}=0$ for all $k$; this can be achieved because $\sum_k \delta_k = 2 i \pi l$. (One obtains that this condition is necessary by summing $\delta_k = \beta_k-\beta_{k+1}$ over all $k$ with $\beta_{k+n} = \beta_k$.) Taking $\beta_1=0$, %\cb{\emph{[add step?]}}
we obtain $\beta_k = \sum_{j=1}^{k-1} \delta_j$. Hence, we can indeed define new jumps
\begin{equation}
	\tilde J_k = \exp\left(i\sum_{j=1}^{k-1} \delta_j\right) J_k
\end{equation}
such that $\mathcal{U}$ acts as a pure translation as required above (without phases).

By a process analogous to Fourier transformation on these new jumps we define ``wave operators'' (directly in terms of the original $J_k$'s):
\begin{equation}
    \hat{J}_l = \sum_k e^{-2 i \pi l k/n} e^{i\sum_{j=1}^{k-1} \delta_j}J_k.
    \label{equ:J-fourier}
\end{equation} 
The transformation from the $J_k$ to the $\hat{J}_l$ is unitary so these new $\hat{J}_l$ define a representation, which is in fact the symmetric representation. It is easily checked that ${\cal U}(\hat{J}_l)=\hat{J}_l e^{{2 i \pi l/n}}$ which establishes the weak symmetry.  The unitary transformation matrix $\textbf{U}$ is obtained by identifying Eq.~\eqref{equ:J-fourier} with $\hat{J}_l = \sum_k \textbf{U}_{lk} J_k$. 

In the Liouville representation $\hat{J}_l$ is a vector with support on a single eigenspace of $\mathcal{U}$. The original $J_k$ are vectors with full support, they correspond to adding up the various $\hat{J}_l$'s with suitable ($k$-dependent) phase factors.

For the spin-chain example given in Sec.~\ref{sec:example_translate}, the jumps \eqref{eq:trans_jumps} obey Symmetry Condition~III when $\theta_\alpha=\Theta\;\forall\alpha$. In this case, no phase appears in the analogue of Eq.~\eqref{eq:reset_SJED_transform} (that is $\delta_j=0$) and hence
\begin{equation}
    \hat{J}_l^{(x)} = \sum_k e^{\frac{-2 i \pi l k}{L}} J_k^x
\end{equation}
for $x=1,2$. %(for each cycle).

\end{appendix}

\twocolumngrid
\bibliography{elusive_symmetry}

\end{document}